\documentclass[aps,prb,onecolumn,longbibliography,nofootinbib,citeautoscript,10pt]{revtex4-2}

\pdfoutput=1
\usepackage[caption=false]{subfig}
\usepackage{array}
\usepackage{slashed,bbold}
\usepackage{physics}
\usepackage{dsfont}
\usepackage{amsmath,amssymb,bm} 
\usepackage{graphicx}
\usepackage{soul}
\usepackage{xcolor}
\usepackage[papersize={8.5in,11in}]{geometry}

\usepackage{color}
\definecolor{darkblue} {rgb} {0.,0.,0.4}
\definecolor{darkred} {rgb} {0.5,0.,0.}
\definecolor{BlueViolet} {RGB} {138,43,226}
\definecolor{SkyBlue} {RGB} {30,144,255}
\definecolor{DarkGreen} {RGB} {0,100,0}
\usepackage[pdftex,colorlinks=true,linkcolor=darkblue,citecolor=blue,urlcolor=darkred]{hyperref}

\def\*#1{\boldsymbol{#1}} 
\def \nn{\nonumber \\}
 
\def\nn{\nonumber \\}

\newcommand\pmtx[1]{\begin{pmatrix}#1\end{pmatrix}} 

\def \be{\begin{equation}}
\def \ee{\end{equation}}
\def \nn{\nonumber \\}

\begin{document}
	
\title{Floquet scattering and Fano resonances in nodal-ring and multi-Weyl semimetals: Role of propagating and evanescent modes} 

\author{Sandip Bera}
\affiliation{Department of Physics, University of Toronto, 60 St. George Street, Toronto, Ontario, Canada M5S 1A7}	 

\author{Ipsita Mandal}
\email{ipsita.mandal@snu.edu.in}
\affiliation{Department of Physics, Shiv Nadar Institution of Eminence (SNIoE), Gautam Buddha Nagar, Uttar Pradesh 201314, India}

\begin{abstract}
We develop a comprehensive Floquet scattering theory for quantum transport in periodically driven nodal-ring semimetals (NRSs) and multi-Weyl semimetals (mWSMs), extending our earlier study reported in Annalen der Physik 535, 2200460 (2023)~\cite{bera2023}, in which evanescent modes were neglected, to a complete formalism that includes both propagating and evanescent channels. By solving the full boundary-value problem, we obtain the complete set of scattering states and show that, although evanescent modes are indispensable for satisfying the matching conditions at the potential interfaces, they carry zero net probability current and do not contribute to any observable transport quantity. We identify a previously unexplored transport regime in NRSs in which two propagating channels coexist and participate in coherent scattering, producing multi-channel quantum interference and Floquet-induced Fano resonances. The transmission, reflection, pumped shot noise, and the associated Fano resonance features are determined entirely by the propagating channels, and the resonance energies coincide with those of the corresponding quasi-bound states of the static potential well. Our results establish a unified framework for Floquet transport in anisotropic topological semimetals, and confirm that the approximation of neglecting evanescent modes in our earlier work is justified for all measurable transport properties.
\end{abstract}

\maketitle

\tableofcontents

\section{Introduction}
\label{secintro}

Quantum pumping has emerged as a powerful technique for inducing directed charge or spin transport in the absence of an external bias, particularly in nanoscale and mesoscopic systems. The concept was first proposed by Thouless in 1983~\cite{thouless83},  who showed that quantised electron transport takes place under a slowly varying adiabatic potential. It was first realised experimentally, a few years later, in quantum dot systems by Gossard and collaborators~\cite{switkes99}. Quantum pumping protocols have since been implemented on several platforms in mesoscopic systems, including quantum dots driven by microwave excitations~\cite{oosterkamp98},  gigahertz pumping~\cite{blumenthal07},  and time-dependent gate voltages~\cite{dicarlo03}. The adiabatic approximation~\cite{brouwer98, Moskalets04} applies only in the low-frequency limit, that is, for scattering by a slowly modulated potential, and beyond this limit the pumping becomes nonadiabatic~\cite{kaestner08, kim2004}. A single-parameter pump, in which current is generated by driving just one system parameter, cannot sustain a finite pump current in the adiabatic regime~\cite{brouwer98, jose2011}. It therefore becomes necessary to consider the nonadiabatic regime~\cite{PhysRevLett.90.210602, PhysRevB.70.155326},  which sustains single-parameter current generation and allows arbitrary driving frequencies beyond the adiabatic limit. Because pumping is associated with an ac gate voltage, it also generates nonequilibrium shot noise that carries signatures of the scattering processes inside the well, information that is washed away in the time-averaged dc current. Shot noise stems from current fluctuations due to the discreteness of electrical charge~\cite{beenakker03}. While the random, independent emission of electrons follows a Poisson distribution in the shot noise spectrum, correlations among electrons reduce the noise below this Poissonian value. The shot noise spectrum can therefore be used to detect the nature of scattering, the amount of transferred charge, or the extent of entanglement, and it can also serve as a proxy for resonance patterns in the transmission spectra, as we explain below.

In the transmission spectrum of a driven potential well, a Fano resonance arises from the interaction between the spatially localised bound states and the propagating modes. It is often interpreted as an interference effect between electron wavefunctions along different quantum paths~\cite{Zhu15},  producing a resonance peak or dip in the transmission spectrum. Such resonances have been reported, at energies in the meV range, for two-dimensional (2d) electron gas and graphene systems~\cite{Zhu15, reichl199, Dai2014},  and more recently for 2d pseudospin-1 quasiparticles~\cite{Zhu17} and quadratic band-touching semimetals~\cite{bera2021},  which encompass both two- and three-dimensional (3d) versions. These examples all involve isotropic dispersions around the band-crossing point. Beyond the familiar Dirac and Weyl nodes, however, semi-Dirac~\cite{pickett09, ips-kush, ips-kush-review, ips-abs-semidirac} and multi-Weyl semimetals (mWSMs)~\cite{pickett09, liu2017predicted, Gang2011} have also been identified, in which the dispersion mixes linear and higher-order terms depending on direction. Owing to this anisotropy, such materials exhibit electromagnetic properties that contrast with those of conventional Dirac and Weyl materials of purely linear dispersion~\cite{jose2011,dietl08}, and the anisotropy is expected to play a significant role in transport characteristics such as quantum tunnelling~\cite{Khokhlov2018, Deng2020, ips-aritra},  thermopower~\cite{ips-kush, ips-kush-review},  chiral photocurrent~\cite{ips-photocurrent}, circular dichroism~\cite{sajid-cd, ips_cd}, the Magnus-Hall effect~\cite{sajid_magnus22}, and magnetoelectric response~\cite{ipsita-shivam, rahul-jpcm, ips-ruiz, ips-tilted, ips-cjp}.

The physics of Floquet scattering rests on photon-assisted tunnelling, in which electrons tunnel through a potential well driven at frequency $\omega$ and exchange energy quanta, in units of $\hbar \, \omega$, in the process. The oscillating modulation introduces Floquet sidebands in the resulting dispersion, with quasienergies $E_n = E + n \, \hbar \, \omega$, where $n \in \mathbb{Z}$ labels the sideband order. For an incoming electron of Fermi energy $E_F$, a Fano resonance appears in the transmission spectrum when the difference between $E_F$ and a bound-state energy of the well, $E_b$, equals the energy of an integer number of photons, each carrying energy $\hbar \, \omega$. The incident electron then emits photons and drops into the bound state, or, conversely, an electron already occupying the bound state absorbs photons and jumps to the incident energy or to one of the Floquet channels $E_F + n \, \hbar \, \omega$. Such resonances leave a sharp peak or dip in the transmission spectrum, and an inflection point in the pumped shot noise. We use the Floquet-scattering matrix formalism~\cite{Zhu15, reichl199, Dai2014,Zhu17, Araujo2021} in the nonadiabatic limit to calculate both quantities.

Floquet scattering has previously been investigated for a variety of isotropic nodal-point semimetals~\cite{Zhu15,reichl199, Dai2014, Zhu17, bera2021, bera2023},  but the corresponding physics of anisotropic systems, such as nodal-ring semimetals (NRSs) and mWSMs, has received far less attention. This gap motivates the present work. For systems with isolated band-crossing points, we consider mWSMs with a linear dispersion along one direction, which we take to be the $z$-direction without loss of generality, and a quadratic or cubic dispersion in the perpendicular ($xy$) plane. An mWSM carries a topological charge $J$ whose magnitude is larger than that of an ordinary Weyl semimetal ($J = 1$): $J = 2$ at a double-Weyl node, realised for example in HgCr$_2$Se$_4$~\cite{Gang2011} and SrSi$_2$~\cite{hasan_mweyl16}, and $J = 3$ at a triple-Weyl node, as in transition-metal monochalcogenides~\cite{liu2017predicted}. We also consider NRSs, in which the band touching occurs along a closed loop in momentum space, the nodal ring~\cite{Khokhlov2018, cheng-nodal, ips-nlsm-ph, ips-dipole-vnr}. Such dispersions arise in materials including Cu$_3$PdN~\cite{xiao_nodal15}, ZrSiS~\cite{fu_nodal19}, and Mg$_3$Bi$_2$~\cite{chang_nodal19}.

\begin{figure}[t!]
	\centering
	\includegraphics[width=0.5 \textwidth]{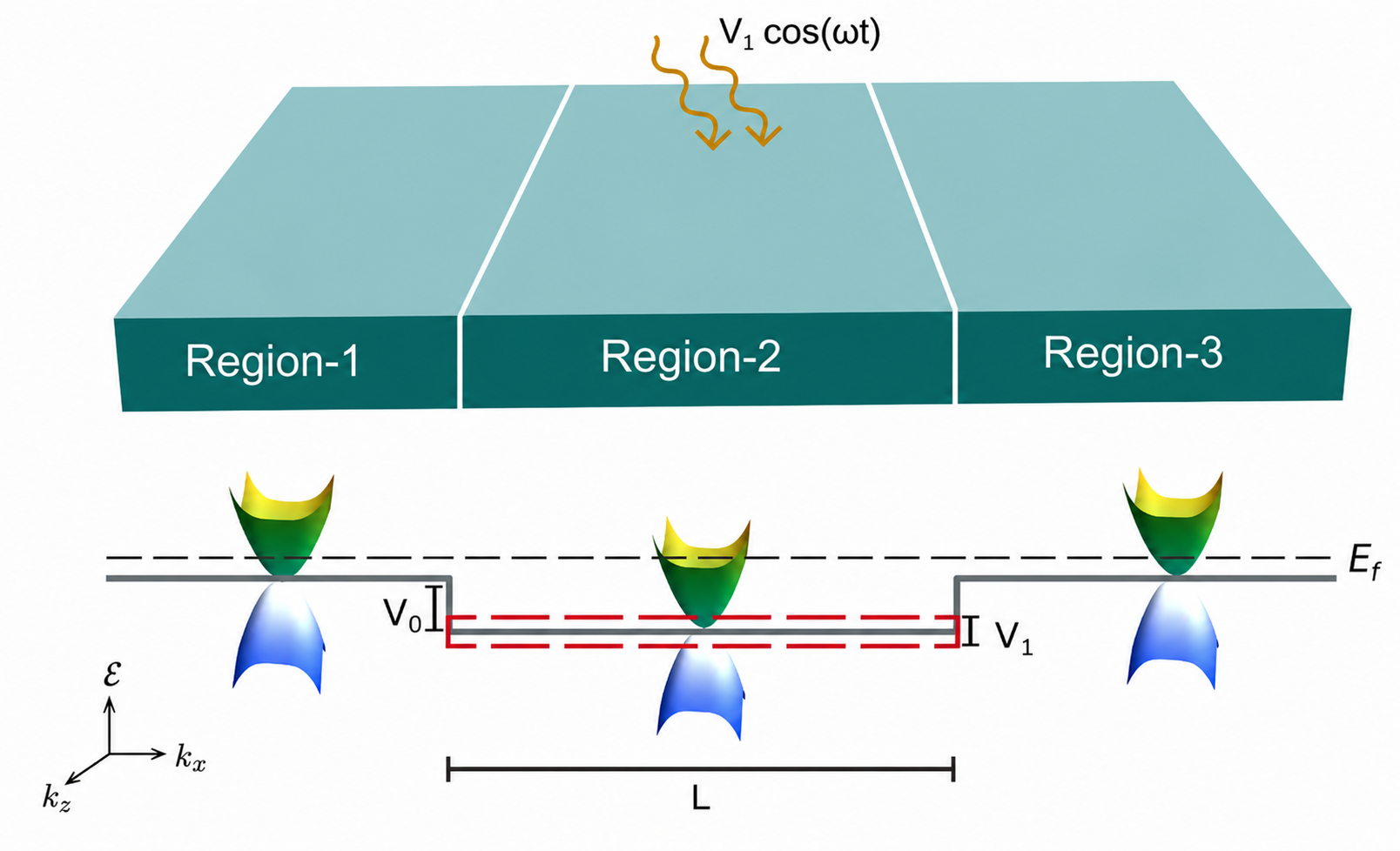}
	\caption{Schematic of the scattering geometry for a quasiparticle incident on a time-periodic potential well in a double-Weyl ($J=2$) mWSM. The potential is oriented along the $x$-axis, dividing the propagation direction into three regions: region I ($x<-L/2$) contains the incident and reflected waves, region II ($-L/2\le x\le L/2$) contains the driven potential well of width $L$ and depth $V_0$, and region III ($x>L/2$) contains the transmitted waves. A time-periodic modulation of amplitude $V_1$ and frequency $\omega$ is applied to the well, indicated by the red dashed lines marking the oscillating potential boundaries. Outside the well, the multi-Weyl cone has a linear dispersion along $k_x$ and is filled up to the Fermi energy $E_F$, marked by the horizontal black dashed line.\label{fig:setup}}
\end{figure}

In our previous work~\cite{bera2023}, we studied Floquet scattering in these systems, but restricted the analysis to propagating modes, neglecting the contribution of evanescent waves. This is justified when the potential well is aligned along the linear-dispersion direction, but it overlooks the role of evanescent modes when the well is oriented along a nonlinear dispersion direction. Here, it involves the $x$-axis for NRSs and the $J=2,3$ cases of mWSMs, where the longitudinal wavevectors admit both propagating and evanescent solutions. A complete scattering description must include the latter to satisfy the boundary conditions at the potential interfaces. In the present work, we go beyond this earlier study by constructing the full scattering matrix for both NRSs and mWSMs, explicitly incorporating the evanescent solutions that arise from imaginary or complex longitudinal wavevectors, alongside the propagating channels, and by assessing their impact on the transport properties. In doing so, we identify a previously unexplored transport regime in NRSs in which two propagating channels coexist and participate in coherent quantum transport, giving rise to distinct Fano resonance features. We show that, although evanescent modes are essential for constructing the complete scattering states and satisfying the boundary-matching conditions at the well interfaces, they carry zero net probability current and do not contribute to the measurable transmission, reflection, shot noise, or the associated Fano resonance features. Consequently, the transport properties reported in our earlier study remain unchanged once the complete set of scattering solutions is included, and this inclusion of evanescent modes validates the approximation made there. These findings provide a unified framework for Floquet transport in anisotropic topological semimetals with coexisting propagating and evanescent scattering channels. A schematic of the scattering geometry is shown in Fig.~\ref{fig:setup},  which depicts the three regions, the driven potential well, and the relevant energy scales.

The paper is organised as follows. We first review the Floquet formalism, scattering matrix theory, and the definition of shot noise in Sec.~\ref{secformalism}. We introduce the model Hamiltonians for mWSMs and NRSs in Sec.~\ref{sec:model_hamiltonian},  with the details for NRSs in Sec.~\ref{subsec:nodal_line} and for mWSMs in Sec.~\ref{sec:multi_weyl}. These subsections also contain the computational details and the final plots, and we compare our numerical results with previous studies for other semimetals. In Sec.~\ref{secresults}, we correlate the existence of the Fano resonances for each case to the energies of the bound states. Finally, we summarise our results, discuss their broader context, and offer some concluding remarks in Sec.~\ref{sec:summary}. Throughout this work, we will henceforth use natural units, thus setting $\hbar = k_B = c = e= 1$. Energies, momenta, and lengths are accordingly expressed in the material-dependent units introduced together with each model Hamiltonian in Sec.~\ref{sec:model_hamiltonian}.

\section{Floquet Formalism and Transport Calculations}
\label{secformalism}

To investigate Floquet scattering and its associated transport properties, we introduce a time-periodic driving potential into the system described by the momentum-space Hamiltonian $\mathcal{H}$. As stated in Sec.~\ref{secintro}, we work in natural units. Thus, the energies, frequencies, and momenta share the same dimension throughout. The spatial orientation of this potential determines which scattering channels are available: applying it along either the $x$- or $y$-direction yields a parameter regime in which two propagating (non-decaying) channels coexist. Because this multi-channel regime is central to our study, we consider a potential well modulated along the $x$-direction, which takes the form~\cite{jose2011, reichl199, bera2021, bera2023}:
\begin{align}
	V(x,t) = \begin{cases} 
		-V_{0} + V_{1}\cos(\omega t) & \text{for } -\frac{L} {2} \le x \le \frac{L} {2} \\ 
		0 & \text{otherwise} 
	\end{cases}.
	\label{eq:potential_well}
\end{align}
Here, $L$ denotes the width of the potential well, $V_0$ its static depth, $V_1$ the driving amplitude, and $\omega = 2\pi/\tau$ the frequency of the periodic modulation. The potential is taken to be spatially homogeneous and infinite along the transverse directions (the $y$- and $z$-axes), corresponding physically to a sufficiently large cross-sectional width $W$. The transverse momentum components are therefore conserved during scattering, reducing the problem to an effective one-dimensional system along the $x$-axis. A schematic of the scattering geometry is shown in Fig.~\ref{fig:setup}.

The dynamics of the system are governed by the time-dependent Schr\"odinger equation:
\begin{align}
i\, \partial_{t}\Psi(\boldsymbol{r} \,, t) = \left[ \mathcal{H} + V(x,t) \right] \Psi(\boldsymbol{r} , t)\,,
\end{align}
where $\Psi(\boldsymbol{r} \,, t)$ is the wavefunction. To solve this equation under periodic driving, we employ Floquet theory, expanding the wavefunction as a superposition of discrete harmonics, the Floquet sidebands, spaced by the driving frequency:
\begin{align}
\Psi(\boldsymbol{r} \,, t) = \sum_{n=-\infty}^{\infty} e^{-i \,E_n \,t} 
	\, e^{i \,k_y \,y} \, e^{i \,k_z \,z}\, \psi_n(x)\,,
\end{align}
where $E_n = E + n\, \omega$ is the quasienergy of the $ n^{\rm th}$ Floquet sideband, and $k_y, k_z$ are the conserved transverse momenta along the unconfined directions. Inside the scattering region (i.e., $-L/2 \le x \le L/2$), the time-periodic part of the wavefunction can be written as a Fourier series in the Floquet modes.

The Floquet-scattering matrix, or $S$-matrix, relates the amplitudes of the incoming and outgoing waves in the asymptotic regions. We construct this matrix by matching the wavefunction, and, where the differential order of the Hamiltonian requires it, its spatial derivatives, at the interfaces $x = \pm L/2$. The resulting linear system fixes the scattering coefficients. The full $S$-matrix contains contributions from both propagating and evanescent modes, but, as we show in this work, only the propagating modes contribute to the asymptotic transport observables: evanescent modes decay exponentially away from the interfaces and carry zero net probability current.

Retaining the evanescent branch in this boundary-value problem is not merely a matter of completeness, but a mathematical necessity whenever the dispersion along the transport direction is nonlinear in $k_x$. For a Hamiltonian linear in $k_x$, as in an ordinary Weyl node, the characteristic equation at fixed transverse momentum and energy is itself linear in $k_x^2$, and matching the two-component spinor wavefunction at each interface closes the linear system using only the pair of propagating roots. Once the dispersion along $x$ becomes nonlinear, however, either through a quadratic curvature term, as in the NRS Hamiltonian of Eq.~\eqref{eq:nl_hamiltonian}, or through the higher powers $k_x^{2J}$ that define the double- and triple-Weyl mWSMs, the characteristic equation for $k_x$ becomes a polynomial of degree higher than two. Its full set of roots, generically a mixture of real and complex-conjugate pairs, is then required to supply enough independent basis solutions to satisfy every matching condition at the interface, including the continuity of the derivatives of increasing order demanded by the higher powers of $k_x$ in the Hamiltonian. Omitting the complex or purely imaginary roots leaves the boundary-value problem under-determined and unable to reproduce a physically consistent wavefunction at the junction, however small the corresponding transport probability eventually turns out to be. This requirement is precisely the one identified in related studies of quasiparticle transmission through rectangular potentials in semimetals with quadratic-in-momentum dispersion~\cite{ips-qbcp-aop}, and in the analogous delta-function-potential junction problem for tilted bands with quadratic dispersion~\cite{ips-qbcp-delta}, where the evanescent solutions generated by the nonlinear dispersion must likewise be retained to close the boundary-matching equations at the junction, even though they do not themselves carry current. The same requirement underlies the static barrier problem for nodal-line semimetals~\cite{Khokhlov2018}, the rectangular-barrier and magnetic-field-assisted tunnelling problems for mWSMs~\cite{Deng2020,ips-aritra}, and the Andreev scattering problem at semi-Dirac Josephson junctions~\cite{ips-abs-semidirac}, in each of which the nonlinear dispersion along the transport direction generates evanescent solutions that must be appended to the propagating ones purely to satisfy the boundary conditions at the interface. Our NRS and mWSM problems extend this principle to a periodically driven barrier, where every Floquet sideband contributes its own set of propagating and evanescent roots, and all of them must be included self-consistently to construct a unitary Floquet-scattering matrix.

To compute the pumped shot noise, we use the zero-temperature, zero-frequency expression for the current fluctuations. For a multi-channel scattering problem, the shot noise spectral density is given by~\cite{Zhu15, reichl199, Dai2014, Zhu17, bera2021, bera2023}:
\begin{align}
	\mathcal{N}_{\alpha \beta} = \frac{1} {4\,\pi} 
	\int_{0}^{\infty} dE \sum_{\gamma,\delta=1}^{\kappa} \sum_{m,n,p=-\infty}^{\infty}  
s^*_{\alpha \gamma}(E,E_n) \,s_{\alpha \delta}(E,E_m) \,s^*_{\beta \delta}(E_p,E_m) \,s_{\beta \gamma}(E_p,E_n) \left[ f_0(E_n) - f_0(E_m) \right]^2,
	\label{eq:shotnoise_general}
\end{align}
where $s_{\alpha \gamma}(E_n, E_m)$ is an element of the scattering matrix connecting the $m^{\rm th} $ incoming Floquet channel to the $n^{\rm th}$ outgoing channel between leads $\alpha$ and $\gamma$. The indices $(m,n,p)$ label the Floquet sidebands, and $f_0(E) = [1 + \exp(E/ T)]^{-1}$ is the Fermi-Dirac distribution function. We take the $T \to 0 $ limit in our computations. The upper summation limit $\kappa$ is the total number of open scattering channels, fixed by the Hamiltonian and the energy window under consideration. The symmetry relations $\mathcal{N}_{LL} = \mathcal{N}_{RR} = - \, \mathcal{N}_{LR} = - \, \mathcal{N}_{RL}$ follow from current conservation, and we restrict our analysis to $\mathcal{N}_{LL}$, without loss of generality.

The scattering matrix is obtained by solving the boundary-value problem for each Floquet sideband, with the wavefunction in each region written as a linear combination of the eigenmodes of the corresponding Hamiltonian. For the NRS, these eigenmodes are set by the four longitudinal wavevectors obtained from the dispersion relation: in the fully propagating regime all four are real, giving two transmission and two reflection channels, while in the mixed regime two of them become imaginary and correspond to evanescent modes. The same procedure applies to the mWSMs, where the number of eigenmodes depends on the topological charge $J$: four for $J=2$ (two propagating and two evanescent), and six for $J=3$ (two propagating and four evanescent). The complete scattering matrix is then constructed by enforcing the continuity conditions at the interfaces, as detailed in the appendices.

The physical, unitary part of this matrix is its flux-normalised propagating block, $s(E_n, E_m) \equiv S_{pp}(E_n, E_m)$, with reflection and transmission amplitudes $r_{nm}^{p}$ and $t_{nm}^{p}$ that set the observable transport coefficients. To confirm that the decaying modes remain decoupled from the asymptotic transport, we also track the amplitude ratios of the two evanescent sectors, $(r_{nm}^{e_1} , \,  t_{nm}^{e_1})$ and $(r_{nm}^{e_2} ,\,  t_{nm}^{e_2})$. Written in terms of $s(E_n,E_m)$, the shot noise of Eq.~\eqref{eq:shotnoise_general} takes the explicit form of
\begin{align}
&	\mathcal{N}_{\alpha \beta} = \frac{1} {4\pi} 
\int_{0}^{\infty} dE \sum_{\gamma,\delta=1}^{\kappa} \sum \limits_{m,n,p=-\infty}^{\infty} 
M_{\alpha \beta \gamma \delta}(E,E_m \,, E_n , E_{p}) \left[ f_{0}(E_{n}) - f_{0}(E_m) \right ]^2, \nn &
M_{\alpha \beta \gamma \delta}(E,E_m \,, E_n \,, E_p) 
= s^*_{\alpha \gamma}(E,E_n) \,s_{\alpha \delta}(E,E_m) \,s^*_{\beta \delta}(E_p, 
E_m) \,s_{\beta \gamma}(E_p\,, E_n) \,,
\label{eq:shotnoise3d}
\end{align}
where $ M_{\alpha \beta \gamma \delta}(E,E_m \,, E_n \,, E_p) $ is the scattering kernel. The upper summation limit for the variable $\kappa$ is $4$ for NRSs and double-Weyl semimetals, and $ 6$ for triple-Weyl semimetals. This follows from matching the number of open channels in each model, as explained in Sec.~\ref{sec:model_hamiltonian}.

\section{Model}
\label{sec:model_hamiltonian}

In order to investigate quantum transport and scattering dynamics across two distinct systems, NRSs and mWSMs, we describe the Hamiltonians of these two systems in this section.

\subsection{Nodal-Ring Semimetals (NRSs)}
\label{subsec:nodal_line}

In NRSs, the upper and lower bands touch along a closed loop in momentum space, rather than at isolated points as in conventional Weyl or Dirac semimetals. We model such a system with the minimal low-energy continuum Hamiltonian~\cite{Khokhlov2018, fang2012multi}
\begin{align}
\mathcal{H}_{\text{nl}} = \left(\mathcal{M} - B \, k_{\perp}^{2}\right) \sigma_{x} + k_z \sigma_{z} \,, 
\quad k_{\perp} \equiv \sqrt{ k_x^2 + k_y^2}\,,
\label{eq:nl_hamiltonian}
\end{align}
where $\mathcal{M}$ is the characteristic mass parameter, $B$ sets the quadratic curvature of the dispersion in the transverse plane, and $\bm{\sigma} = (\sigma_x, \sigma_y, \sigma_z)$ is the vector of Pauli matrices acting on the pseudospin degrees of freedom. This Hamiltonian supports a closed nodal line in the $k_z = 0$ plane, defined by the ring radius $  \mathcal{M}/B$. The energy eigenvalues are $\mathcal{E}^{\pm}_{\text{nl}}(\boldsymbol{k}) = \pm \,\sqrt{\left(\mathcal{M} - B \, k_{\perp}^{2}\right)^2 + k_z^2}$, describing a gapless ring in momentum space surrounded by gapped bulk excitations.

For numerical convenience we express the Hamiltonian in dimensionless form by scaling all terms by the mass parameter $\mathcal{M}$. Dividing Eq.~\eqref{eq:nl_hamiltonian} by $\mathcal{M}$ gives
\begin{align}
\frac{\mathcal{H}_{\text{nl}}} {\mathcal{M}} 
= \left[1 - B\, \mathcal{M} \left(\frac{k_{\perp}} {\mathcal{M}}\right)^2\right]
 \sigma_{x} + \frac{k_z} {\mathcal{M}} \, \sigma_{z} \,, 
\end{align}
so that all energy and momentum scales are implicitly expressed in units of $\mathcal{M}$. Under this convention, the product $B\mathcal{M}$ is a dimensionless parameter, set to unity in our simulations ($B = \mathcal{M}^{-1}$), and all lengths are measured in units of $\mathcal{M}^{-1}$.

For a fixed energy $\mathcal{E}$ and transverse momenta $(k_y, k_z)$, the dispersion relation gives four longitudinal wavevector solutions ($k_x$) along the propagation direction. The wavefunction component $\psi_{n}(x,t)$ in each spatial region is accordingly expanded as a linear combination of these four eigenmodes:
\begin{align} 
\label{eq:nodal_two_propagator}
\pmtx{\psi_{n,1}  \\ \psi_{n,2} } =
\begin{cases}
& A_{1,n}^{i}(t) \,e^{ i\, k_n^+ \, x} \pmtx{ f_1  \\  f_2 }
+ A_{1,n}^{o}(t) \,e^{- i\, k_n^+ \, x} \pmtx{ f_1  \\  f_2 }
+ A_{2,n}^{i}(t) \,e^{ i\, k_n^- \, x} \pmtx{- f_1  \\  f_2 }
+ A_{2,n}^{o}(t) \,e^{- i\, k_n^- \, x} \pmtx{- f_1  \\  f_2 }
 \text{ for } x< -\frac{L} {2}   \\ & \\
& \sum \limits_{m=-\infty}^\infty
\Bigl[ \alpha_{1,m}(t) \,e^{i q_m^+ x} \pmtx{ g_1  \\ g_2}
+ \beta_{1,m}(t) \, e^{-\,i \, q_m^+ \, x} \pmtx{ g_1  \\ g_2}
+ \alpha_{2,m}(t) \,e^{i q_m^- x} \pmtx{- g_1  \\ g_2}
+ \beta_{2,m}(t) \,e^{-i q_m^- x} \pmtx{- g_1  \\ g_2} \Bigr] \\
& \qquad \qquad \times \, J_{n-m} ( {V_{1}} / {\omega} ) \, \Theta(-E_m-V_0)
 \text{ for } -\frac{L} {2} \le x \le \frac{L} {2}   \\ & \\
& B_{1,n}^{i}(t) \,e^{- i\, k_n^+ \, x} \pmtx{ f_1  \\  f_2 }
+ B_{1,n}^{o}(t) \,e^{ i\, k_n^+ \, x} \pmtx{ f_1  \\  f_2 }
+ B_{2,n}^{i}(t) \,e^{- i\, k_n^- \, x} \pmtx{- f_1  \\  f_2 }
+ B_{2,n}^{o}(t) \,e^{ i\, k_n^- \, x} \pmtx{- f_1  \\  f_2 }
 \text{ for } x> \frac{L} {2}
\end{cases}
\end{align}
where
\begin{align}
\label{eqnodalx}
&  f_1  = -\,\frac{1} {n_1} \, \sqrt{\frac{E_n +k_z} {E_n -k_z}} \,,  \quad
 f_2  = \frac{1} {n_1} \,, \quad  g_1  =  -\,\frac{ 1} {n_2} \,\sqrt{\frac{E_m +V_0 +k_z} {E_m+ V_0 -k_z}} \,,  \quad
g_2 = \frac{1} {n_2} \,,  \quad 
n_1 = \sqrt{\frac{E_n +k_z} {E_n-k_z}+1} \,, \nn &
n_2 = \sqrt{\frac{E_m +V_0 +k_z} {E_m +V_0-k_z}+1} \,,  \quad 	 
k_n ^{\pm} = \sqrt{\frac{1-B\, k_y^2 \pm \sqrt{E_n^2-k_z^2}} {B}} \,,  \quad
q_m ^{\pm}= \sqrt{\frac{1-B\, k_y^2 \pm \sqrt{(E_m+V_0)^2-k_z^2}} {B}}.
\end{align}
For brevity, we introduce the shorthand $k^{+} \equiv k_1$, $k^{-} \equiv k_2$, $q^{+} \equiv q_1$, and $q^{-} \equiv q_2$. Here $k_1$ and $k_2$ are the longitudinal wavevectors outside the potential well (Regions I and III), while $q_1$ and $q_2$ are the corresponding wavevectors inside the time-periodically driven potential barrier (region II). The coefficients $A_{\nu,n}^{i/o}$ and $B_{\nu,n}^{i/o}$ (with $\nu \in \{1,2\}$) are the incoming and outgoing wave amplitudes in the incident (left) and transmitted (right) regions. Inside the well, the coefficients $\alpha_{\nu,m}$ and $\beta_{\nu,m}$ are the amplitudes of the forward- and backward-propagating Floquet sidebands, modulated by the Bessel functions of the first kind, $J_{n-m}(V_1/\omega)$, arising from the periodic drive.

The physical nature of these wavevectors depends on the chosen energy and momentum. The modes $k_{1,n}$ and $q_{1,n}$ remain strictly real across the entire parameter space investigated, representing robust propagating channels. The longitudinal wavevectors $k_{2,n}$ and $q_{2,n}$, by contrast, remain real only within the regime defined by $\sqrt{E_n^2-k_z^2} \geq 1-B \, k_y^2$ and $\sqrt{(E_m+V_0)^2-k_z^2} \geq 1-B \,k_y^2$, where $E_n = \mathcal{E} + n\omega$ is the energy of the $ n^{\rm th}$ Floquet sideband. Outside this regime, $k_{2,n}$ and $q_{2,n}$ become purely imaginary, describing decaying evanescent modes.

The physical origin of this threshold lies in the momentum-dependent mass term of Eq.~\eqref{eq:nl_hamiltonian}. For fixed transverse momenta $(k_y,k_z)$, the Hamiltonian along $x$ reduces to an effective one-dimensional two-band problem with an $x$-independent but $k_y$-dependent mass, $\mathcal{M}-B \, k_\perp^2 $, whose sign and magnitude set a local gap around the nodal ring. When the sideband energy $E_n$ is large enough relative to this local gap, both roots of the quadratic equation for $k_x^2$ are positive, and the $k_1$ and $k_2$ branches both describe propagating waves. Once $E_n$ falls below the threshold fixed by $k_y$, one root of $k_x^2$ turns negative, and the corresponding wavevector becomes purely imaginary: the wave can no longer propagate and instead decays exponentially away from the interface. This is precisely the mechanism identified by Khokhlov et al.~\cite{Khokhlov2018} for a static rectangular potential barrier in nodal-line semimetals, where an incident wave scattering off a step or barrier generates a purely real wavevector solution only up to a threshold energy set by the transverse momentum, beyond which an exponentially decaying solution must be retained to complete the set of scattering states. Our Floquet problem extends this static picture to a driven potential: because every Floquet sideband $E_n$ enters the threshold condition independently, a given wavevector branch can be propagating for the central band ($n=0$) while its $n\ne 0$ replicas remain evanescent, or vice versa, depending on how each sideband energy compares with the local gap $1-Bk_y^2$. The two transport regimes analysed below, hosting either two or four real wavevectors among the central Floquet channels, are simply the two ways this threshold condition can be satisfied.

The evanescent modes decay over a characteristic length set by the imaginary part of $k_{2,n}$ or $q_{2,n}$: deeper inside the classically forbidden window, $|\text{Im}(k_{2,n})|$ grows and the mode is confined ever closer to the interface, in direct analogy with tunnelling under a potential barrier in the ordinary Dirac equation, of which the nodal-line Hamiltonian of Eq.~\eqref{eq:nl_hamiltonian} is a two-band anisotropic generalisation. As we show explicitly in Sec.~\ref{secformalism}, such a mode carries zero net probability current asymptotically and therefore cannot itself constitute a transport channel. It remains indispensable locally, however, since it is required, exactly as in the static barrier problem of Ref.~\cite{Khokhlov2018}, to satisfy the wavefunction and derivative matching conditions at the interfaces of the driven well and to construct a complete, physically consistent set of scattering states. This interplay between an exponentially confined evanescent mode and the current-carrying propagating channels is precisely what allows the boundary-value problem of Sec.~\ref{subsubsec:one_propagating_one_evanescent} to be solved uniquely even when only one propagating channel is open. This behaviour delineates two distinct transport regimes: a fully propagating regime, where all four wavevectors are real, and a mixed regime with one propagating and one evanescent channel. We analyse the quantum transport characteristics of these two regimes in turn.

\subsubsection{Regime with Two Propagating Wavevectors}
\label{subsubsec:two_propagating}

In the fully propagating regime, all four longitudinal wavevectors are real, giving four scattering channels for an incident electron: two associated with forward transmission and two with backward reflection. To identify which wavevectors correspond to transmission and which to reflection, we evaluate the expectation value of the probability current density along the propagation direction, defined via the velocity operator $v_x = \partial_{k_x} \mathcal{H}_{\text{nl}}  = -2 \,B \,k_x  \, \sigma_x$. The average current density is
\begin{align}
\langle j_x \rangle  & =  \Big\langle \Psi \Big| -2 \,B\, k_x \,\sigma_x \Big| 
\, \Psi \Big\rangle 
= 2 \,B\, f_1 \, f_2 \left[ k_1 \left( A_{o \,, 1}^2 - A_{i \,, 1}^2 \right)
- k_2 \left( A_{o \,, 2}^2 - A_{i \,, 2}^2 \right) \right],
\label{eq:current_density}
\end{align}
where $f_1$ and $f_2$ are normalisation factors specific to the spinor eigenstates. This shows that the wavevectors $k_{1,n}$ and $-k_{2,n}$ correspond to transmission channels, since they give $\langle j_x \rangle > 0$, a net positive current in the forward direction, while $-k_{1,n}$ and $k_{2,n}$ correspond to reflection channels, with $\langle j_x \rangle < 0$ directed backward.

The scattering geometry is divided into three regions along the propagation axis: region I ($x < -L/2$), the incident region, region II ($-L/2 \le x \le L/2$), the central potential barrier, and region III ($x > L/2$), the transmitted region. Inside the barrier there are again four characteristic wavevectors, and the wavefunction is a linear combination of the corresponding exponential modes. To determine the unknown scattering coefficients, we enforce continuity of the wavefunction and its first spatial derivative at the two boundaries, $x = -L/2$ and $x = L/2$. At the first interface, the matching conditions between region I and region II are
\begin{align}
\label{eq:boundary_nl_x1_psi}
\Psi_{\text{I}} (-\frac{L} {2} \,,  y, z ) = \Psi_{\text{II}} (-\frac{L} {2} \,,  y, z ), \quad
 \partial_x \Psi_{\text{I}}(x, y, z)  \big |_{x=-\frac{L} {2}} 
=  \partial_x \Psi_{\text{II}}(x, y, z)  \big |_{x=-\frac{L} {2}}.
\end{align}
Similarly, at the second interface ($x = \frac{L} {2}$), the continuity conditions between region II and region III are
\begin{align}
	\label{eq:boundary_nl_x2_psi}
	\Psi_{\text{II}}\left(\frac{L} {2} \,,  y, z\right) = \Psi_{\text{III}}\left(\frac{L} {2} \,,  y, z\right), \quad
	\partial_x \Psi_{\text{II}}(x, y, z)  \big |_{x=\frac{L} {2}} 
	=  \partial_x \Psi_{\text{III}}(x, y, z)  \big |_{x=\frac{L} {2}}.
\end{align}
To confirm that this matching procedure fully determines the scattering problem, it is instructive to first count the unknown coefficients in the undriven, static limit ($V_1=0$, so that only the $n=m=0$ Floquet channel is populated), following the same counting argument used for the analogous static rectangular-barrier problems in nodal-line and multi-Weyl semimetals~\cite{Khokhlov2018,Deng2020,ips-aritra}. In this limit, Eq.~\eqref{eq:nodal_two_propagator} contains twelve undetermined amplitude coefficients in total: four in region I ($A_1^i, A_1^o, A_2^i, A_2^o$), four inside the well ($\alpha_1,\beta_1,\alpha_2,\beta_2$), and four in region III ($B_1^i, B_1^o, B_2^i, B_2^o$). Since the two-component wavefunction and its first derivative must be continuous at each interface, we obtain $2\times2=4$ matrix equations at $x=-L/2$ [Eq.~\eqref{eq:boundary_nl_x1_psi}] and a further four at $x=L/2$ [Eq.~\eqref{eq:boundary_nl_x2_psi}], eight scalar equations in total. Fixing the four independently specifiable incident-wave amplitudes as external inputs therefore leaves exactly eight coefficients to be determined by these eight equations, so that the linear system is exactly determined. The same eight-equation, eight-unknown counting applies separately within every Floquet sideband once the periodic drive is switched on, and the resulting linear system can be written compactly in matrix form, relating the incoming, reflected and transmitted wave amplitudes across the spatial regions and Floquet sidebands:
\begin{align}
	\begin{pmatrix} 
		A_{1,n}^{o} \\ A_{2,n}^{i} \\ B_{1,n}^{o} \\ B_{2,n}^{i} 
	\end{pmatrix} 
	= \sum_{m=-\infty}^{\infty} \mathcal{S}_{nm} 
	\begin{pmatrix} 
		A_{1,m}^{i} \\ A_{2,m}^{o} \\ B_{1,m}^{i} \\ B_{2,m}^{o} 
	\end{pmatrix}.
	\label{eq:floquet_S_matriX_relation}
\end{align}
Here $\mathcal{S}_{nm}$ is the global Floquet-scattering matrix, encoding the transition amplitudes between the incoming sideband channel $m$ and the outgoing sideband channel $n$. Its detailed analytical structure, derived from the interface matching conditions, is given in Appendix~\ref{appendiX_nodal_two}.

Physically, $\mathcal{S}_{nm}$ gives the probability amplitude for an electron entering via the $m^{\rm th}$ Floquet channel to be scattered into the $ n^{\rm th}$ channel. A strictly unitary scattering matrix retains only the propagating modes, those with purely real longitudinal wavevectors within the active energy window, since evanescent modes carry zero net probability current and do not contribute directly to asymptotic transport. Extracting the propagating sector from $\mathcal{S}_{nm}$ gives the current-normalised scattering matrix $s(E_n, E_m)$, connecting the asymptotically open incoming and outgoing channels at sideband energies $E_m$ and $E_n$. Its matrix-elements are
\begin{align}
s(E_n , E_m)  & =  
\begin{pmatrix} 
	\chi_{1,nm} \, S_{11}(E_n , E_m) & \chi_{3,nm} \, S_{12}(E_n , E_m) & 
	\chi_{1,nm} \, S_{13}(E_n , E_m) & \chi_{3,nm} \, S_{14}(E_n , E_m) \\ 
	\chi_{2,nm} \, S_{21}(E_n , E_m) & \chi_{4,nm} \, S_{22}(E_n , E_m) & 
	\chi_{2,nm} \, S_{23}(E_n , E_m) & \chi_{4,nm} \, S_{24}(E_n , E_m) \\ 
	\chi_{1,nm} \, S_{31}(E_n , E_m) & \chi_{3,nm} \, S_{32}(E_n , E_m) & 
	\chi_{1,nm} \, S_{33}(E_n , E_m) & \chi_{3,nm} \, S_{34}(E_n , E_m) \\ 
	\chi_{2,nm} \, S_{41}(E_n , E_m) & \chi_{4,nm} \, S_{42}(E_n , E_m) & 
	\chi_{2,nm} \, S_{43}(E_n , E_m) & \chi_{4,nm} \, S_{44}(E_n , E_m)
\end{pmatrix} \nn & \equiv 
\begin{pmatrix} 
	s_{11}(E_n , E_m) & s_{12}(E_n , E_m) & s_{13}(E_n , E_m) & s_{14}(E_n , E_m) \\ 
	s_{21}(E_n , E_m) & s_{22}(E_n , E_m) & s_{23}(E_n , E_m) & s_{24}(E_n , E_m) \\ 
	s_{31}(E_n , E_m) & s_{32}(E_n , E_m) & s_{33}(E_n , E_m) & s_{34}(E_n , E_m) \\ 
	s_{41}(E_n , E_m) & s_{42}(E_n , E_m) & s_{43}(E_n , E_m) & s_{44}(E_n , E_m)
\end{pmatrix}  =  
\begin{pmatrix} 
	r_{nm}^{+} & \mathfrak{r}_{nm}^{+} & \tilde{t}_{nm}^{+} & \tilde{\mathfrak{t}}_{nm}^{+} \\ 
	r_{nm}^{-} & \mathfrak{r}_{nm}^{-} & \tilde{t}_{nm}^{-} & \tilde{\mathfrak{t}}_{nm}^{-} \\ 
	t_{nm}^{+} & \mathfrak{t}_{nm}^{+} & \tilde{r}_{nm}^{+} & \tilde{\mathfrak{r}}_{nm}^{+} \\ 
	t_{nm}^{-} & \mathfrak{t}_{nm}^{-} & \tilde{r}_{nm}^{-} & \tilde{\mathfrak{r}}_{nm}^{-}
\end{pmatrix} , 
\label{eq:unitary_s_matriX_elements}
\end{align}
where 
\begin{align}
\chi_{1,nm} = \sqrt{\frac{\text{Re}[k_{1,n}]} {\text{Re}[k_{1,m}]}} \,,  \quad
\chi_{2,nm} = \sqrt{\frac{\text{Re}[k_{2,n}]} {\text{Re}[k_{1,m}]}} \,,  \quad
\chi_{3,nm} = \sqrt{\frac{\text{Re}[k_{1,n}]} {\text{Re}[k_{2,m}]}} \,,  \quad 
\chi_{4,nm} = \sqrt{\frac{\text{Re}[k_{2,n}]} {\text{Re}[k_{2,m}]}}.
\end{align}
All scattering amplitudes obtained from the boundary conditions enter this framework, giving the multi-channel transmission and reflection coefficients explicitly. For an incident electron injected exclusively via the $k_{1,m}$ channel, the scattered coefficients $A_{1,n}^{o}$ and $B_{1,n}^{o}$, normalised to the incoming amplitude $A_{1,m}^{i}$, give the reflection and transmission amplitudes for the $k_{1,n}$ sideband sub-channels. Similarly, $A_{2,n}^{i}$ and $B_{2,n}^{i}$ give the reflection and transmission amplitudes into the secondary $k_{2,n}$ channel. Summing over all active sidebands (each indexed by $m$), the total transmission and reflection  probabilities satisfy the unitarity condition, $T_{+} + T_{-} + R_{+} + R_{-} = 1$,  with the individual channel components $T_{+} = \sum_m |t_{0m}^{+}|^2$,  $T_{-} = \sum_m |t_{0m}^{-}|^2$, $R_{+} = \sum_m |r_{0m}^{+}|^2$, and  $R_{-} = \sum_m |r_{0m}^{-}|^2$. Here, $T_{\pm}$ and $R_{\pm}$ are the total  transmission and reflection probabilities through the two open physical Floquet  channels for modes incident from the left. Correspondingly, $\tilde{\mathfrak{t}}^{\pm}_{0m}$ and $\tilde{\mathfrak{r}}^{\pm}_{0m}$ represent the transmission and reflection amplitudes for modes incident from the right, which similarly serve to construct the corresponding right-incident transmission and reflection probabilities.

\subsubsection{Regime with One Propagating and One Evanescent Wavevector}\label{subsubsec:one_propagating_one_evanescent}
In the alternative energy window identified in Sec.~\ref{subsec:nodal_line} \,,  the system enters a mixed transport regime hosting exactly one propagating and one evanescent mode. While the real longitudinal wavevectors $k_1$ and $q_1$ retain the forms discussed for the fully propagating case, the remaining two wavevectors become purely imaginary. To handle these decaying channels, we define the real-valued evanescent attenuation constants $\zeta_1$ and $\zeta_2$ through $\zeta_1 =  i\, k_2$ and $\zeta_2 = i \,q_2$.

In the semi-infinite Regions I ($x \rightarrow -\infty$) and III ($x \rightarrow +\infty$), the physical boundary condition requires the wavefunction to remain finite at infinity. Consequently, the coefficients of the exponentially diverging solutions must vanish identically. Thus, we set $A_2^o = 0$ in region I and $B_2^o = 0$ in region III, leaving only the physically admissible evanescent modes that decay away from the scattering region for the boundary-matching procedure, under which
the scattering wavefunctions continue to retain the structural form of Eq.~\eqref{eq:nodal_two_propagator}.

The global $4 \times 4$ Floquet-scattering matrix therefore keeps the same structure as in the fully propagating case, now comprising distinct propagating and evanescent sectors. Although the evanescent modes are necessary to satisfy the interface boundary-matching conditions and are included in the complete wavefunctions within the central region, they carry zero net probability current along the transport direction. We therefore enforce the same boundary conditions as in Eq.~\eqref{eq:boundary_nl_x1_psi} to solve the linear system. To examine the coupling between these open and closed channels, we partition the full scattering matrix into its constituent blocks:
\begin{align}
\mathcal{S}_{nm} = 
\begin{pmatrix}
	S_{pp} & S_{pe} \\ S_{ep} & S_{ee}
\end{pmatrix} , 
\label{eq:partitioneD_S_matrix}
\end{align}
where each block is a $2 \times 2$ sub-matrix. The block $S_{pp}$ governs scattering restricted entirely to open propagating channels, while $S_{pe}$, $S_{ep}$ and $S_{ee}$ describe the coupling paths involving the localised evanescent modes. The physical, unitary scattering matrix needed for the transport coefficients is the flux-normalised propagating sector, $s(E_n, E_m) \equiv S_{pp}(E_n, E_m)$.

In this mixed regime, the scattering dynamics separate conceptually into distinct sectors. The propagating-propagating sector describes the ordinary transmission and reflection between open channels, satisfying strict current conservation. The cross-coupling between open and closed channels sets how evanescent modes influence the propagating wavefunctions near the interfaces, where localised modes can match onto or disperse back into propagating states. The evanescent-evanescent sector captures the localised behaviour within the barrier region, characterised by exponential decay factors of the form $e^{-\zeta_i x}$ (with $i \in \{1,2\}$).

The reflection and transmission amplitudes from the open propagating block $S_{pp}$ are denoted $r_{nm}^{p}$ and $t_{nm}^{p}$, and set the observable transport properties, with total coefficients $T_p = \sum_m |t_{0m}^{p}|^2$ and $R_p = \sum_m |r_{0m}^{p}|^2$. Current conservation gives $T_p + R_p = 1$. To examine the role of the closed channels, we also track the amplitude ratios of the evanescent sector $S_{ee}$, defined as $T_e = \sum_m |t_{0m}^{e}|^2$ and $R_e = \sum_m |r_{0m}^{e}|^2$. Because evanescent modes carry zero net time-averaged probability current asymptotically, $T_e$ and $R_e$ are not observable transport probabilities, but rather a measure of the relative amplitude of the localised evanescent fields generated at the heterojunction interfaces. As our numerical results show, these quantities remain negligible over the parameter space investigated and have no influence on the open transport coefficients $T_p$ and $R_p$.

A general remark is in order regarding the treatment of evanescent modes. In the semi-infinite Regions I ($x \rightarrow -\infty$) and III ($x \rightarrow +\infty$), the physically admissible solutions strictly require that the scattering wavefunctions remain finite for large $|x|$. Consequently, regardless of the specific forms of the Hamiltonian or wave-amplitudes, the coefficients corresponding to unphysical spatially-diverging components must vanish identically. These are $\propto e^{-\zeta \,x}$ and $\propto e^{-\beta \,x}$ in region I, or $\propto e^{ \zeta \,x}$ and $\propto e^{ \beta \,x}$ in region III. While these diverging channels may be formally retained in the generalised boundary-matching matrices to preserve structural symmetry, they are explicitly set to zero when evaluating the physical quantities. As a result, the calculated transmission and reflection coefficients remain entirely independent of these unphysical amplitudes.

\subsection{Multi-Weyl Semimetals (mWSMs)}\label{sec:multi_weyl}
We consider the low-energy continuum model of a Weyl or mWSM near a single isolated Weyl node. The effective low-energy Hamiltonian is~\cite{liu2017predicted, lundgren2014thermo, fang2012multi, Gang2011, Mukherjee2018}
\begin{align}
\mathcal{H}_{\text{mw}} = \alpha_J \left( k_-^J \, \sigma_+ + k_+^J  \,\sigma_- \right) 
+ \xi \,v_z\, k_z\, \sigma_z\,, \quad
k_\pm = k_x \pm i\, k_y \,, \quad \sigma_\pm = \frac{\sigma_x \pm i \, \sigma_y} {2}\,,
\quad \alpha_J = \frac{v_\perp} {k_0^{J-1}}\,,
\label{eq:mweyl_ham}
\end{align}
where $\xi = \pm 1$ is the chirality of the Weyl node. Crystalline symmetry makes the two chiralities equivalent for transport - hence we restrict our analysis to $\xi = 1$ without loss of generality. Here, $v_\perp$ and $v_z$ denote the Fermi velocities perpendicular and parallel to the $k_z$-momentum, and $k_0$ is a material-dependent momentum scale.

The integer exponent $J$ is the monopole charge of the Weyl node. Crystal rotational symmetry restricts this topological index to $J \le 3$, giving a conventional Weyl semimetal ($J=1$), a double-Weyl semimetal ($J=2$), or a triple-Weyl semimetal ($J=3$). Diagonalising Eq.~\eqref{eq:mweyl_ham} gives the bulk energy spectrum $\mathcal{E}_{\text{mw}}^{\pm}(\boldsymbol{k}) = \pm\sqrt{\alpha_J^2 \, k_\perp^{2J} + v_z^2 \, k_z^2}$. The conduction and valence bands touch at the nodal point $\boldsymbol{k} = 0$, giving a gapless dispersion. For $J=1$ the spectrum is linear in all directions, describing a conventional isotropic Weyl semimetal, while for $J=2$ and $J=3$ the dispersion becomes quadratic and cubic in the transverse ($xy$) plane, respectively, remaining linear along $z$. Because the differential operators along the transport direction scale with $J$, solving the dispersion relation at fixed energy and transverse momentum gives more longitudinal wavevector solutions for $J \ge 2$.

To numerically simulate the system, we scale our Hamiltonian by $v_\perp k_0$:
\begin{align}
\frac{\mathcal{H}_{\text{mw}}} {v_\perp k_0} = \left(\frac{k_{-}} {k_0}\right)^J \sigma_+ 
+ \left(\frac{k_{+}} {k_0}\right)^J \sigma_- + \chi \, \frac{v_z\, k_z} {v_\perp \, k_0} \,\sigma_z \,,
\end{align}
such that all momentum components are measured in units of $k_0$, energy is measured in units of $v_\perp k_0$, and the length scales are in units of $1/k_0$. For calculational simplicity, we set $v_z = v_\perp$ for all $J$ values.

The physical origin of the evanescent channels that appear for $J\ge 2$ traces to the same momentum-dependent mass mechanism identified for the nodal-ring case in Sec.~\ref{subsec:nodal_line}, now generalised to a $J$-fold anisotropic band touching. For fixed transverse momentum $k_y$ and longitudinal momentum $k_z$, the dispersion relation $\alpha_J^2\,k_\perp^{2J} + v_z^2 \, k_z^2 = \mathcal{E}^2$ defines an effective one-dimensional problem along $x$, whose characteristic polynomial in $k_x$ has degree $2J$. For the isotropic Weyl case ($J=1$), this polynomial is simply quadratic, and its two roots are always either both real or both purely imaginary: an incident electron at a given $(k_y, \,k_z, \,\mathcal{E})$ either propagates freely or is fully reflected by an evanescent barrier, with no coexistence of the two behaviours. Once $J\ge 2$, the higher-order term $k_\perp^{2J}$ raises the polynomial to a higher degree, and some of its roots can remain real and propagating while the rest turn complex or purely imaginary at the very same energy and transverse momentum, exactly as found for a rectangular potential barrier in mWSMs~\cite{Deng2020} and, in the presence of an applied magnetic field~\cite{ips-aritra}. The number of evanescent branches grows correspondingly with the topological charge $J$: a single complex-conjugate pair for the double-Weyl node ($J=2$), and two complex-conjugate pairs for the triple-Weyl node ($J=3$), as we derive explicitly below.

This proliferation of evanescent channels with increasing $J$ is a generic feature of higher-order anisotropic band-touching materials, not restricted to mWSMs. A comparable structure, in which a nonlinear dispersion along one direction generates evanescent solutions distinct from the propagating ones supported along the linear direction, appears in Andreev scattering at Josephson junctions of semi-Dirac semimetals~\cite{ips-abs-semidirac}, where the quadratic-dispersion axis likewise admits states that decay into the junction rather than propagate through it. As in the nodal-ring case, these evanescent modes carry zero net probability current asymptotically and cannot themselves constitute an open transport channel. They remain indispensable, however, for satisfying the boundary conditions at the interfaces of the driven potential well, and, as we show in Sec.~\ref{subsubsec:double_weyl_roots} and Sec.~\ref{subsubsec:triple_weyl_roots} below, combine with the propagating channels to determine the complete multi-channel Floquet-scattering matrix for each value of $J$. We now analyse the multi-channel scattering structures for the $J=2$ and $J=3$ cases in turn.

\subsubsection{Double-Weyl mWSM ($J=2$)}
\label{subsubsec:double_weyl_roots}

\begin{figure}[t!]
\centering
\subfloat[\label{kz01ky01_q_vector}]{\includegraphics[scale=0.32]{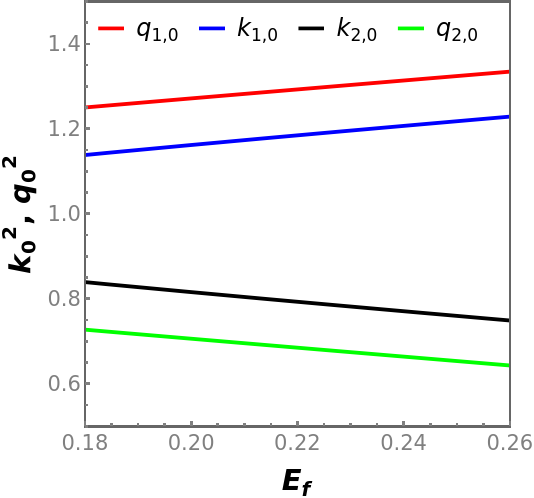}} \quad
\subfloat[\label{kz014ky012_q_vector}]{\includegraphics[scale=0.32]{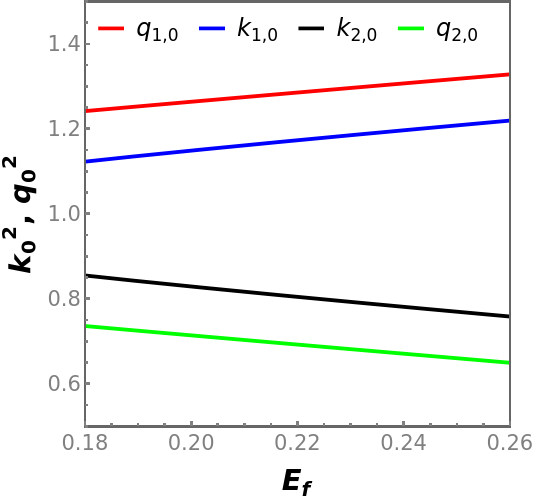}} \quad
\subfloat[\label{kz014ky01_q_vector}]{\includegraphics[scale=0.32]{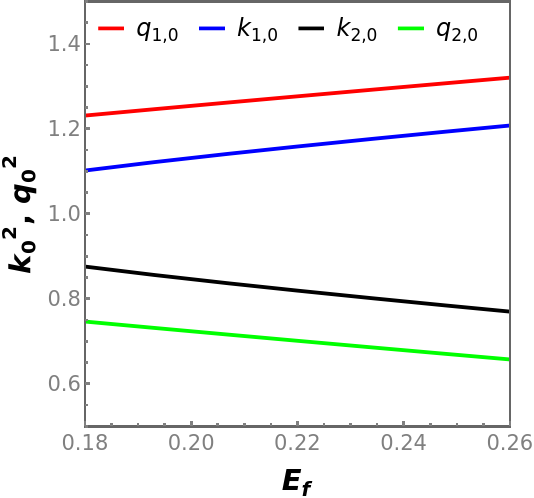}}
\caption{Variation of the squared longitudinal wavevectors as a function of Fermi energy for the undriven case ($m=n=0$). Subfigures (a), (b), and (c) show the squares of the wavevectors against the Fermi energy ($E_f$) for $k_z = 0.1\,\mathcal{M}$, $0.12\,\mathcal{M}$, and $0.14\,\mathcal{M}$, respectively, with $k_y$ fixed at $0.1\,\mathcal{M}$. Within the displayed energy-window, the squared wavevectors remain positive, indicating that all four channels are propagating and contribute directly to transport. This is the regime of complete multi-channel propagation. The plot legends encode the magnitudes of the individual momentum-vectors. All wavevectors, energy scales, and inverse lengths are expressed in units of $\mathcal{M}$. The remaining parameters are fixed at $B = \mathcal{M}^{-1}$, $\omega = 0.05\, \mathcal{M}$, $V_0 = 0.1 \, \mathcal{M}$, $V_1 = 0.025 \,\mathcal{M}$, and $L = 25 \, \mathcal{M}^{-1}$.\label{q_vector}}
\end{figure}

For a double-Weyl mWSM ($J=2$), the dispersion relation gives four longitudinal wavevector solutions ($k_x$) along the transport direction. Provided the system parameters satisfy $\mathcal{E}^2 - k_z^2 \ge 0$ and $k_y^2 < \sqrt{\mathcal{E}^2 - k_z^2}$, the solutions separate into propagating and evanescent channels. The propagating modes have a pair of purely real longitudinal wavevectors, $\pm  \, \sqrt{-k_y^2 + \sqrt{\mathcal{E}^2 - k_z^2}}$, while the remaining two branches give a pair of purely imaginary roots, defining the evanescent modes $\pm \, i \,\sqrt{k_y^2 + \sqrt{\mathcal{E}^2 - k_z^2}}$. In each spatial region across the heterojunction, the longitudinal scattering wavefunction $\psi_n(x)$ is built as a superposition of these four propagating and evanescent eigenstates:
\begin{align} 
\label{eq1:double_psi}
\pmtx{\psi_{n,1} \\ \psi_{n,2} } =
\begin{cases}
& A_{1,n}^{i}(t) \,e^{i \, k_1 \, x} \pmtx{f_{11} \\ f_{12}}
	+ A_{1,n}^{o}(t) \,e^{- \,i \, k_1 \, x} \pmtx{f_{21} \\ f_{22}}
	+ A_{2,n}^{i}(t) \,e^{ -\, \zeta_1 \, x} \pmtx{f_{31} \\ f_{32}}
	+ A_{2,n}^{o}(t) \,e^{ \zeta_1 \, x} \pmtx{f_{41} \\ f_{42}}
 \text{ for } x< -\frac{L} {2} \\ & \\
& \sum \limits _{m=-\infty}^\infty
\Bigl[ \alpha_{1,m}(t) \,e^{i \, q_1\, x} \pmtx{g_{11} \\ g_{12}}
	+ \beta_{1,m}(t) \,e^{-i \, q_1 \,x} \pmtx{g_{21} \\ g_{22}}
	+ \alpha_{2,m}(t) \,e^{-\, \zeta_2 \,x} \pmtx{g_{31} \\ g_{32}}
	+ \beta_{2,m}(t) \,e^{ \zeta_2 \, x} \pmtx{g_{41} \\ g_{42}} \Bigr]
\\ & \qquad \quad \times J_{n-m} ( {V_{1}} /{\omega} ) \, \Theta(-E_m-V_0)
\text{ for } -\frac{L} {2} \le x \le \frac{L} {2} \\ & \\
&	B_{1,n}^{i}(t) \,e^{-i \, k_1 \, x} \pmtx{f_{21} \\ f_{22}}
	+ B_{1,n}^{o}(t) \,e^{i \, k_1 \, x} \pmtx{f_{11} \\ f_{12}}
	+ B_{2,n}^{i}(t) \,e^{ \zeta_1 \, x} \pmtx{f_{41} \\ f_{42}}
	+ B_{2,n}^{o}(t) \,e^{  - \,\zeta_1 \, x} \pmtx{f_{31} \\ f_{32}}
 \text{ for } x> \frac{L} {2}
\end{cases}
\end{align}
where 
\begin{align}
&k_1 =  \sqrt{ -k_y^2 + \sqrt{\mathcal{E}^2 - k_z^2} } \,,  \quad 
\zeta_1 =  \sqrt{ k_y^2 + \sqrt{\mathcal{E}^2 - k_z^2} } \,,  \quad 
n_1=\sqrt{\frac{2E_n} { (E_n-k_z)}} \,,  \nn  &
q_1 =  \sqrt{ -k_y^2 + \sqrt{(\mathcal{E}+V_0)^2 - k_z^2} } \,,  \quad 
\zeta_2 =  \sqrt{ k_y^2 + \sqrt{(\mathcal{E} +V_0)^2 - k_z^2} } \,,  
\quad n_2=\sqrt{\frac{2(E_m+V_0)} { (E_m+V_0-k_z)}} \,, 
\end{align}
\begin{align}
&	f_{11}= \frac{(k_1 - i\, k_y)^2} {n_1 (E_n -k_z)} \,,  \quad 	f_{12}=\frac{1} {n_1} \,,  
\quad 	f_{21}= \frac{(k_1 + i\, k_y)^2} {n_1 (E_n -k_z)} \,,  \quad 	f_{22}=\frac{1} {n_1} \,,  \nn &
f_{31}= -\frac{(\zeta_1 - k_y)^2} {n_1 (E_n -k_z)} \,,  \quad 	f_{32}=\frac{1} {n_1} \,,  
\quad 	f_{41}= -\frac{(\zeta_1 + k_y)^2} {n_1 (E_n -k_z)} \,,  \quad 	f_{42}=\frac{1} {n_1} \,,  \nn  &
g_{11}= \frac{(q_1 - i\, k_y)^2} {n_2(E_n+V_0 -k_z)} \,,  \quad 	g_{12}=\frac{1} {n_2} \,,  
\quad 	g_{21}= \frac{(q_1 + i\, k_y)^2} {n_2(E_n+V_0 -k_z)} \,,  \quad 	g_{22}=\frac{1} {n_2} \,,  \nn &
g_{31}= -\frac{(\zeta_2 - k_y)^2} {n_2(E_n+V_0 -k_z)} \,,  \quad 	
g_{32}=\frac{1} {n_2} \,,  \quad 	
g_{41}= -\frac{(\zeta_2 + k_y)^2} {n_2(E_n+V_0 -k_z)} \,,  \quad g_{42}=\frac{1} {n_2}.
\end{align}
To ensure normalisability in the semi-infinite regions, the coefficients of the spatially diverging exponential solutions are again set to zero. Consequently, we enforce $A_2^i = 0$ in region I ($x \rightarrow -\infty$) and $B_2^i = 0$ in region III ($x \rightarrow +\infty$), retaining only the decaying evanescent channels for boundary matching.

Since the Hamiltonian is quadratic in the longitudinal momentum $k_x$, exactly as for the NRS case, matching the wavefunction at each interface requires continuity of $\Psi$ and of its first spatial derivative $\partial_x \Psi$, with no higher derivatives entering the boundary conditions. At the first interface ($x = -L/2$), the matching conditions between region I and region II are
\begin{align}
\label{eq:boundary_j2_x1_psi}
\Psi_{\text{I}} (-\frac{L} {2} \,,  y, z ) = \Psi_{\text{II}} (-\frac{L} {2} \,,  y, z ), \quad
 \partial_x \Psi_{\text{I}}(x, y, z)  \big |_{x=-\frac{L} {2}} 
=  \partial_x \Psi_{\text{II}}(x, y, z)  \big |_{x=-\frac{L} {2}}.
\end{align}
Similarly, at the second interface ($x = \frac{L} {2}$), the continuity conditions between region II and region III are
\begin{align}
	\label{eq:boundary_j2_x2_psi}
	\Psi_{\text{II}}\left(\frac{L} {2} \,,  y, z\right) = \Psi_{\text{III}}\left(\frac{L} {2} \,,  y, z\right), \quad
	\partial_x \Psi_{\text{II}}(x, y, z)  \big |_{x=\frac{L} {2}} 
	=  \partial_x \Psi_{\text{III}}(x, y, z)  \big |_{x=\frac{L} {2}}.
\end{align}
As in the NRS case, each interface contributes four linear constraints, two for the wavefunction and two for its first derivative, owing to the two-component spinor structure, giving eight boundary conditions in total across the heterojunction. In the static limit ($n=m=0$), Eq.~\eqref{eq1:double_psi} likewise contains twelve undetermined coefficients, four in each of the three spatial regions, so that once the four independently specifiable incident amplitudes are fixed, the remaining eight coefficients are uniquely determined by the eight boundary equations above, exactly as for the analogous static rectangular-barrier problem in mWSMs~\cite{Deng2020,ips-aritra} and in nodal-line semimetals~\cite{Khokhlov2018}.

Here, $k_1$ and $q_1$ are the real longitudinal wavevectors for the propagating modes outside and inside the potential barrier, while $\zeta_1$ and $\zeta_2$ are the corresponding wavevector magnitudes for the evanescent modes in those two regions. Since this configuration hosts exactly one propagating and one evanescent channel, the boundary-matching calculations, the block-partitioned scattering matrix, and the resulting transmission and reflection coefficients are identical to those in Sec.~\ref{subsubsec:one_propagating_one_evanescent}. The full algebraic derivation of the multi-channel Floquet-scattering matrix-elements for $J=2$ is given in Appendix~\ref{appendiX_Weyl_j2}.

\subsubsection{Triple-Weyl mWSM ($J=3$)} \label{subsubsec:triple_weyl_roots}

For a triple-Weyl mWSM ($J=3$), the dimensionless dispersion relation from the low-energy effective Hamiltonian is $\left(k_x^2+k_y^2\right)^3 = \mathcal{E}^2 - k_z^2$. Introducing the characteristic momentum scale $\left(\mathcal{E}^2 - k_z^2\right)^{1/3}$, the dispersion equation can be written compactly in terms of its complex roots as $k_x^2 + k_y^2 = \left(\mathcal{E}^2 - k_z^2\right)^{1/3} e^{i 2\pi n / 3}$, with $n \in \{0, 1, 2\}$. The primary branch, $n=0$, gives a pair of purely real longitudinal wavevectors, $k_x = \pm \sqrt{\left(\mathcal{E}^2 - k_z^2\right)^{1/3} - k_y^2}$, representing two open propagating modes. These channels are well defined provided the transverse momentum satisfies $k_y^2 \le \left(\mathcal{E}^2 - k_z^2\right)^{1/3}$.

The remaining two branches ($n=1,2$) give four complex-conjugate solutions corresponding to the closed evanescent sectors, $k_x = \pm \left(\sqrt{\frac{\sqrt{k_y^4 + \left(\mathcal{E}^2 - k_z^2\right)^{2/3}} - k_y^2} {2}} \pm i 
\, \sqrt{\frac{\sqrt{k_y^4 + \left(\mathcal{E}^2 - k_z^2\right)^{2/3}} + k_y^2} {2}}\,\right)$. The longitudinal momentum space for $J=3$ thus admits a six-component solution basis, one pair of real roots (propagating channels) and two complex-conjugate pairs (evanescent channels), provided $\mathcal{E}^2-k_z^2 \ge 0$ and $k_y^2 \le (\mathcal{E}^2-k_z^2)^{1/3}$.

\begin{figure}[t!]
\centering
\subfloat[\label{ky01kz01_rnp}]{
	\includegraphics[scale=0.32]{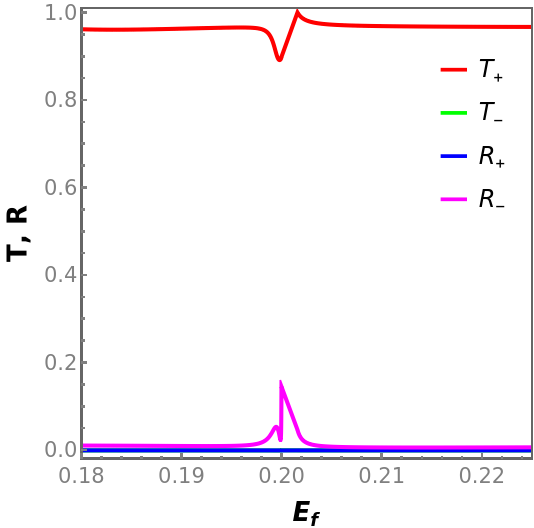}} \quad
\subfloat[\label{ky01kz012_rnp}]{
	\includegraphics[scale=0.32]{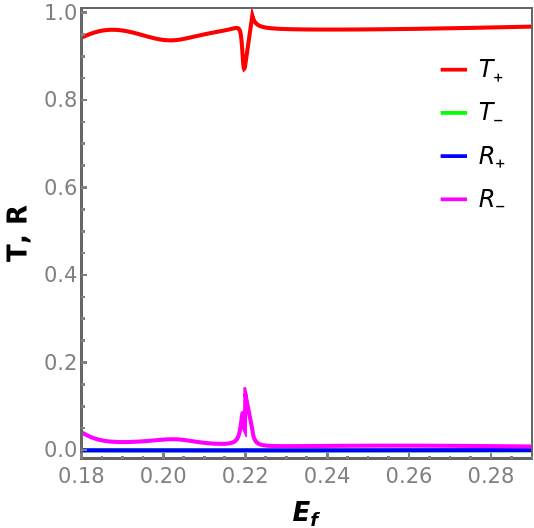}} \quad
\subfloat[\label{ky01kz014_rnp}]{
	\includegraphics[scale=0.32]{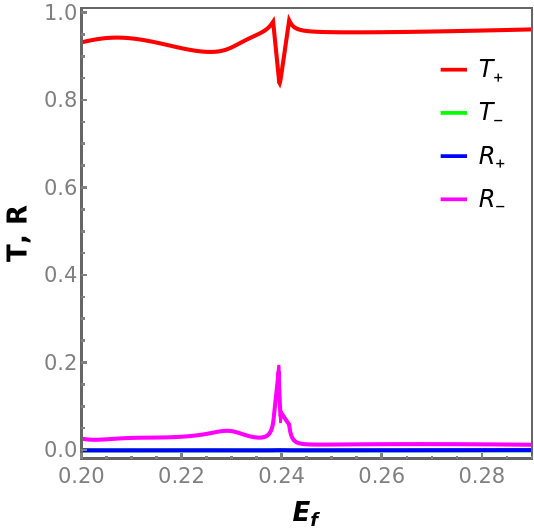}}\\
\subfloat[\label{log_ky01kz01_rnp}]{
	\includegraphics[scale=0.33]{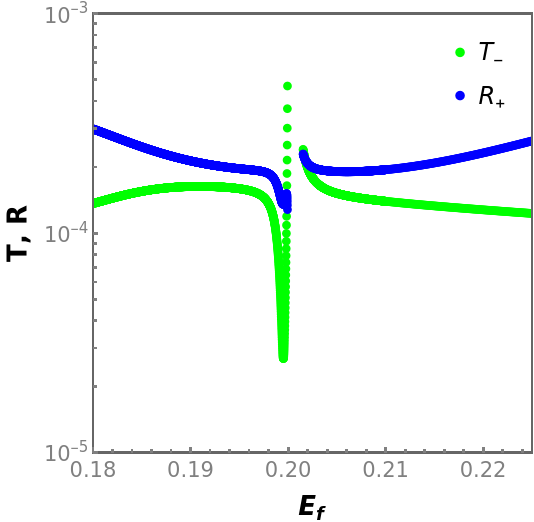}}\quad
\subfloat[\label{log_ky01kz012_rnp}]{
	\includegraphics[scale=0.33]{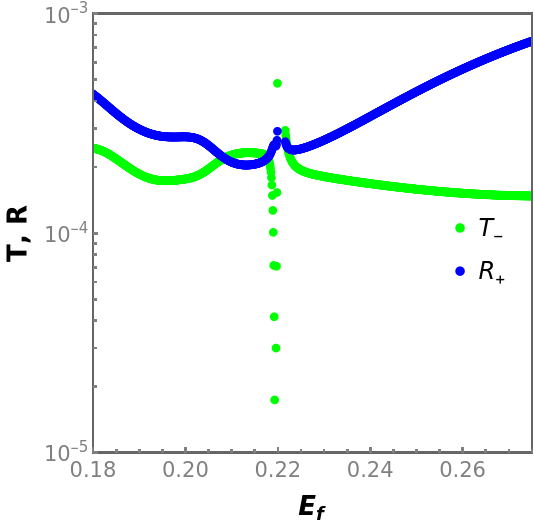}} \quad
\subfloat[\label{log_ky01kz014_rnp}]{\includegraphics[scale=0.33]{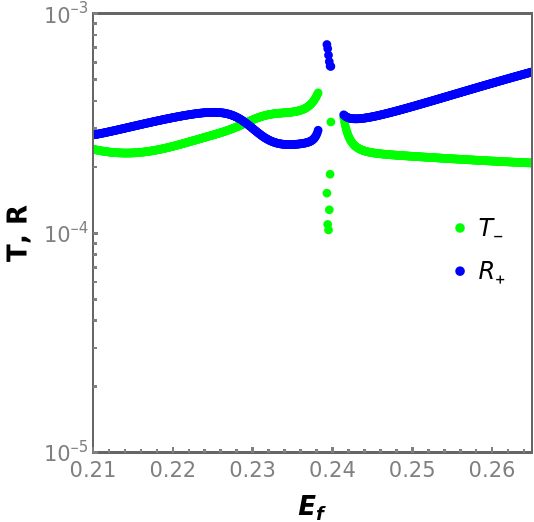}}
\caption{Reflection and transmission coefficients as functions of Fermi energy for different values of longitudinal momentum. Subfigures (a), (b), and (c) show the Fermi energy dependence of the reflection ($R_+, R_-$) and transmission ($T_+, T_-$) coefficients for three values of $k_z$, $0.1\,\mathcal{M}$, $0.12\,\mathcal{M}$, and $0.14\,\mathcal{M}$, with $k_y = 0.1\mathcal{M}$. The lower subfigures show the same data on a logarithmic scale, which highlights subtle features, such as sharp dips or small oscillations, that are obscured on a linear scale. All wavevectors, energy scales, and inverse lengths are expressed in units of $\mathcal{M}$. The remaining parameters are fixed at $B = 1\,\mathcal{M}^{-1}$, $\omega = 0.05\,\mathcal{M}$, $V_0 = 0.1\,\mathcal{M}$, $V_1 = 0.025\,\mathcal{M}$, and $L = 25\,\mathcal{M}^{-1}$.\label{rrttcoefficient}}
\end{figure}

\begin{figure}[t!]
\centering
\subfloat[\label{ky01kz02_rnp}]{
	\includegraphics[scale=0.32]{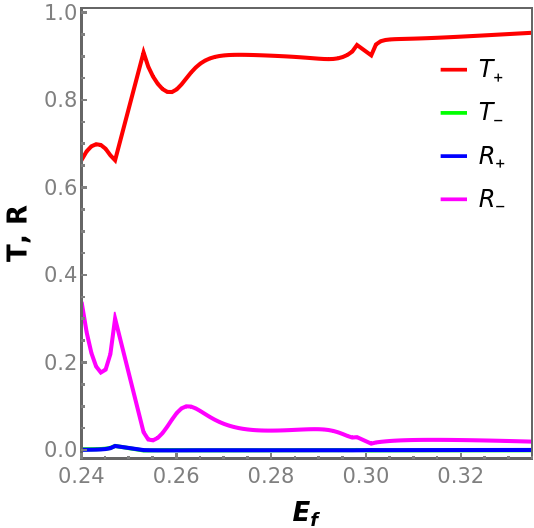}} \quad
\subfloat[\label{ky03kz02_rnp}]{
	\includegraphics[scale=0.32]{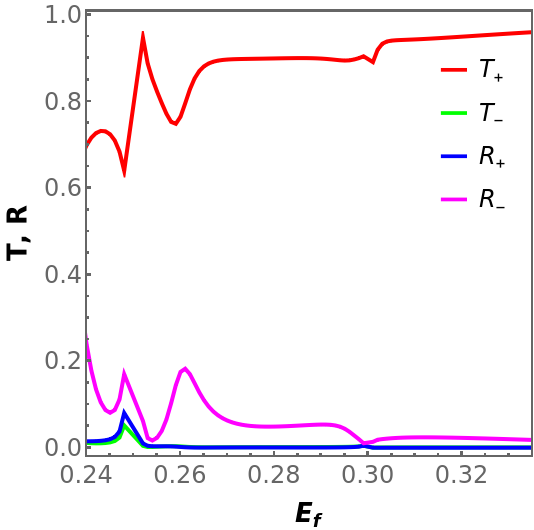}} \quad
\subfloat[\label{ky05kz02_rnp}]{
	\includegraphics[scale=0.32]{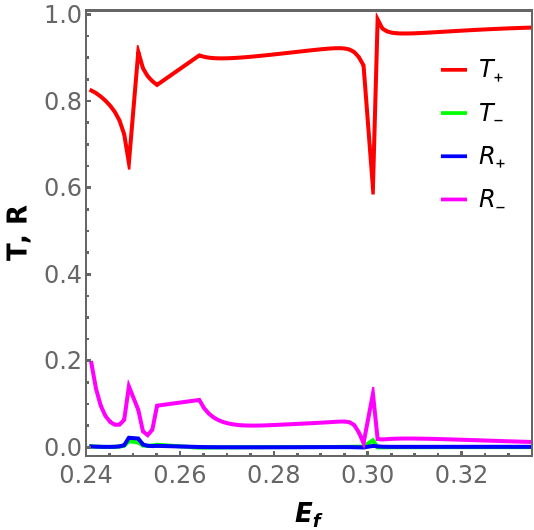}}\\
\subfloat[\label{log_ky01kz02_rnp}]{
	\includegraphics[scale=0.32]{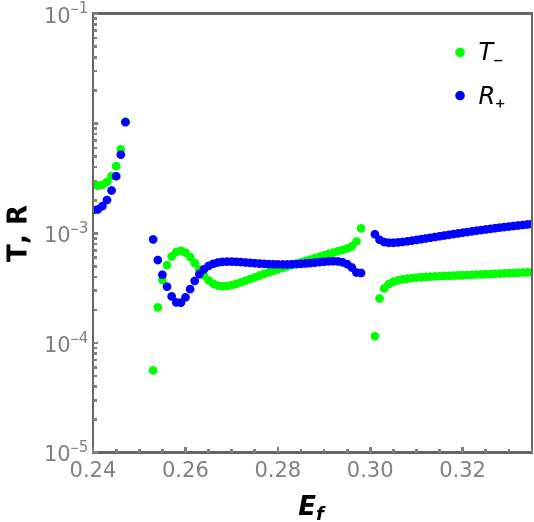}} \quad
\subfloat[\label{log_ky03kz02_rnp}]{
	\includegraphics[scale=0.32]{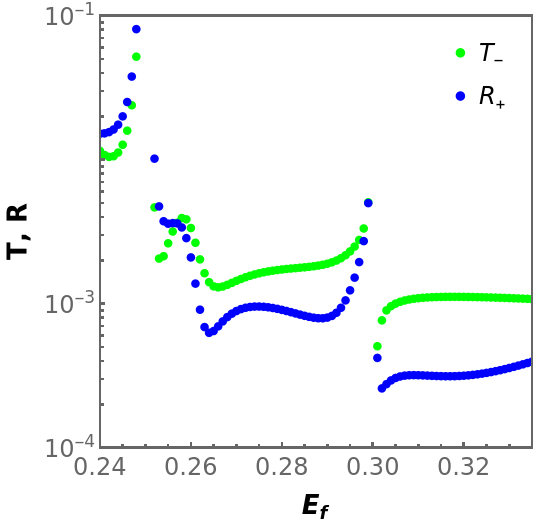}} \quad
\subfloat[\label{log_ky05kz02_rnp}]{
	\includegraphics[scale=0.32]{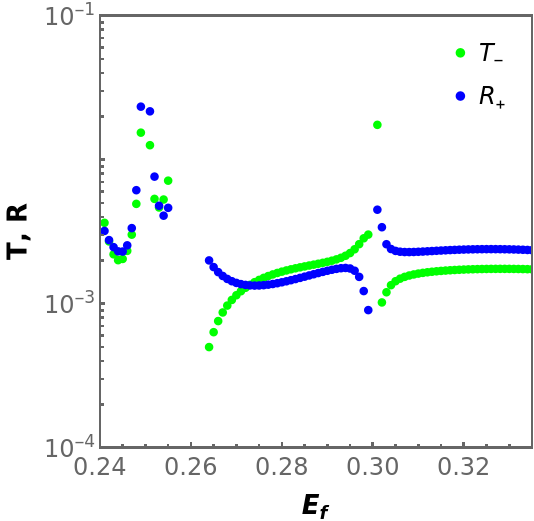}}
\caption{Reflection and transmission coefficients as functions of incident energy for different values of transverse momentum. Subfigures (a), (b), and (c) show the energy dependence of the reflection ($R_+, R_-$) and transmission ($T_+, T_-$) coefficients for three values of transverse momentum, $k_y = 0.1\,\mathcal{M}$, $0.3\,\mathcal{M}$, and $0.5\,\mathcal{M}$. Unlike Fig.~\ref{rrttcoefficient},  where $k_y$ is fixed and $k_z$ varies, here $k_z$ is kept constant at $0.2\,\mathcal{M}$ to isolate the effect of $k_y$ on the scattering characteristics. Increasing $k_y$ raises the transverse kinetic energy and changes the coupling between Floquet sidebands, shifting the resonance patterns in both channels. The lower subfigures show the same data on a logarithmic scale, which resolves fine features such as narrow resonances or heavily suppressed transmission. All wavevectors, energy scales, and inverse lengths are expressed in units of $\mathcal{M}$. The remaining parameters are fixed at $B = \mathcal{M}^{-1}$, $\omega = 0.05\,\mathcal{M}$, $V_0 = 0.1\,\mathcal{M}$, $V_1 = 0.025\,\mathcal{M}$, and $L = 25\,\mathcal{M}^{-1}$.\label{rrttcoefficientky}}
\end{figure}

Unlike the $J=2$ case, the $J=3$ Hamiltonian contains third-order differential operators along the transport direction, so the boundary conditions require continuity of the wavefunction and of its first and second spatial derivatives at each interface. At the first interface ($x = -L/2$), the matching conditions between region I and region II are
\begin{align}
\label{eq:boundary_j3_x1_psi}
\Psi_{\text{I}} (-\frac{L} {2} \,,  y, z ) = \Psi_{\text{II}} (-\frac{L} {2} \,,  y, z), \quad \partial_x \Psi_{\text{I}}(x, y, z)   \big |_{x=-\frac{L} {2}} 
= \partial_x \Psi_{\text{II}}(x, y, z)  \big |_{x=-\frac{L} {2}} \,,  \quad
 \partial_x^2 \Psi_{\text{I}}(x, y, z)  \big |_{x=-\frac{L} {2}} 
 = \partial_x^2 \Psi_{\text{II}}(x, y, z)  \big |_{x=-\frac{L} {2}}.
\end{align}
Similarly, at the second interface ($x = \frac{L} {2}$), the continuity conditions between region II and region III are
\begin{align}
	\label{eq:boundary_j3_x2_psi}
	\Psi_{\text{II}}\left(\frac{L} {2} \,,  y, z\right)  & =  
	\Psi_{\text{III}}\left(\frac{L} {2} \,,  y, z\right), \quad
\partial_x \Psi_{\text{II}}(x, y, z)  \big |_{x=\frac{L} {2}} =  
\partial_x \Psi_{\text{III}}(x, y, z)  \big |_{x=\frac{L} {2}} \,,  \quad
\partial_x^2 \Psi_{\text{II}}(x, y, z)  \big |_{x=\frac{L} {2}} = 
\partial_x^2 \Psi_{\text{III}}(x, y, z)  \big |_{x=\frac{L} {2}}.
\end{align}
Since the wavefunction is a two-component spinor, each differential constraint contributes two components, giving six matching equations at each interface and twelve boundary conditions in total across the heterojunction. In the static limit ($n=m=0$), the corresponding wavefunction, given explicitly below, contains eighteen undetermined coefficients in total: six in each of the three spatial regions, since each region hosts three channels (one propagating and two evanescent), and every channel carries an incoming and an outgoing amplitude. Fixing the six independently specifiable incident-wave amplitudes, three from each side, as external inputs leaves exactly twelve coefficients to be determined by the twelve boundary equations above, so that, exactly as for the analogous static rectangular-barrier problem in mWSMs~\cite{Deng2020,ips-aritra}, the linear system is exactly determined. The boundary-value problem for $J=3$ therefore involves a larger linear system than for $J=2$, though it remains equally well posed. Following the same procedure as for $J=2$, the scattering wavefunction is a linear combination of all six eigenmodes in each spatial region, including both propagating and evanescent solutions. Imposing these boundary conditions gives the complete Floquet-scattering matrix, containing both propagating and evanescent sectors. The corresponding wavefunction is
\begin{align} 
\label{eq:psi_j3}
\pmtx{\psi_{n,1}  \\ \psi_{n,2} } =
\begin{cases}
& A_{1,n}^{i}(t) \,e^{i \, k_1 \, x} \pmtx{f_{11} \\ f_{12}}
	+ A_{1,n}^{o}(t) \,e^{- \, i \, k_1 \, x} \pmtx{f_{21} \\ f_{22}}
	+ A_{2,n}^{i}(t) \,e^{ i \,k_2\, x} \pmtx{f_{31} \\ f_{32}}
	+ A_{2,n}^{o}(t) \,e^{ -\,i \,k_2 \, x} \pmtx{f_{41} \\ f_{42}}
\\ & 	+ \, A_{3,n}^{i}(t) \,e^{ i \,k_3 \, x} \pmtx{f_{51} \\ f_{52}}
	+ A_{3,n}^{o}(t) \,e^{ - \, i \,k_3 \, x} \pmtx{f_{61} \\ f_{62}}
\text{ for } x< -\frac{L} {2} \\ & \\
& \sum \limits _{m=-\infty}^\infty 
\Bigl[ \alpha_{1,m}(t) \,e^{i \, q_1 \, x} \pmtx{g_{11} \\ g_{12}}
	+ \beta_{1,m}(t) \,e^{-i \, q_1 \, x} \pmtx{g_{21} \\ g_{22}}
	+ \alpha_{2,m}(t) \,e^{i \, q_2 \, x} \pmtx{g_{31} \\ g_{32}}
	+ \beta_{2,m}(t) \,e^{- \,i \, q_2 \, x} \pmtx{g_{41} \\ g_{42}}
\\ & \qquad \quad \;	+\, \alpha_{3,m}(t) \,e^{i \, q_3 \, x} \pmtx{g_{51} \\ g_{52}}
	+ \beta_{3,m}(t) \,e^{-\, i \, q_3 \, x} \pmtx{g_{61} \\ g_{62}} \Bigr]
J_{n-m} ({V_{1}} / {\omega}) \, \Theta(-E_m-V_0)
 \text{ for } -\frac{L} {2} \le x \le \frac{L} {2} \\ & \\
&	B_{1,n}^{o}(t) \,e^{i \, k_1 \, x} \pmtx{f_{11} \\ f_{12}}
	+ B_{1,n}^{i}(t) \,e^{-i \, k_1 \, x} \pmtx{f_{21} \\ f_{22}}
	+ B_{2,n}^{o}(t) \,e^{ i \,k_2 \, x} \pmtx{f_{31} \\ f_{32}}
	+ B_{2,n}^{i}(t) \,e^{ -i \,k_2 \, x} \pmtx{f_{41} \\ f_{42}}
\\ & 	+ \, B_{3,n}^{o}(t) \,e^{ i \,k_3 \, x} \pmtx{f_{51} \\ f_{52}}
	+ B_{3,n}^{i}(t) \,e^{ -i \,k_3 \, x} \pmtx{f_{61} \\ f_{62}}
 \text{ for } x> \frac{L} {2}
\end{cases}.
\end{align}
where 
\begin{align}
& k_1=\sqrt{\delta_1-k_y^2} \,, \quad k_2 = \sqrt{-k_y^2
+\frac{1} {2} \left(-1+i \, \sqrt{3}\right) \delta_1} \,,  
\quad k_3=\sqrt{-k_y^2+\frac{1} {2} \left(-1-i \, \sqrt{3}\right) \delta_1} \,, 
\nn & 
q_1=\sqrt{\delta_2-k_y^2} \,, \quad q_2=\sqrt{-k_y^2+\frac{1} {2} 
\left(-1+i \, \sqrt{3}\right) \delta_2} \,,  \quad q_3=\sqrt{-k_y^2+\frac{1} {2}
 \left(-1-i \, \sqrt{3}\right) \delta_2} \,,  \nn 
&\delta_1=\sqrt[3]{E_n^2-k_z^2} \,,  \quad 
\delta_2=\sqrt[3]{(E_m+V_0)^2-k_z^2} \,,  \quad n_1=\sqrt{2 \, E_n \, (E_n-k_z)} \,,  
\quad n_2=\sqrt{2\,(E_m+V_0) \, (E_m+V_0-k_z)}\,,
\end{align}
\begin{align}
& f_{11}= \frac{(k_1 - i\, k_y)^3} {n_1 } \,,  \quad 	
f_{12}=\frac{(E_n -k_z)} {n_1} \,,  \quad 	f_{21}= \frac{(k_1 + i\, k_y)^3} {n_1 } \,,  
\quad f_{22} = \frac{(E_n -k_z)} {n_1} \,,  \nn &
f_{31}= \frac{(k_2 - i\, k_y)^3} {n_1 } \,,  \quad 	f_{32}=\frac{E_n -k_z} {n_1} \,,  
\quad 	f_{41}= \frac{(k_2 + i\, k_y)^3} {n_1 } \,,  \quad 	f_{42}=\frac{E_n -k_z} {n_1} \,,  \nn &
f_{51}= \frac{(k_3 - i\, k_y)^3} {n_1 } \,,  \quad 	f_{52}=\frac{ E_n -k_z } {n_1} \,,  \quad 	
f_{61}= \frac{(k_3 + i\, k_y)^3} {n_1 } \,,  \quad 	f_{62}=\frac{E_n -k_z} {n_1} \,,
\end{align}
\begin{align}
&g_{11}= \frac{(q_1 - i\, k_y)^3} {n_2} \,,  \quad 	g_{12}= \frac{ E_m+V_0 -k_z }  {n_2} \,,  \quad 	g_{21}= \frac{(q_1 + i\, k_y)^3} {n_2} \,,  \quad 	g_{22}= \frac{ E_m+V_0 -k_z }  {n_2} \,,  \nn &
g_{31}= \frac{(q_2 - i\, k_y)^3} {n_2} \,,  \quad 	g_{32}= \frac{ E_m+V_0 -k_z }  {n_2} \,,  \quad 	g_{41}= \frac{(q_2 + i\, k_y)^3} {n_2} \,,  \quad 	g_{42}= \frac{ E_m+V_0 -k_z }  {n_2} \,,  \nn &
g_{51}= \frac{(q_3 - i\, k_y)^3} {n_2} \,,  \quad 	g_{52}= \frac{ E_m+V_0 -k_z }  {n_2} \,,  \quad 	g_{61}= \frac{(q_3 + i\, k_y)^3} {n_2} \,,  \quad 	g_{62}= \frac{ E_m+V_0 -k_z }  {n_2}\,.
\end{align}
Here, $k_1$ and $q_1$ are the real longitudinal wavevectors for the propagating modes outside and inside the potential barrier. The pairs $(k_2, k_3)$ and $(q_2, q_3)$ form complex-conjugate longitudinal wavevectors representing oscillatory evanescent channels. To ensure physical normalisability at infinity, the coefficients corresponding to the exponentially growing components of these complex pairs must again be eliminated. We therefore enforce $A_2^i = 0$ and $A_3^i = 0$ in region I ($x \rightarrow -\infty$), and $B_2^i = 0$ and $B_3^i = 0$ in region III ($x \rightarrow +\infty$), leaving only the physically admissible channels for the boundary-matching procedure.

Inside the central potential barrier, the longitudinal wavefunction is a linear combination of all six eigenmodes. The unknown coefficients follow from imposing continuity of the two-component spinor wavefunction, and of its first and second spatial derivatives, at the interface boundaries ($x = -L/2$ and $x = \frac{L} {2}$), matching the conditions of Eq.~\eqref{eq:boundary_j3_x1_psi}. Each interface gives six linear constraints ($2$ spinor components $\times$ $3$ differential constraints), or twelve boundary equations in total across the heterojunction. This linear system determines the global Floquet-scattering matrix, relating the incoming, reflected and transmitted amplitudes across all open propagating and closed evanescent sideband channels (see Appendix~\ref{appendiX_Weyl_j3}):
\begin{align}
\begin{pmatrix}
A_{1,n}^{o} \\
A_{2,n}^{i} \\
A_{3,n}^{i} \\
B_{1,n}^{o} \\
B_{2,n}^{i} \\
B_{3,n}^{i}
\end{pmatrix}
= \sum_m \mathcal{S}_{nm} \begin{pmatrix}
A_{1,m}^{i} \\
A_{2,m}^{o} \\
A_{3,m}^{o} \\
B_{1,m}^{i} \\
B_{2,m}^{o} \\
B_{3,m}^{o}
\end{pmatrix}.
\label{eq:j3_global_S_matrix}
\end{align}
The global scattering matrix $\mathcal{S}_{nm}$ couples both the propagating and evanescent sectors. Although the evanescent modes are necessary to satisfy the higher-order boundary-matching conditions and are included in the complete wavefunctions within region II, they carry zero net time-averaged probability current asymptotically along the transport direction, and so do not contribute directly to measurable transport observables. For clarity, we partition the full multi-channel scattering matrix into its open and closed sectors:
\begin{align}
\mathcal{S}_{nm} = \begin{pmatrix}
S_{pp} & S_{pe_1} & S_{pe_2} \\
S_{e_1p} & S_{e_1e_1} & S_{e_1e_2} \\
S_{e_2p} & S_{e_2e_1} & S_{e_2e_2}
\end{pmatrix} , 
\label{eq:j3_partitioneD_S_matrix}
\end{align}
where each block is a $2 \times 2$ sub-matrix. Here $S_{pp}$ governs scattering restricted entirely to open propagating channels, while the remaining blocks describe the cross-coupling between the propagating and evanescent sectors, together with the scattering confined within the closed evanescent sector.

\section{Numerical results}
\label{secresults}

\subsection{NRSs: Regime with Two Propagating Wavevectors}

Having established the theoretical framework for Floquet scattering in NRSs, we now turn to the numerical evaluation of transmission properties and noise characteristics. In our model, four scattering channels arise due to the presence of two distinct wavevectors in each region of space. Analytical expressions for these wavevectors – $k_{1,n}$, $k_{2,n}$, $q_{1,m}$ and $q_{2,m}$ – reveal a finite energy window within which all four become real. Specifically, this interval is bounded by $k_z<E<\sqrt{(1-B\, k_y^2)^2 +k_z^2}-V_0$. The lower limit originates from the realness condition of $k_n$, while the upper limit arises from the condition on $q_m$. Within this regime, all four coefficients $T_-$, $T_+$, $R_-$ and $R_+$ are finite, indicating the presence of two active transmission and two reflection channels.

In a typical NRS such as ZrSiS~\cite{fu_nodal19} \,,  the band-crossing region is governed by a characteristic energy mass scale of $\mathcal{M} \approx 100\text{ meV}$ to $300$ meV, alongside a characteristic Fermi velocity of $v_F \approx 5 \times 10^5\text{ m/s}$~\cite{fu_nodal19, schilling2017flat}. In our numerical simulations, we choose a representative parameter configuration given by $B = \mathcal{M}^{-1}$, $\omega = 0.05 \,\mathcal{M}$, $V_0 = 0.1\,\mathcal{M}$, $V_1 = 0.025 \,\mathcal{M}$, and $L = 25\,\mathcal{M}^{-1}$, with a typical transverse momentum scale of $k \sim 0.1\,\mathcal{M}$. By taking a baseline value of $\mathcal{M} = 100$ meV, the parameters in actual units become: $\omega = 5.0$ meV, $V_0 = 10.0$ meV, $V_1 = 2.5$ meV, $L \approx 82.3$ nm, and a momentum $0.003 \text{ \AA}^{-1}$. The number of Floquet sidebands $N$ included in the simulation is determined by the drive strength, with the condition $N>V_1/ \omega$. Accordingly, we choose $N = 2$ since $V_1/\omega=0.5$.

We consider three distinct values of the longitudinal momentum, namely $k_z = 0.1\,\mathcal{M}$, $0.12\,\mathcal{M}$, and $0.14\,\mathcal{M}$, for our numerical calculations, while keeping the transverse momentum fixed at $k_y = 0.1\,\mathcal{M}$. Under this parameter configuration, the squares of all longitudinal wavevectors are plotted across a specific energy interval from $0.18\,\mathcal{M}$ to $0.26\,\mathcal{M}$. Within this window, all squared wavevectors remain strictly positive, as illustrated in Fig.~\ref{q_vector} \,,  confirming that all channels correspond to propagating modes.

\begin{figure}[t!]
	\begin{center}
		\subfloat[\label{l30}]{
			\includegraphics[scale=0.32]{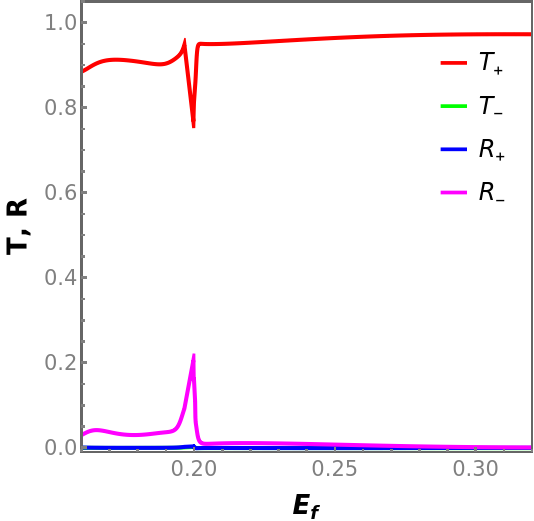}}\qquad
		\subfloat[\label{l60}]{
			\includegraphics[scale=0.32]{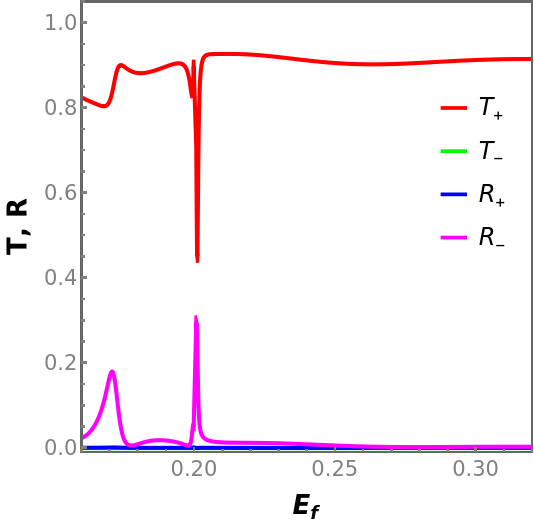}}
	\end{center}
	\caption{Reflection and transmission coefficients as functions of incident energy for different barrier widths. All wavevectors $k$, energy scales, and inverse lengths are expressed in units of $\mathcal{M}$. Subfigures (a) and (b) show the reflection ($R_+, R_-$) and transmission ($T_+, T_-$) coefficients for barrier widths of $L = 30\,\mathcal{M}^{-1}$ and $L = 60\,\mathcal{M}^{-1}$, respectively. The remaining parameters are fixed at $k_y = 0.1\mathcal{M}$, $k_z = 0.1\mathcal{M}$, $B = 1\,\mathcal{M}^{-1}$, $\omega = 0.05\,\mathcal{M}$, $V_0 = 0.1\,\mathcal{M}$, and $V_1 = 0.025\,\mathcal{M}$.}
	\label{rrttcoefficientwithl}
\end{figure}

\subsubsection{Scattering Coefficients as a Function of Energy}

Fig.~\ref{rrttcoefficient} shows the full set of scattering coefficients, the transmission ($T_+, T_-$) and reflection ($R_+, R_-$) probabilities, for these three representative values of $k_z$ with $k_y = 0.1\,\mathcal{M}$, illustrating how each transport coefficient varies with the incident energy under the Floquet-driven modulation. Among the open pathways, the transmission coefficient $T_+$ consistently has the highest amplitude across the selected parameter space, a dominance we attribute to its stronger coupling matrix-elements with the incident mode, favoured by the symmetry of the Floquet sideband structure and the underlying energy-momentum matching conditions. The coefficients $T_-$ and $R_+$, by contrast, are significantly weaker, indicating heavily suppressed transition probabilities along those pathways. To make these suppressed features visible, the lower subfigures (d), (e), and (f) show the same data on a logarithmic scale, which resolves the fine resonance structure and the relative weight of each coefficient, and highlights the asymmetry in channel coupling introduced by the combination of the periodic drive and the anisotropic band dispersion.

We also examine how the transmission and reflection coefficients respond to variations in the transverse momentum $k_y$, keeping $k_z=0.2\,\mathcal{M}$ fixed. Fig.~\ref{rrttcoefficientky} shows $T_+,\, T_-,\, R_+,\, R_-$ for $k_y=0.1\,\mathcal{M}$, $0.3\,\mathcal{M}$, and $0.5\,\mathcal{M}$. As before, $T_+$ dominates across all three values of $k_y$, confirming its robustness as the primary transmission channel. An important distinction emerges here, however: as $k_y$ increases, the transmission spectra develop multiple resonance points. These additional resonances likely arise from enhanced interference effects, driven by the increased phase space and stronger coupling to higher-order Floquet sidebands. This multiplicity of resonances illustrates the tunable nature of the system, and shows how the transverse momentum can be used to control the spectral complexity and channel selectivity of Floquet-driven transport.

\subsubsection{Scattering Coefficients as a Function of Barrier Width}

Fig.~\ref{rrttcoefficientwithl} shows how the transmission and reflection coefficients depend on the barrier width. The position of the primary Fano resonance remains nearly unchanged as $L$ varies, but its amplitude grows more pronounced for larger $L$, as seen in subfigure (b). For wider barriers, such as $L=60\,\mathcal{M}^{-1}$, additional resonance peaks emerge at lower Fermi energies $E_f$, marking the appearance of higher-order resonance modes. This trend follows from enhanced constructive interference between the incident and reflected waves: a wider barrier supports more quasi-bound states within the well, and hence a larger number of resonance conditions. This behaviour is discussed in detail in our earlier work~\cite{bera2023},  where the $L$-dependence of the oscillatory factors $e^{iq^{\pm}_m L}$ and $e^{ik^{\pm}_n L}$ is shown to set the density of resonance points, giving a richer, more finely spaced resonance spectrum as $L$ increases.

\subsubsection{Shot noise and its derivative}

In a quantum well, a Fano resonance arises when a bound state interacts with the first-order Floquet sidebands. At the resonance point, the transmission and reflection coefficients display a characteristic symmetry, reminiscent of that seen in pseudospin-1 Dirac-Weyl systems~\cite{Zhu17} but distinct from the behaviour of conventional electron gas and graphene systems~\cite{reichl199,Zhu15}. Away from the resonance, the transmission coefficient dominates, whereas exactly at the resonance the system shows either a transmission dip or a reflection peak, depending on the parity of the bound state involved.

\begin{figure}[t!]
	\begin{center}
		\subfloat[\label{n11_ky01}]{
			\includegraphics[scale=0.28]{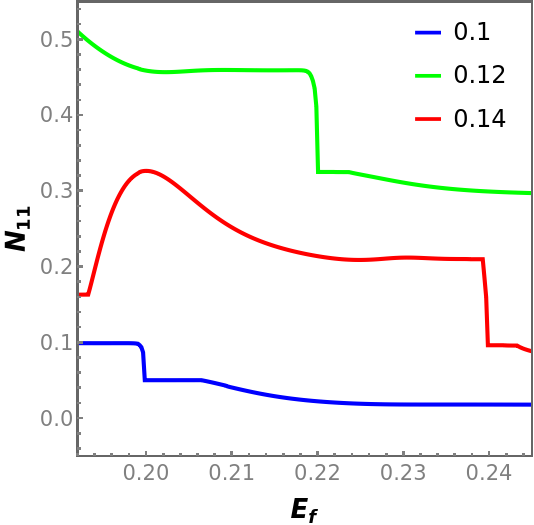}}
		\subfloat[\label{n12_ky01}]{
			\includegraphics[scale=0.29]{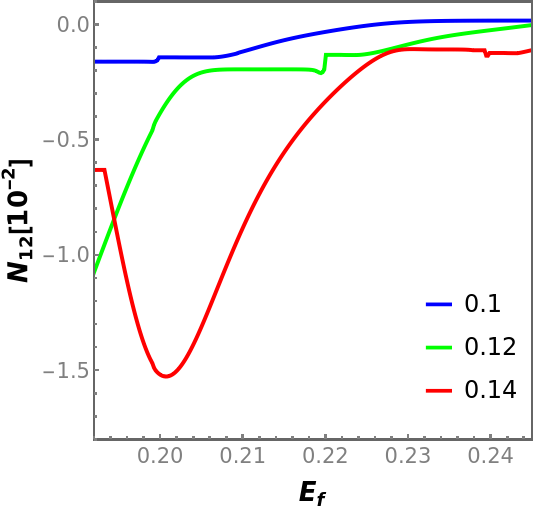}}
		\subfloat[\label{n21_ky01}]{
			\includegraphics[scale=0.28]{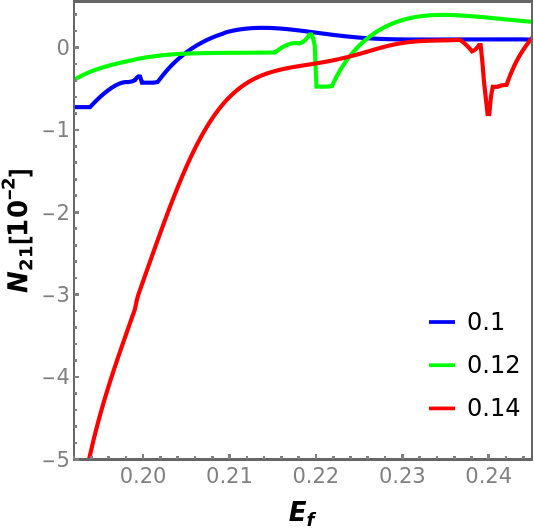}}
		\subfloat[\label{n22_ky01}]{
			\includegraphics[scale=0.28]{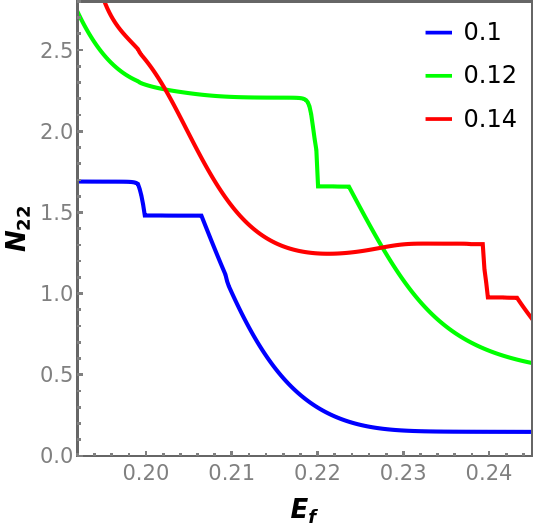}}\\
		\subfloat[\label{n11_deri_ky01}]{
			\includegraphics[scale=0.28]{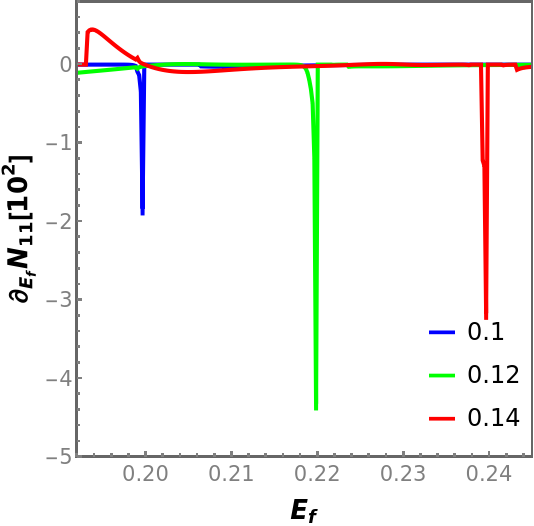}}
		\subfloat[\label{n12_deri_ky01}]{
			\includegraphics[scale=0.28]{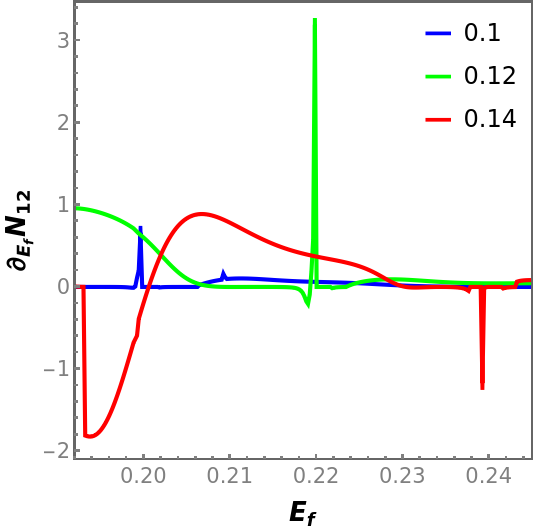}}
		\subfloat[\label{n21_deri_ky01}]{
			\includegraphics[scale=0.28]{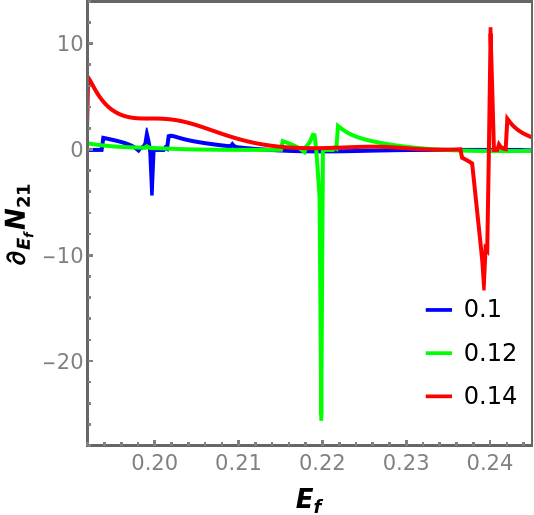}}
		\subfloat[\label{n22_deri_ky01}]{
			\includegraphics[scale=0.28]{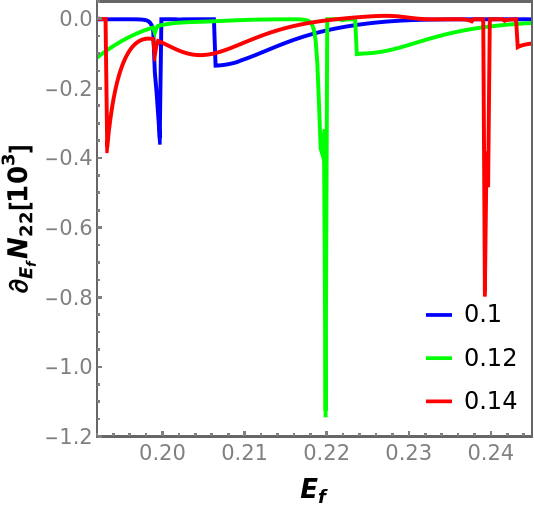}}
	\end{center}
	\caption{Variation of the components of pumped shot noise $\mathcal{N}_{LL}$ and their derivatives. Subfigures (a)–(d) show four components of the pumped shot noise $\mathcal{N}_{LL}$ (in units of $2\pi\mathcal{M}$), while the lower subfigures (e)–(h) show their derivatives (in units of $2\pi$). The curves correspond to three values of $k_z$, as indicated in the legend. All wavevectors, energy scales, and inverse lengths are expressed in units of $\mathcal{M}$. The remaining parameters are fixed at $B = 1\,\mathcal{M}^{-1}$, $\omega = 0.05\,\mathcal{M}$, $V_0 = 0.1\,\mathcal{M}$, $V_1 = 0.025\,\mathcal{M}$, and $L = 25\,\mathcal{M}^{-1}$.}
	\label{shotnoiserrtt}
\end{figure}

Fig.~\ref{shotnoiserrtt}(a)–(d) shows the individual components of the pumped shot noise, while subfigures (e)–(h) present the corresponding derivatives, all for different values of $k_z$ with $k_y=0.1\,\mathcal{M}$ fixed. The associated transmission and reflection behaviour for these parameters is shown in Fig.~\ref{rrttcoefficient} \,,  where, as noted earlier, the transmission coefficient $T_+$ dominates over the other scattering channels. Because the total shot noise $\mathcal{N}_{\alpha\beta}$ is built from the full Floquet-scattering matrix, each scattering pathway contributes to the noise profile: for the left terminal ($\alpha=\beta=L$), the total noise $\mathcal{N}_{LL}$ consists of four terms, $\mathcal{N}_{11} \,,  \mathcal{N}_{12} \,,  \mathcal{N}_{21}$, and $\mathcal{N}_{22}$, which are plotted individually. Among these, $\mathcal{N}_{11}$ contributes the most, consistent with the dominance of the $T_+$ channel in transmission, whereas $\mathcal{N}_{22}$ remains almost negligible, reflecting minimal participation from its associated scattering processes. The energy derivatives in subfigures (e)–(h) highlight sharp features, particularly resonance peaks, more clearly, and provide a more precise identification of the Fano resonances in the system.

\begin{figure}[t!]
	\begin{center}
		\subfloat[\label{2D_quasi_nodal_kX_ev_im}]{
			\includegraphics[scale=0.35]{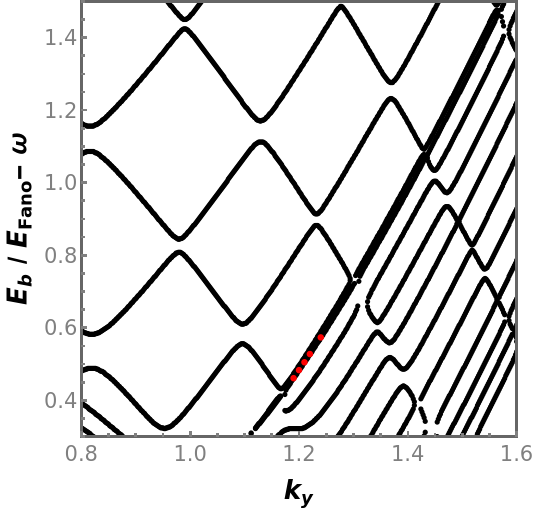}}
		\subfloat[\label{2D_quasi_nodal_ky115_im}]{
			\includegraphics[scale=0.35]{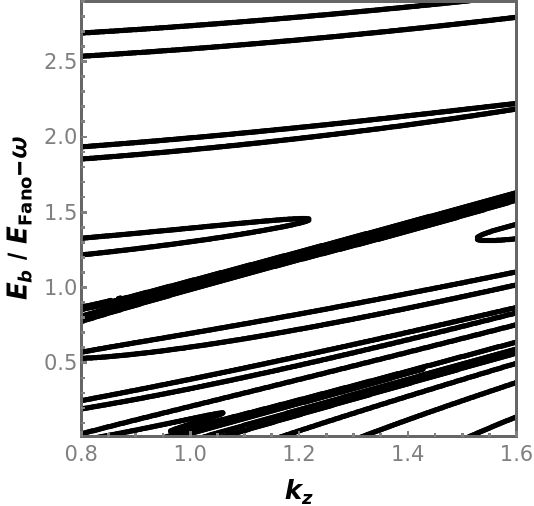}}
	\end{center}
	\caption{Evolution of bound state energies as a function of momentum and comparison with Floquet-Fano resonances. All wavevectors, energy scales, and inverse lengths are expressed in units of $\mathcal{M}$. Subfigures (a) and (b) show the bound state energy in a static quantum well versus $k_y$ (with $k_z = 0.2\,\mathcal{M}$) and $k_z$ (with $k_y = 1.15\,\mathcal{M}$), respectively. Black dots denote the exact numerical bound state energies for the undriven system. Red dots represent the bound state energies extracted from the driven system through the Fano resonance condition, $E_b = E_{\text{Fano}} - \omega$, where $E_{\text{Fano}}$ is the resonance energy of the first-order Floquet sideband. The close agreement between the two datasets confirms that the Fano resonances originate from the underlying static bound states. The remaining parameters are fixed at $B = \mathcal{M}^{-1}$, $\omega = 0.25\,\mathcal{M}$, $V_0 = 1\,\mathcal{M}$, $V_1 = 0.025\,\mathcal{M}$, and $L = 15\,\mathcal{M}^{-1}$.}
	\label{evanescentquasi}
\end{figure}

\subsubsection{Identifying Fano Resonances with (Quasi)Bound States: Two-Propagating-Channel Regime}

\begin{figure}[t!]
	\begin{center}
		\subfloat[\label{2D_quasi_nodal_kX_tnP_im}]{
			\includegraphics[scale=0.32]{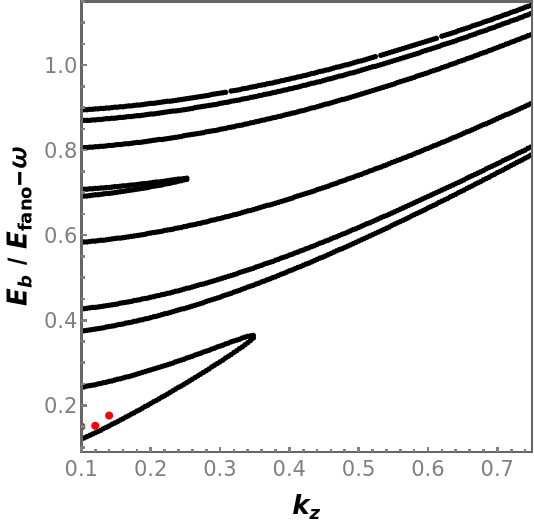}}\qquad
		\subfloat[\label{quasi_kZ_02}]{
			\includegraphics[scale=0.32]{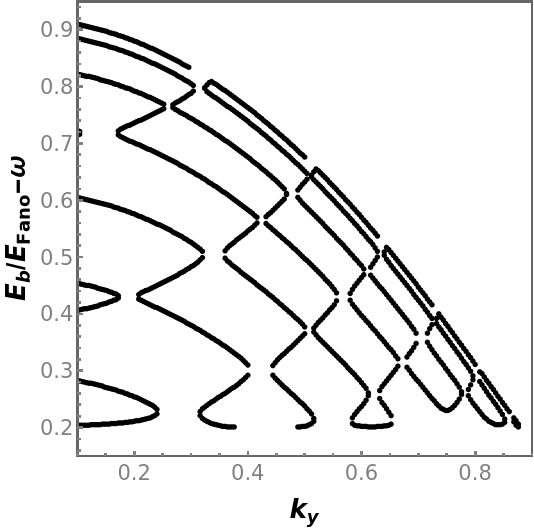}}
	\end{center}
	\caption{Evolution of bound state energies as a function of momentum and comparison with Floquet-Fano resonances. Subfigures (a) and (b) show the bound-state energies in a static quantum well versus $k_z$ (with $k_y = 0.1\,\mathcal{M}$) and $k_y$ (with $k_z = 0.2\,\mathcal{M}$), respectively. Black dots denote the exact numerical bound state energies for the undriven system. Red dots represent the bound state energies extracted from the driven system through the Fano resonance condition, $E_b = E_{\text{Fano}} - \omega$, where $E_{\text{Fano}}$ is the resonance energy of the first-order Floquet sideband. The close agreement between the two datasets confirms the correspondence between the Fano resonance and the underlying static bound state. The remaining parameters are fixed at $B = \mathcal{M}^{-1}$, $\omega = 0.05\mathcal{M}$, $V_0 = 0.1\,\mathcal{M}$, and $V_1 = 0.025\,\mathcal{M}$.}
	\label{quasirrtt}
\end{figure}

We begin by examining the bound-state solutions of a static quantum well formed in an NRS. This corresponds to setting the Floquet indices to $m=n=0$ in Eq.~\eqref{eq:nodal_two_propagator}. Enforcing continuity of the wavefunction at the potential interfaces gives the following secular equation governing the bound-state conditions:
\begin{align}\label{boundrrtt}
	\left(
	\begin{array} {cccccccc}
		f_1 \, e^{-\frac{1} {2}  i\, k_1\, L} &  f_1  \left(-e^{\frac{ i\, k_2 \,L} {2}}\right) &  g_1  \left(-e^{-\frac{i\, L} {2} q_1}\right) &  g_1  \left(-e^{\frac{i L\,q_1} {2}}\right) &  g_1  e^{-\frac{i\, L} {2} q_2} &  g_1  e^{\frac{i L\,q_2} {2}} & 0 & 0 \\
		 f_2  e^{-\frac{1} {2}  i\, k_1 \,L} &  f_2  e^{\frac{ i\, k_2 \,L} {2}} & g_2 \left(-e^{-\frac{i\, L} {2} q_1}\right) & g_2 \left(-e^{\frac{i L\,q_1} {2}}\right) & g_2 \left(-e^{-\frac{i\, L} {2} q_2}\right) & g_2 \left(-e^{\frac{i L\,q_2} {2}}\right) & 0 & 0 \\
		i  f_1  k_1 e^{-\frac{1} {2}  i\, k_1 \,L} & i  f_1  k_2 e^{\frac{ i\, k_2 \,L} {2}} & -i  g_1  q_1 e^{-\frac{i\, L} {2} q_1} & i  g_1  q_1 e^{\frac{i L\,q_1} {2}} & i  g_1  q_2 e^{-\frac{i\, L} {2} q_2} & -i  g_1  q_2 e^{\frac{i L\,q_2} {2}} & 0 & 0 \\
		i  f_2  k_1 e^{-\frac{1} {2}  i\, k_1 \,L} & -i  f_2  k_2 e^{\frac{ i\, k_2 \,L} {2}} & i g_2 q_1 e^{-\frac{i\, L} {2} q_1} & i g_2 q_1 e^{\frac{i L\,q_1} {2}} & -i g_2 q_2 e^{-\frac{i\, L} {2} q_2} & i g_2 q_2 e^{\frac{i L\,q_2} {2}} & 0 & 0 \\
		0 & 0 &  g_1  e^{\frac{i L\,q_1} {2}} &  g_1  e^{-\frac{i\, L} {2} q_1} &  g_1  \left(-e^{\frac{i L\,q_2} {2}}\right) &  g_1  \left(-e^{-\frac{i\, L} {2} q_2}\right) &  f_1  \left(-e^{-\frac{1} {2}  i\, k_1\,L}\right) & f_1 \, e^{\frac{ i\, k_2 \,L} {2}} \\
		0 & 0 & g_2 e^{\frac{i L\,q_1} {2}} & g_2 e^{-\frac{i\, L} {2} q_1} & g_2 e^{\frac{i L\,q_2} {2}} & g_2 e^{-\frac{i\, L} {2} q_2} &  f_2  \left(-e^{-\frac{1} {2}  i\, k_1\,L}\right) &  f_2  \left(-e^{\frac{ i\, k_2 L} {2}}\right) \\
		0 & 0 & i  g_1  q_1 e^{\frac{i L q_1} {2}} & -i  g_1  q_1 e^{-\frac{i\, L} {2} q_1} & -i  g_1  q_2 e^{\frac{i L q_2} {2}} & i  g_1  q_2 e^{-\frac{i\, L} {2} q_2} & i  f_1  k_1 e^{-\frac{1} {2}  i\, k_1 L} & i  f_1  k_2 e^{\frac{ i\, k_2 L} {2}} \\
		0 & 0 & i g_2 q_1 e^{\frac{i L q_1} {2}} & -i g_2 q_1 e^{-\frac{i\, L} {2} q_1} & i g_2 q_2 e^{\frac{i L q_2} {2}} & -i g_2 q_2 e^{-\frac{i\, L} {2} q_2} & i  f_2  k_1 e^{-\frac{1} {2}  i\, k_1 L} & -i  f_2  k_2 e^{\frac{ i\, k_2 L} {2}} \\
	\end{array}
	\right)=0.
\end{align}
Solving the secular equation gives the bound-state energy levels, denoted by $E_b$. The numerical solutions, obtained using the same parameters as in Fig.~\ref{rrttcoefficient} \,,  are shown as black dots in Fig.~\ref{quasirrtt}. These parameters ensure that all four wavevectors remain real, a condition necessary for coherent transport, and they also set the allowed energy window for bound states, $\sqrt{(1-B\, k_y^2)^2 +k_z^2}-V_0<E_b<\sqrt{(1-B\, k_y^2)^2 +k_z^2}$, within which bound states can exist and contribute to the observed resonance features.

Fano resonances arise from the interference between a discrete bound state and a continuum of extended states. In Floquet scattering, such a resonance occurs when the energy of a bound state inside the potential well coincides with the energy of a Floquet sideband of the incident electron. This condition is expressed as $E_b=E_{\text{Fano}} -n\omega$, where $E_b$ is the bound-state energy, $E_{\text{Fano}}$ is the Fermi energy of the incoming electron, $\omega$ is the driving frequency, and $n$ is the Floquet sideband index. When the first-order sideband ($n=1$) satisfies this condition, the resulting Fano resonance falls within the energy window of practical interest. In Fig.~\ref{quasirrtt} \,,  the red dotted lines mark the bound-state energies $E_b$ that correspond to the Fano resonance points identified earlier in Fig.~\ref{rrttcoefficient}. For example, at $k_z =0.1\,\mathcal{M}$, a prominent Fano resonance appears at $E_{\text{Fano}}=0.2497\,\mathcal{M}$ (Fig.~\ref{rrttcoefficient}a). Applying the resonance condition for $n=1$ gives a bound-state energy of $E_b=E_{\text{Fano}} -\omega =0.199\,\mathcal{M}$, which agrees well with the numerically obtained value $E_b=0.199\,\mathcal{M}$ shown in Fig.~\ref{quasi_kZ_02} \,,  confirming that the Fano resonance originates from bound-continuum interference.

\begin{figure}[t!]
	\begin{center}
		\subfloat[\label{ev_kz02ky119}]{
			\includegraphics[scale=0.32]{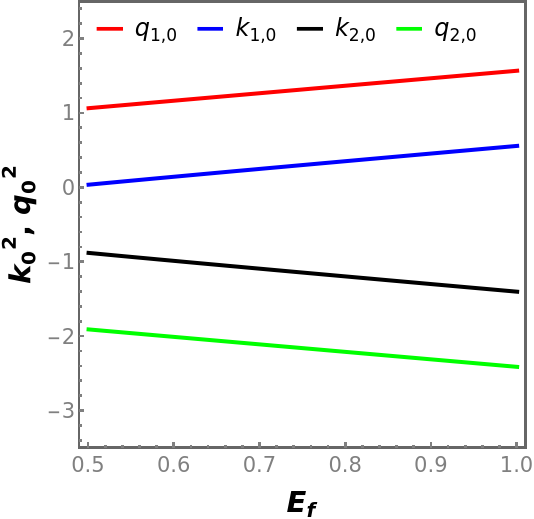}}\quad
		\subfloat[\label{ev_kz02ky120}]{
			\includegraphics[scale=0.32]{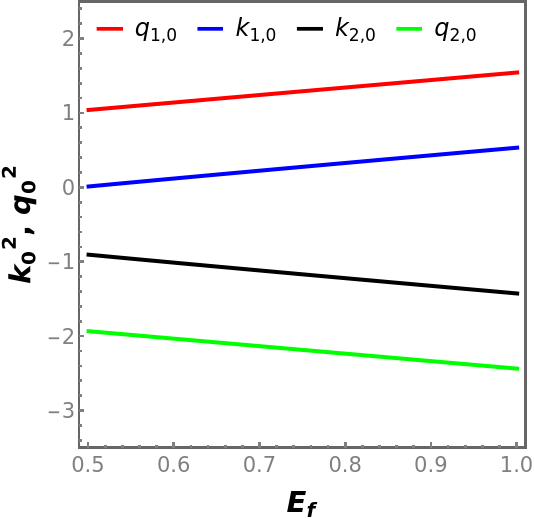}}\quad
		\subfloat[\label{ev_kz02ky121}]{
			\includegraphics[scale=0.32]{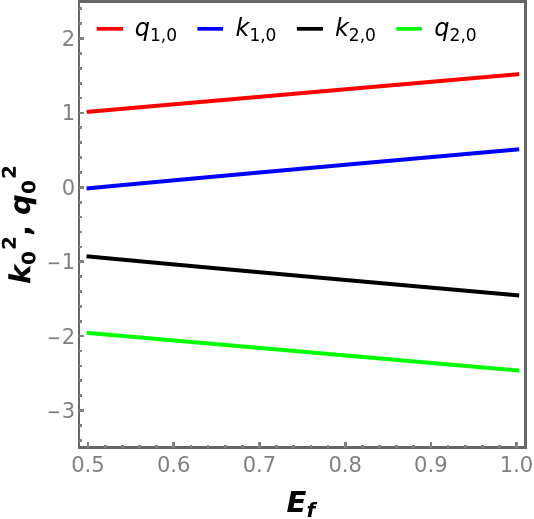}}
	\end{center}
	\caption{Variation of the squared longitudinal wavevectors as a function of Fermi energy for the undriven case ($m=n=0$). Subfigures (a), (b), and (c) correspond to three values of $k_y$. Within the energy range considered, the squares of $k_{1,n}$ and $q_{1,n}$ remain positive, representing propagating modes, whereas the squares of $k_{2,n}$ and $q_{2,n}$ become negative, indicating purely imaginary evanescent modes. All wavevectors, energy scales, and inverse lengths are expressed in units of $\mathcal{M}$. The parameters used are $B = 1\,\mathcal{M}^{-1}$, $\omega = 0.25\,\mathcal{M}$, $V_0 = 1.0\,\mathcal{M}$, $V_1 = 0.025\,\mathcal{M}$, and $L = 15\,\mathcal{M}^{-1}$.}
	\label{q_vector2}
\end{figure}

\subsection{NRSs: Regime with One Propagating and One Evanescent Wavevector}

In our numerical simulations, we choose a representative parameter configuration given by $B = \mathcal{M}^{-1}$, $\omega = 0.25\,\mathcal{M}$, $V_0 = 1\,\mathcal{M}$, $V_1 = 0.025\,\mathcal{M}$, and $L = 15\,\mathcal{M}^{-1}$, with a typical transverse momentum scale of $k \sim 1.0\,\mathcal{M}$. Taking a baseline value of $\mathcal{M} = 100$ meV from the experimental literature~\cite{fu_nodal19, schilling2017flat} \,,  these parameters correspond to $\omega = 25$ meV, $V_0 = 100$ meV, $V_1 = 2.5$ meV, $L \approx 49.4$ nm, and a momentum component of $k \sim 0.03\text{ \AA}^{-1}$. The number of Floquet sidebands $N$ included in the computation is set by the drive strength, through the cutoff condition $N > V_1/\omega$. We accordingly choose $N = 2$, consistent with the dimensionless driving ratio $V_1/\omega = 0.1$.

Before evaluating the transport coefficients, we verify that the chosen parameters correspond to a regime with exactly one propagating and one evanescent channel. Figure~\ref{q_vector2} plots the squared longitudinal wavevectors against the Fermi energy $E_f$ for three transverse momenta, $k_y = 1.19\,\mathcal{M}$, $1.20\,\mathcal{M}$, and $1.21\,\mathcal{M}$. As shown, $k_1^2$ and $q_1^2$ remain positive throughout, confirming that these modes are propagating states outside and inside the well, respectively, whereas $k_2^2 \equiv -\zeta_1^2$ and $q_2^2 \equiv -\zeta_2^2$ are negative, confirming that the corresponding modes are evanescent. This parameter choice therefore isolates a single propagating and a single evanescent scattering channel, which defines the transport regime analysed throughout this section.

\begin{figure}[t!]
\centering
\subfloat[\label{kz02ky120_ev}]{\includegraphics[scale=0.32]{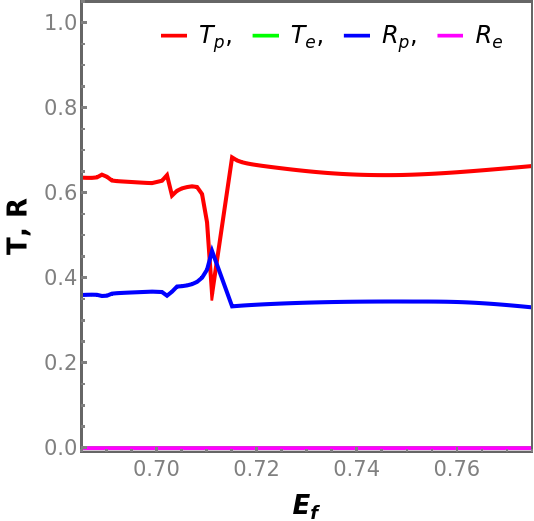}}\quad
\subfloat[\label{kz02ky122_ev}]{\includegraphics[scale=0.32]{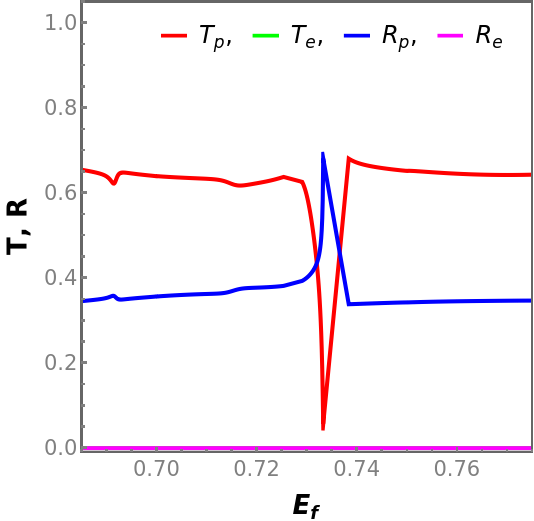}}\quad
\subfloat[\label{kz02ky124_ev}]{\includegraphics[scale=0.32]{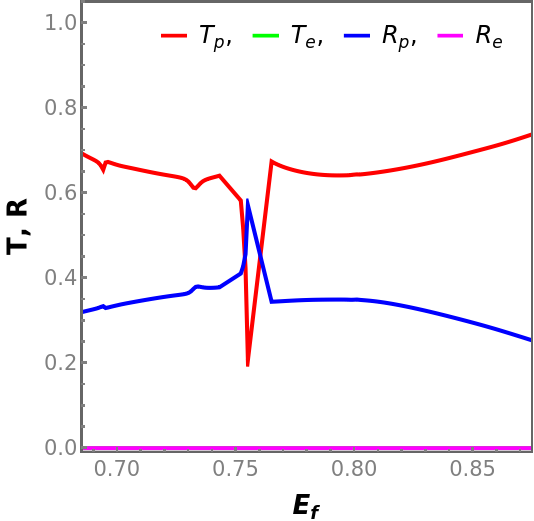}}
\caption{Reflection and transmission coefficients as functions of Fermi energy for different values of longitudinal momentum. Subfigures (a), (b), and (c) show the Fermi energy dependence of the reflection ($R_p, R_e$) and transmission ($T_p, T_e$) coefficients for three values of $k_y$, $1.19\,\mathcal{M}$, $1.2\,\mathcal{M}$, and $1.21\,\mathcal{M}$, with the transverse momentum fixed at $k_z = 0.2\,\mathcal{M}$. These coefficients give the scattering probabilities into the different Floquet channels, showing the onset of resonances and channel-opening thresholds. The reflection $R_e$ and transmission $T_e$ contributions from evanescent waves are negligible compared with those from propagating waves ($R_p, T_p$). All wavevectors, energy scales, and inverse lengths are expressed in units of $\mathcal{M}$. The remaining parameters are fixed at $B = \mathcal{M}^{-1}$, $\omega = 0.25\,\mathcal{M}$, $V_0 = 1\,\mathcal{M}$, $V_1 = 0.025\,\mathcal{M}$, and $L = 15\,\mathcal{M}^{-1}$.\label{evanescentky}}
\end{figure}

\begin{figure}[t!]
\centering
\subfloat[\label{kz01ky115_ev}]{\includegraphics[scale=0.32]{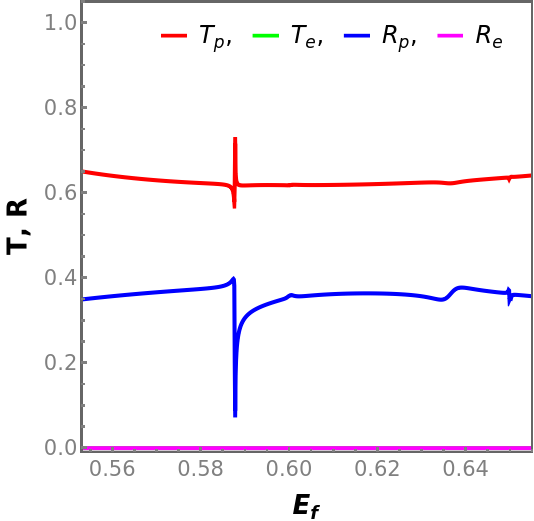}}\quad
\subfloat[\label{kz012ky115_ev}]{\includegraphics[scale=0.32]{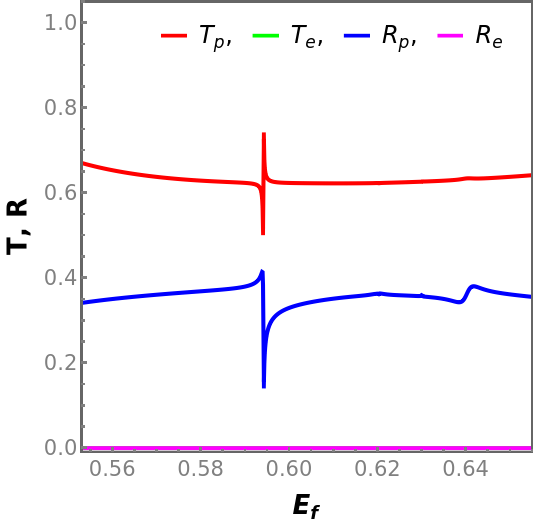}} \quad
\subfloat[\label{kz014ky115_ev}]{
\includegraphics[scale=0.32]{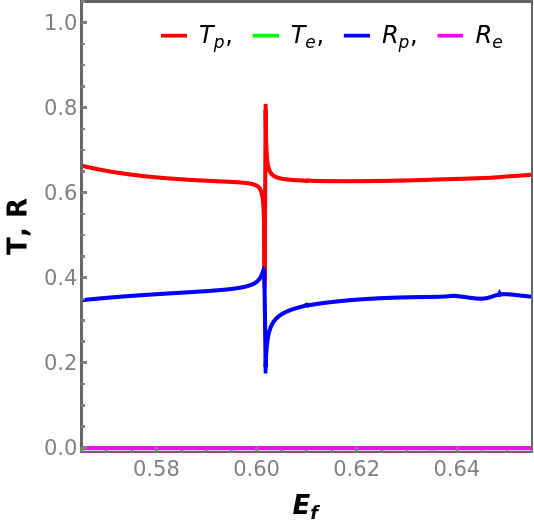}}
\caption{Reflection and transmission coefficients as functions of Fermi energy for different values of longitudinal momentum. Subfigures (a), (b), and (c) show the Fermi energy dependence of the reflection ($R_p, R_e$) and transmission ($T_p, T_e$) coefficients for three values of $k_z$, $0.1\,\mathcal{M}$, $0.12\,\mathcal{M}$, and $0.14\,\mathcal{M}$, with the transverse momentum fixed at $k_y = 1.15\,\mathcal{M}$. These coefficients give the scattering probabilities into the different Floquet channels, showing the onset of resonances and channel-opening thresholds. The reflection $R_e$ and transmission $T_e$ contributions from evanescent waves are negligible compared with those from propagating waves ($R_p, T_p$). All wavevectors, energy scales, and inverse lengths are expressed in units of $\mathcal{M}$. The remaining parameters are fixed at $B = \mathcal{M}^{-1}$, $\omega = 0.25\,\mathcal{M}$, $V_0 = 1\,\mathcal{M}$, $V_1 = 0.025\,\mathcal{M}$, and $L = 15\,\mathcal{M}^{-1}$.\label{evanescentkz}}
\end{figure}

\subsubsection{Reflection and transmission amplitudes}

The transmission and reflection amplitudes, $T_p$, $T_e$, $R_p$, and $R_e$, are shown in Fig.~\ref{evanescentky} for three values of $k_y=1.19\,\mathcal{M}$, $1.20\,\mathcal{M}$, and $1.21\,\mathcal{M}$ at fixed $k_z=0.2\,\mathcal{M}$. The evanescent-sector contributions, $T_e$ and $R_e$, remain negligible for all parameters considered. The measurable transport properties are therefore determined entirely by the propagating-sector contributions, $T_p$ and $R_p$. The transmission and reflection spectra display the characteristic Fano resonance features, with a symmetric line shape similar to that reported in pseudospin-1 Dirac-Weyl systems \cite{Zhu17} \,,  in contrast to the asymmetric behaviour seen in electron-gas and graphene systems \cite{reichl199, Zhu15}. Away from the resonance, the transmission coefficient remains dominant over the other scattering contributions, whereas near the resonance the reflection contribution becomes significant.

The states $\psi_{1}$ and $\psi_2$ correspond to odd and even parities, respectively. Fano resonances mediated by even-parity bound states feature perfect transmission followed by complete reflection, whereas those mediated by odd-parity bound states show the opposite behaviour. The resonance energy increases with $k_y$, as shown in Fig.~\ref{evanescentky}. This trend matches the behaviour of pseudospin-1 Dirac-Weyl systems \cite{Zhu17} but contrasts with that of graphene \cite{reichl199, Zhu15}. The resonance bandwidth also broadens as $k_y$ increases.

Figure~\ref{evanescentkz} shows the variation of the reflection and transmission coefficients for three values, $k_z = 0.1\,\mathcal{M}$, $0.12\,\mathcal{M}$, and $0.14\,\mathcal{M}$, at $k_y = 1.15\,\mathcal{M}$. The reflection and transmission coefficients display the same symmetric behaviour at the Fano resonance as in Fig.~\ref{evanescentky} \,,  with the resonance energy increasing with $k_z$. The peak amplitude also grows with $k_z$. The shift in resonance energy with $k_y$ is markedly smaller than that with $k_z$.

\begin{figure}
	\begin{center}
		\subfloat[\label{n11_kz02}]{
			\includegraphics[scale=0.29]{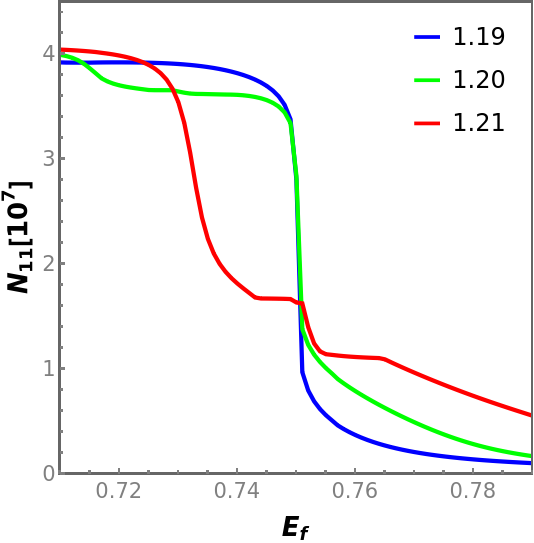}}
		\subfloat[\label{n12_kz02}]{
			\includegraphics[scale=0.29]{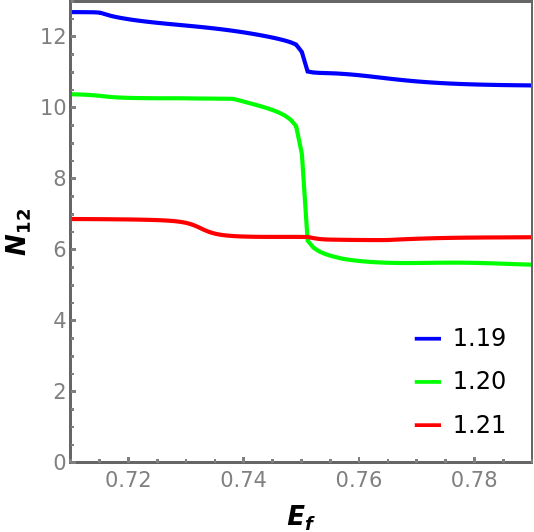}}
		\subfloat[\label{n21_kz02}]{
			\includegraphics[scale=0.29]{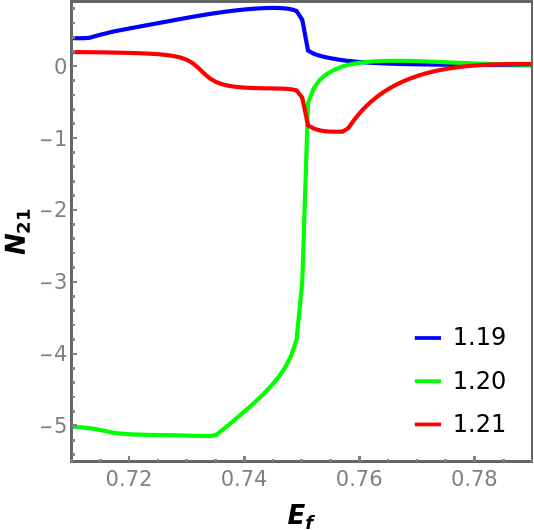}}
		\subfloat[\label{n22_kz02}]{
			\includegraphics[scale=0.29]{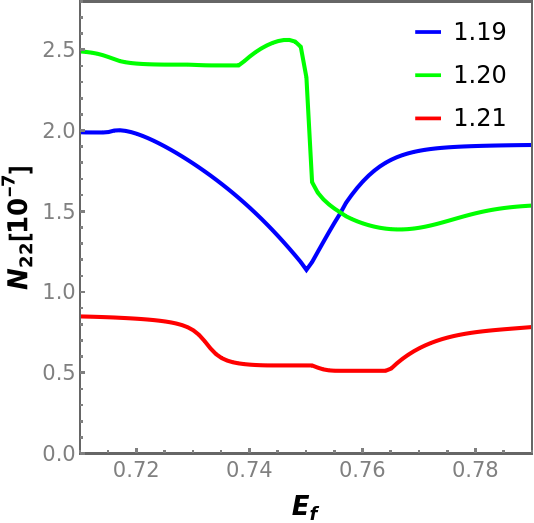}}	\\	
		\subfloat[\label{deri_n11_kz02}]{
			\includegraphics[scale=0.3]{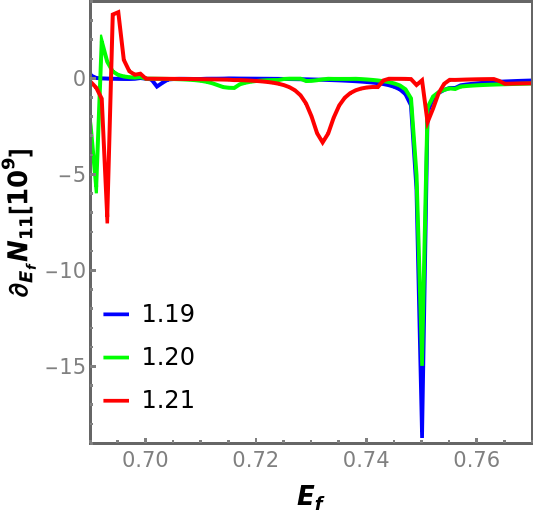}}
		\subfloat[\label{deri_n12_kz02}]{
			\includegraphics[scale=0.29]{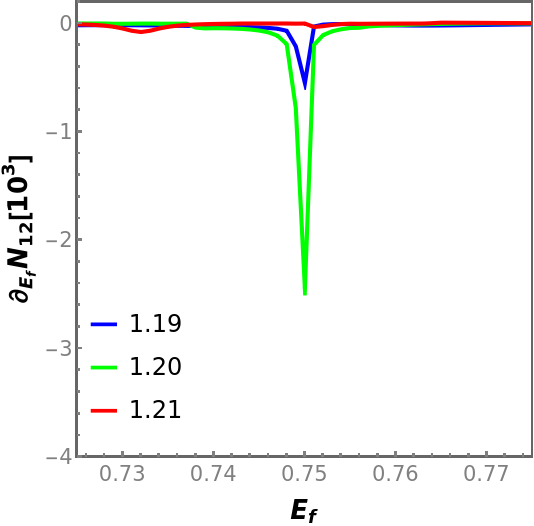}}
		\subfloat[\label{deri_n21_kz02}]{
			\includegraphics[scale=0.3]{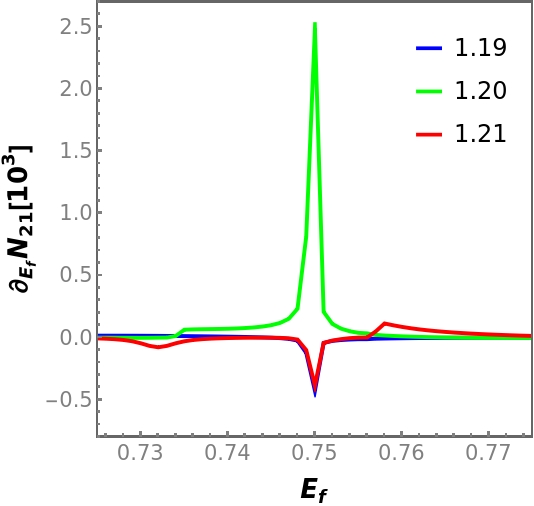}}
		\subfloat[\label{deri_n22_kz02}]{
			\includegraphics[scale=0.29]{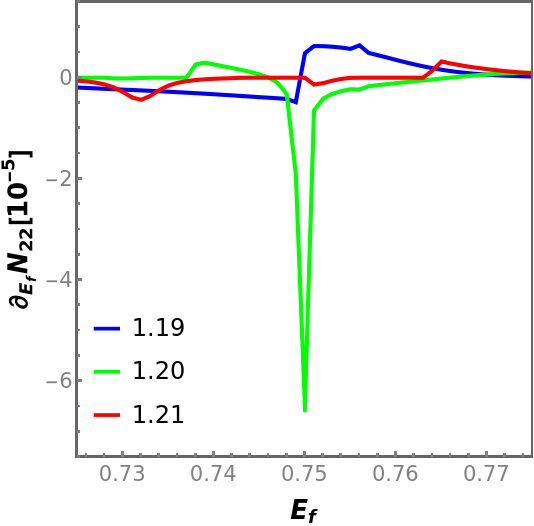}}
	\end{center}
	\caption{Variation of the propagating, evanescent, and cross-sector pumped shot-noise contributions and their derivatives as a function of Fermi energy. Subfigures (a)–(d) show the Fermi energy dependence of the pumped shot-noise components $\mathcal{N}_{11}$ (propagating sector), $\mathcal{N}_{12}$ (mixed cross-sector), $\mathcal{N}_{21}$ (mixed cross-sector), and $\mathcal{N}_{22}$ (evanescent sector), for three values of $k_y$. Subfigures (e)–(h) show the corresponding derivatives. The noise components are expressed in units of $2\pi\mathcal{M}$ and their derivatives in units of $2\pi$. The sharp dips in the shot noise and the associated peaks in their derivatives mark the onset of Fano resonances. All wavevectors, energy scales, and inverse lengths are expressed in units of $\mathcal{M}$. The remaining parameters are fixed at $B = \mathcal{M}^{-1}$, $\omega = 0.25\,\mathcal{M}$, $V_0 = 1\,\mathcal{M}$, $V_1 = 0.025\,\mathcal{M}$, and $L = 15\,\mathcal{M}^{-1}$.}
	\label{evanescentfano}
\end{figure}

\subsubsection{Shot Noise and its Derivative}\label{subsubsec:shot_noise_results}

Figures~\ref{evanescentfano}(a–d) and (e–h) show the pumped shot noise components and their energy derivatives, respectively, for three transverse momenta, $k_y = 1.19\,\mathcal{M}$, $1.20\,\mathcal{M}$, and $1.21\,\mathcal{M}$, at fixed $k_z = 0.2\,\mathcal{M}$. To identify the origin of the current fluctuations, we separate the noise contributions into distinct matrix sectors: $\mathcal{N}_{11}$ corresponds to the propagating sector, $\mathcal{N}_{22}$ to the evanescent sector, and $\mathcal{N}_{12}$ and $\mathcal{N}_{21}$ to the mixed cross-sectors coupling propagating and evanescent states. As shown in Figs.~\ref{evanescentfano}(a–d), the propagating component $\mathcal{N}_{11}$ dominates across the entire energy window, while the evanescent contribution $\mathcal{N}_{22}$ and the mixed cross-sector terms remain negligible. Figures~\ref{evanescentfano}(e–h) show the corresponding energy derivatives of these shot noise components. Sharp derivative peaks mark the onset of Fano resonances and provide a clear proxy for locating the resonance energies. The near-zero background of the evanescent-containing sectors ($\mathcal{N}_{22}$, $\mathcal{N}_{12}$, $\mathcal{N}_{21}$), in both the raw shot noise spectra and their derivatives, confirms that the resonant transport properties originate entirely from the propagating scattering channels.

\subsubsection{Identifying Fano Resonances with (Quasi)Bound States: Mixed Regime}

The numerical solutions of the longitudinal wavevector roots for the parameters used in Fig.~\ref{evanescentky} are shown as black dots in Fig.~\ref{evanescentquasi}. A Fano resonance occurs when a bound state becomes energetically aligned with a Floquet sideband of the incident channel, satisfying the condition $E_b=E_{\text{Fano}}-n\,\omega$, where $n$ is the Floquet sideband index. In the present case, the resonance features within the considered energy range arise from the first-order sideband. The bound-state energies extracted from the Fano resonance positions in Fig.~\ref{evanescentky} are marked by the red dotted lines in Fig.~\ref{evanescentquasi}.

For example, at $k_y=1.2\,\mathcal{M}$, one resonance energy is found at $E_{\text{Fano}}=0.733\,\mathcal{M}$ (Fig.~\ref{evanescentky}(b)). Using the resonance condition with $\omega=0.25\,\mathcal{M}$ gives a bound-state energy of $E_b=E_{\text{Fano}}-\omega=0.483\,\mathcal{M}$, which agrees well with the value obtained directly from Fig.~\ref{evanescentquasi} \,,  $E_b=0.482\,\mathcal{M}$. The same correspondence holds for all other resonance points marked by red dots. The number of bound states also increases with $k_y$, reflecting the growing availability of quasi-bound states responsible for the Fano resonances. The dependence of the bound-state spectrum on $k_z$ is shown in Fig.~\ref{2D_quasi_nodal_ky115_im}.

\subsection{mWSMs: Double-Weyl Semimetal ($J=2$)}

\begin{figure}[t!]
\begin{center}
\subfloat[]{
	\includegraphics[scale=0.32]{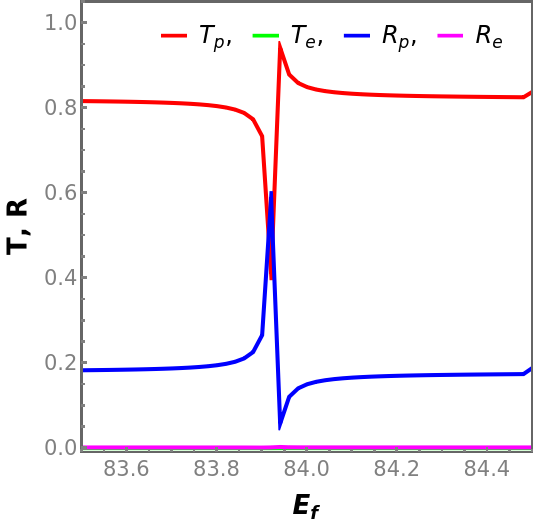}} 
\subfloat[]{
	\includegraphics[scale=0.32]{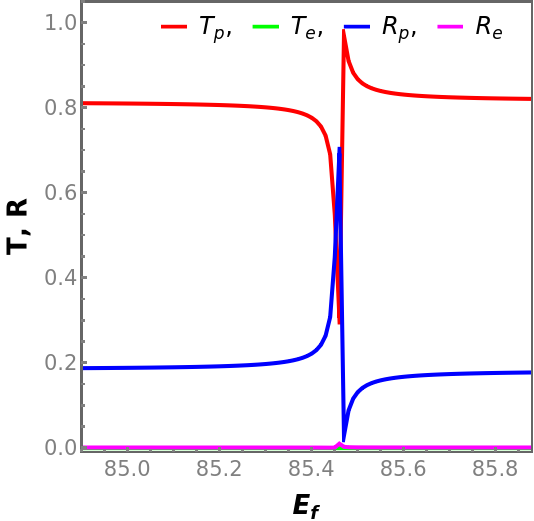}} 
\subfloat[]{
	\includegraphics[scale=0.32]{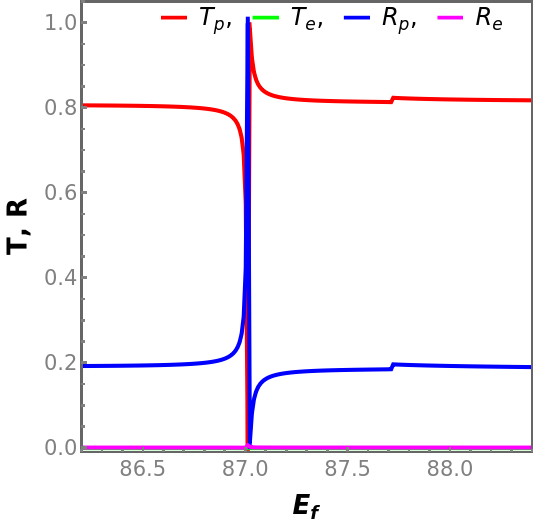}} 
	\subfloat[]{
	\includegraphics[scale=0.32]{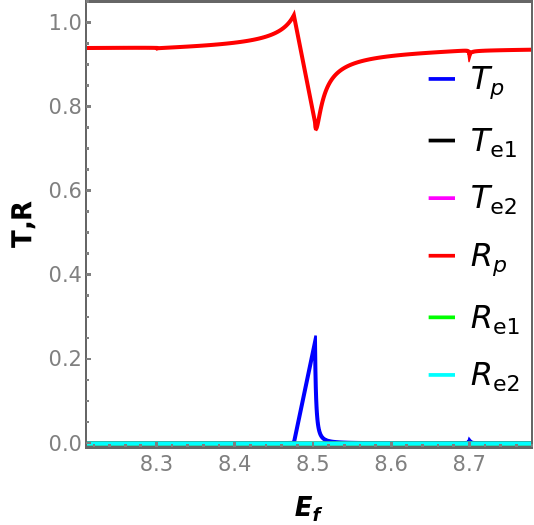}}
\end{center}
	\caption{Reflection and transmission coefficients as functions of incident energy for different values of transverse momentum in mWSMs. Subfigures (a)–(c) show the energy dependence of the coefficients for a double-Weyl ($J=2$) semimetal across three transverse wavevectors, $k_y = 8.0\,k_0$, $8.1\,k_0$, and $8.2\,k_0$, with $k_z = 8.0\,k_0$ fixed. Subfigure (d) shows the corresponding transport coefficients for a triple-Weyl ($J=3$) semimetal at fixed wavevectors $k_y = 0.2\,k_0$ and $k_z = 0.2\,k_0$. The scattering states are constructed by including both propagating and evanescent modes in the boundary-matching procedure, while the transport coefficients are obtained strictly from the propagating channels. For both topological charges ($J=2,3$), the evanescent-sector amplitudes remain numerically negligible over the entire energy range investigated and do not modify the transmission and reflection probabilities. All wavevectors, energy scales, and inverse lengths are expressed in units of $k_0$, $v_\perp k_0$, and $k_0^{-1}$, respectively. For subfigures (a)–(c), the remaining parameters are fixed at $\omega = 20.0\,v_\perp k_0$, $V_0 = 80.0\,v_\perp k_0$, $V_1 = 2.0\,v_\perp k_0$, and $L = 0.55\,k_0^{-1}$. For subfigure (d), the parameters are fixed at $\omega = 5.0\,v_\perp k_0$, $V_0 = 1.5\,v_\perp k_0$, $V_1 = 0.5\,v_\perp k_0$, and $L = 1.5\,k_0^{-1}$.}
	\label{TRforj2}
\end{figure}

We numerically evaluate the scattering amplitudes for the parameter configuration established earlier. The resulting transmission and reflection probabilities are shown in Fig.~\ref{TRforj2}(a–c) as a function of the incident energy for three values of the transverse wavevector, $k_y = 8.0\,k_0$, $8.1\,k_0$, and $8.2\,k_0$. Both propagating and evanescent modes are included in the boundary-matching procedure to construct the complete scattering states. The quantities $T_e$ and $R_e$ denote the transport coefficients of the evanescent sector, and these remain zero or numerically negligible over the entire energy range considered, in all three subfigures. The propagating channel contributions, $T_p$ and $R_p$, by contrast, show the characteristic asymmetric line shapes of Fano resonances, closely mirroring the behaviour established in our earlier work. These results confirm that, although evanescent modes are needed to satisfy the boundary conditions and construct the complete wavefunctions at the interfaces, they make no direct contribution to the measurable transport signatures. The observed Fano resonances originate entirely from the propagating scattering channels.

\subsection{mWSMs: Triple-Weyl Semimetal ($J=3$)}

Following the same procedure as for $J=2$, we numerically evaluate the scattering amplitudes by including both propagating and evanescent modes. The resulting transmission and reflection probabilities are shown in Fig.~\ref{TRforj2}(d) as a function of the incident energy for a transverse wavevector of $k_y = 0.2\,k_0$.

The evanescent-sector transport coefficients, $T_{e_1}$, $T_{e_2}$, $R_{e_1}$, and $R_{e_2}$, remain zero or numerically negligible over the entire energy range considered. The propagating channel contributions, $T_p$ and $R_p$, by contrast, show the characteristic asymmetric line shapes of Fano resonances, closely mirroring the behaviour established in the preceding sections.

These results confirm that, although evanescent modes are needed to satisfy the boundary conditions and construct the complete wavefunctions at the interfaces, they do not affect the measurable transport signatures. The observed Fano resonances originate entirely from the propagating scattering channels.

\section{Summary, Discussion, and Concluding Remarks}\label{sec:summary}

In this work, we have developed a comprehensive Floquet scattering theory for periodically driven NRSs and mWSMs, incorporating both propagating and evanescent modes into the global scattering states and the multi-channel Floquet-scattering matrix. Building on our earlier study~\cite{bera2023},  which treated only the propagating sector, we have solved the complete boundary-value problem at each interface of the driven potential well, retaining every admissible longitudinal wavevector solution for the NRS and for the double- and triple-Weyl mWSMs. We identified a previously unexplored transport regime in NRSs in which two propagating channels coexist, giving rise to genuine multi-channel quantum interference and to Floquet-induced Fano resonances that have no counterpart in the single-channel problem. By analysing the longitudinal dispersion relation for each topological charge $J$, we derived a complete scattering formalism that accounts for every admissible propagating and evanescent wavevector solution, together with the associated pumped shot noise.

Our numerical results show a consistent picture across both material families. In the two-propagating-channel regime of the NRS, the channel $T_+$ dominates the transport at essentially all energies and transverse momenta considered, while $T_-$ and $R_+$ remain strongly suppressed except in narrow windows around the Fano resonances, where the roles of transmission and reflection are locally inverted. The associated pumped shot noise inherits this hierarchy, with the component $\mathcal{N}_{11}$ tracking the dominant channel and the remaining components contributing only weakly. In the mixed regime, and throughout the double- and triple-Weyl mWSMs, the evanescent-sector coefficients $T_e$ and $R_e$ (and their multi-channel analogues $T_{e_1} \,,  T_{e_2} \,,  R_{e_1} \,,  R_{e_2}$ for $J=3$) remain zero or numerically negligible over every energy window we examined, even though the corresponding evanescent modes are indispensable for satisfying the boundary conditions at the interfaces. The transmission, reflection, pumped shot noise, and the associated Fano lineshapes are therefore determined entirely by the propagating channels. We further verified, for both the two-propagating and the mixed regime of the NRS, that the Fano resonances coincide with the (quasi-)bound states of the corresponding static potential well through the condition $E_b = E_{\text{Fano}} - n\omega$: the bound-state energies extracted independently from the secular equation agree with those inferred from the resonance positions in the driven transmission spectra to within our numerical precision, confirming that these resonances originate from the coherent coupling between the Floquet sidebands of the incident electron and the localised states of the well.

These findings sharpen the picture obtained for isotropic band-touching systems. Fano resonances arising from bound-continuum interference are by now a familiar feature of mesoscopic transport, having been documented in two-dimensional electron gases and graphene~\cite{Zhu15,reichl199,Dai2014,Tekman93} \,,  in pseudospin-1 Dirac-Weyl systems~\cite{Zhu17} \,,  and in quadratic band-touching semimetals~\cite{bera2021} \,,  and the underlying interference mechanism traces back to Fano's original analysis of configuration interaction~\cite{Fano61}. The same physics recurs, with different microscopic origins, in photonic and plasmonic nanostructures~\cite{Cui2019,Sun2014,Karmakar19} \,,  which underlines the generality of the interference mechanism across otherwise unrelated wave platforms. What distinguishes NRSs and mWSMs is the anisotropy of their dispersion: because the band touching extends along a ring or is stretched by the topological charge $J$, a potential well oriented along a nonlinear dispersion direction opens more than one propagating channel and admits genuinely multi-channel Fano physics, together with evanescent sectors of higher dimension for $J=2,3$ that have no analogue in the isotropic case. Our result that these evanescent modes are dynamically inert, despite their formal necessity, is consistent with the general expectation that only current-carrying modes contribute to asymptotic transport~\cite{beenakker03} \,,  and it extends the validation of this expectation from adiabatic and single-channel pumps~\cite{brouwer98,avron01,kamenev00} to the nonadiabatic, multi-channel, and topologically anisotropic setting studied here. Practically, it also means that the transport predictions of our earlier propagating-only treatment~\cite{bera2023} were correct for the observables computed there, and it justifies neglecting the evanescent sector in future calculations of the same type, at a substantial saving in computational and analytical complexity.

The pumped shot noise offers a particularly sensitive diagnostic of the propagating-evanescent separation established here. Because the noise kernel $M_{\alpha\beta\gamma\delta}$ is quartic in the scattering amplitudes, any residual leakage between the propagating and evanescent sectors would be amplified relative to the linear-order transmission and reflection coefficients, yet the cross-sector components $\mathcal{N}_{12}$ and $\mathcal{N}_{21}$ remain as negligible as the pure evanescent component $\mathcal{N}_{22}$ throughout our simulations. This is consistent with related studies of pumped and driven noise in other mesoscopic systems, including adiabatically modulated graphene double barriers~\cite{Zhu_2011},  charge pumping in driven quantum wires~\cite{Zhu02},  spin-dependent pumping through magnetic domain walls~\cite{Zhu10},  harmonically driven ballistic graphene~\cite{shot_noise},  and the edge states of inverted-band HgTe quantum wells~\cite{shot_hg},  where noise likewise proves to be a robust probe of the underlying scattering channels. The energy derivative of the noise, which we used throughout to pinpoint resonance positions, could equally serve as a practical experimental signature: a sharp derivative feature in the noise spectrum is, in principle, easier to isolate from background than a modest dip or peak in the transmission itself, and pumped noise measurements of the kind envisaged for periodically driven mesoscopic conductors~\cite{platero2004,kamenev00} are within reach of existing techniques.

Several of the material candidates already invoked to motivate our choice of parameters suggest that the regimes studied here are experimentally accessible. NRSs with band-crossing energy scales of a few hundred meV, such as ZrSiS~\cite{fu_nodal19},  together with Cu$_3$PdN~\cite{xiao_nodal15} and Mg$_3$Bi$_2$~\cite{chang_nodal19},  provide a natural setting for the two-propagating-channel regime identified in Sec.~\ref{secresults}, while double- and triple-Weyl nodes have been predicted or observed in HgCr$_2$Se$_4$ and SrSi$_2$~\cite{Gang2011, hasan_mweyl16} and in transition-metal monochalcogenides~\cite{liu2017predicted},  with the broader family of monopnictide Weyl semimetals (TaAs, TaP, NbAs, and NbP) offering further scope for anisotropic multi-Weyl physics once tilt and higher-order corrections are accounted for~\cite{grassano18, grassano20}. Realising the driven potential well of Eq.~\eqref{eq:potential_well} experimentally would call for a gate-defined barrier modulated at radio to microwave frequencies, of the kind already used to demonstrate photon-assisted tunnelling and quantum pumping in conventional 2d electron-systems~\cite{oosterkamp98, blumenthal07, dicarlo03, switkes99}. The principal experimental challenge is likely to be resolving the comparatively small pumped noise signal against thermal and $1/f$ backgrounds at the meV drive amplitudes considered here, which will require operation at low temperature and long averaging times, as is standard in shot-noise measurements of mesoscopic conductors~\cite{beenakker03}.

Several extensions of the present formalism suggest themselves. We have worked throughout in natural units at zero temperature and within a single-particle picture. Incorporating a finite temperature into the Fermi-Dirac factors of Eq.~\eqref{eq:shotnoise_general} is straightforward, but a more substantial extension would be to ask how electron-electron interactions and disorder, which are known to reshape the low-energy physics of quadratic and Luttinger band-touching semimetals into non-Fermi-liquid or nematic phases~\cite{Sun09,  ips-seb, cho16, ips-hermann-review}, modify the Floquet-Fano resonances and the evanescent-propagating separation established here. A magnetic field applied along or transverse to the barrier would couple to the pseudospin structure of both Hamiltonians and could be treated within the same scattering framework, in the spirit of related studies of Klein tunnelling and magnetotransport in mWSMs~\cite{mansoor, rahul-jpcm, ips-ruiz, ips-cjp, ips-tilted, ipsita-shivam}. It would also be natural to move beyond the two-terminal geometry considered here to multi-terminal devices and to the full counting statistics of the transferred charge, extending the noise formalism used throughout this work~\cite{avron01, kamenev00},  and to examine other transport observables sensitive to the anisotropic, higher-$J$ dispersion, such as the thermopower~\cite{ips-kush, lundgren2014thermo, ips-kush-review}, the chiral photocurrent~\cite{ips-photocurrent, kozii}, circular dichroism~\cite{sajid-cd, ips_cd}, and the Magnus-Hall response~\cite{sajid_magnus22},  all of which are already known to be sensitive to the same anisotropy that drives the multi-channel physics reported here.

Collectively, our results establish a consistent and unified theoretical framework for multi-channel Floquet scattering in anisotropic topological semimetals. They confirm that the transport properties and Fano features reported in our earlier study remain unchanged once the full set of scattering modes is included, while extending that work by uncovering a genuinely new multi-channel transport regime in NRSs with coexisting propagating states. We hope that the formalism and the diagnostic role of the pumped shot noise developed here will prove useful in future studies of Floquet transport in nodal-ring and multi-Weyl semimetals, and more broadly in other anisotropic topological materials where multiple propagating channels and evanescent sectors coexist.

\appendix

\section{Detailed calculation of the scattering matrix for nodal-ring semimetals}
\label{appendiX_nodal_two}

In this appendix we derive the full Floquet-scattering matrix for the nodal-ring semimetal, including both propagating and evanescent modes. In each spatial region the complete wavefunction is formed from four independent solutions: two propagating modes with real longitudinal wavevectors and two evanescent modes with imaginary longitudinal wavevectors. The continuity of the two-component spinor wavefunction and its first spatial derivative at the two interfaces, $x=-L/2$ and $x=L/2$, yields a set of linear equations that relate the scattering amplitudes in the three regions. (Note that the auxiliary matrices $U$, $V$, and, in Appendices~\ref{appendiX_Weyl_j2} and~\ref{appendiX_Weyl_j3}, $P$, $D$, $Z$, $X$, $M$, $N$, are defined independently in each appendix and do not carry the same meaning across them; each appendix is self-contained.)

At the left- and right-interfaces, the boundary conditions give
\begin{align}
\begin{pmatrix}
A_1^i \\ A_2^o \\ B_1^i \\ B_2^o \end{pmatrix} = U \begin{pmatrix}
\alpha_1 \\\beta_1 \\ \alpha_2 \\
\beta_2 \end{pmatrix} \text{ and }
\begin{pmatrix} A_1^o \\ A_2^i \\ B_1^o \\ B_2^i
\end{pmatrix} = V \begin{pmatrix}
\alpha_1 \\ \beta_1 \\ \alpha_2 \\
\beta_2 \end{pmatrix} , 
\end{align}
respectively. Here, $A_1^i$, $ A_1^o$, $ B_1^i$, and $ B_1^o$ denote the amplitudes of the propagating modes, while $A_2^i$, $ A_2^o$, $ B_2^i$, and $ B_2^o$ represent the amplitudes of the evanescent modes. The symbols $\alpha_1$, $\beta_1$, $\alpha_2$, and $ \beta_2$ represent the modes inside the driven potential well. Eliminating the internal amplitudes yields
\begin{align}
\begin{pmatrix} A_1^o \\ A_2^i \\ B_1^o \\ B_2^i \end{pmatrix}
=V \,U^{-1}\begin{pmatrix}
A_1^i \\ A_2^o \\ B_1^i \\ B_2^o
\end{pmatrix}.
\end{align}
$U$ and $V$ are $4\times 4$ matrices constructed from the boundary conditions at $x=-L/2$ and $x=L/2$, respectively. Their explicit elements are as follows:
\begin{align}
U_{11} & = \frac{1} {4} e^{\frac{i\, L} {2} ( k_1-q_1)}  
\left(\frac{ g_1 } {f_1} +\frac{g_2} {f_2}\right) \left(\frac{q_1} {k_1}+1\right)
J_{n-m} ( V_1 / \omega )  \,, \quad
U_{12} =\frac{1} {4} e^{\frac{i\, L} {2} ( k_1+q_1)}  
\left(\frac{ g_1 } {f_1} +\frac{g_2} {f_2}\right) 
\left(1-\frac{q_1} {k_1}\right) J_{n-m} ( V_1 / \omega )  \,, \nn
U_{13} & = \frac{1} {4} e^{\frac{i\, L} {2} ( k_1-q_2)} 
 \left(-\frac{ g_1 } {f_1} +\frac{g_2} {f_2}\right) 
 \left(\frac{q_2} {k_1}+1\right) J_{n-m} ( V_1 / \omega )  \,, \quad
U_{14} =\frac{1} {4} e^{\frac{i\, L} {2} ( k_1+q_2)}  
\left(-\frac{ g_1 } {f_1} +\frac{g_2} {f_2}\right) 
\left(1-\frac{q_2} {k_1}\right) J_{n-m} ( V_1 / \omega )  \,, \nn
U_{21} & = \frac{1} {4} e^{-\frac{i\, L} {2} ( k_2+q_1)}  
\left(-\frac{ g_1 } {f_1} +\frac{g_2} {f_2}\right) 
\left(1-\frac{q_1} {k_2}\right) J_{n-m} ( V_1 / \omega )  \,, \quad
U_{22} =\frac{1} {4} e^{-\frac{i\, L} {2} ( k_2-q_1)}  
\left(-\frac{ g_1 } {f_1} +\frac{g_2} {f_2}\right) 
\left(\frac{q_1} {k_2}+1\right) J_{n-m} ( V_1 / \omega )  \,, 
\nn
U_{23} & = \frac{1} {4} e^{-\frac{i\, L} {2} ( k_2+q_2)}  \left(\frac{ g_1 } {f_1} +\frac{g_2} {f_2}\right) \left(1-\frac{q_2} {k_2}\right) J_{n-m} ( V_1 / \omega )  \,, \quad
U_{24}  = \frac{1} {4} e^{-\frac{i\, L} {2} ( k_2-q_2)}  
\left(\frac{ g_1 } {f_1} +\frac{g_2} {f_2}\right) \left(\frac{q_2} {k_2}+1\right) J_{n-m} ( V_1 / \omega )  \,, \nn
U_{31} & = \frac{1} {4} e^{\frac{i\, L} {2} ( k_1+q_1)}  \left(\frac{ g_1 } {f_1} +\frac{g_2} {f_2}\right) \left(1-\frac{q_1} {k_1}\right) J_{n-m} ( V_1 / \omega )  \,, \quad
U_{32}  = \frac{1} {4} e^{\frac{i\, L} {2} ( k_1-q_1)} 
\left(\frac{ g_1 } {f_1} +\frac{g_2} {f_2}\right) \left(\frac{q_1} {k_1}+1\right) J_{n-m} ( V_1 / \omega )  \,,\nn
U_{33} & = \frac{1} {4} e^{\frac{i\, L} {2} ( k_1+q_2)}  
\left(-\frac{ g_1 } {f_1} +\frac{g_2} {f_2}\right) \left(1-\frac{q_2} {k_1}\right) J_{n-m} ( V_1 / \omega )  \,, \quad
U_{34}  = \frac{1} {4} e^{\frac{i\, L} {2} ( k_1-q_2)} 
\left(-\frac{ g_1 } {f_1} +\frac{g_2} {f_2}\right)
 \left(\frac{q_2} {k_1}+1\right) J_{n-m} ( V_1 / \omega )  \,,  \nn
U_{41} & = \frac{1} {4} e^{-\frac{i\, L} {2} ( k_2-q_1)} 
 \left(-\frac{ g_1 } {f_1} +\frac{g_2} {f_2}\right) \left(\frac{q_1} {k_2}+1\right) J_{n-m} ( V_1 / \omega )  \,, \quad
U_{42}  = \frac{1} {4} e^{-\frac{i\, L} {2} ( k_2+q_1)} 
\left(-\frac{ g_1 } {f_1} +\frac{g_2} {f_2}\right) 
\left(1-\frac{q_1} {k_2}\right) J_{n-m} ( V_1 / \omega )  \,, \nn
U_{43} & = \frac{1} {4} e^{-\frac{i\, L} {2} ( k_2-q_2)}  
\left(\frac{ g_1 } {f_1} +\frac{g_2} {f_2}\right) 
\left(\frac{q_2} {k_2}+1\right) J_{n-m} ( V_1 / \omega )  \,, \quad
U_{44}  = \frac{1} {4} e^{-\frac{i\, L} {2} ( k_2+q_2)}  \left(\frac{ g_1 } {f_1} +\frac{g_2} {f_2}\right) \left(1-\frac{q_2} {k_2}\right) J_{n-m} ( V_1 / \omega )  \,,
\end{align}
\begin{align}
V_{11}  & = \frac{1} {4} e^{-\frac{i\, L} {2} ( k_1+q_1)}  \left(\frac{ g_1 } {f_1} +\frac{g_2} {f_2}\right) \left(1-\frac{q_1} {k_1}\right) J_{n-m} ( V_1 / \omega )  \,, \quad
V_{12} =\frac{1} {4} e^{-\frac{i\, L} {2} ( k_1-q_1)}  \left(\frac{ g_1 } {f_1} +\frac{g_2} {f_2}\right) \left(\frac{q_1} {k_1}+1\right) J_{n-m} ( V_1 / \omega )  \,, \nn
V_{13} & = \frac{1} {4} e^{-\frac{i\, L} {2} ( k_1+q_2)}  \left(-\frac{ g_1 } {f_1} +\frac{g_2} {f_2}\right) \left(1-\frac{q_2} {k_1}\right) J_{n-m} ( V_1 / \omega )  \,,  \quad
V_{14} =\frac{1} {4} e^{-\frac{i\, L} {2} ( k_1-q_2)}  \left(-\frac{ g_1 } {f_1} +\frac{g_2} {f_2}\right) \left(\frac{q_2} {k_1}+1\right) J_{n-m} ( V_1 / \omega )  \,, \nn
V_{21} & = \frac{1} {4} e^{\frac{i\, L} {2} ( k_2-q_1)}  \left(-\frac{ g_1 } {f_1} +\frac{g_2} {f_2}\right) \left(\frac{q_1} {k_2}+1\right) J_{n-m} ( V_1 / \omega )  \,,  \quad
V_{22}=\frac{1} {4} e^{\frac{i\, L} {2} (k_2+q_1)} 
\left(-\frac{ g_1 } {f_1} +\frac{g_2} {f_2}\right) \left(1-\frac{q_1} {k_2}\right) J_{n-m} ( V_1 / \omega )  \,, \nn
V_{23} & = \frac{1} {4} e^{\frac{i\, L} {2} (k_2-q_2)}  
\left(\frac{ g_1 } {f_1} +\frac{g_2} {f_2}\right) 
\left(\frac{q_2} {k_2}+1\right) J_{n-m} ( V_1 / \omega )  \,,  \quad
V_{24}=\frac{1} {4} e^{\frac{i\, L} {2} ( k_2+q_2)}  
\left(\frac{ g_1 } {f_1} +\frac{g_2} {f_2}\right) 
\left(1-\frac{q_2} {k_2}\right) J_{n-m} ( V_1 / \omega )  \,, \nn
V_{31} & = \frac{1} {4} e^{-\frac{i\, L} {2} ( k_1-q_1)}  
\left(\frac{ g_1 } {f_1} +\frac{g_2} {f_2}\right)
\left(\frac{q_1} {k_1}+1\right) J_{n-m} ( V_1 / \omega )  \,,  \quad
V_{32}=\frac{1} {4} e^{-\frac{i\, L} {2} ( k_1+q_1)} 
 \left(\frac{ g_1 } {f_1} +\frac{g_2} {f_2}\right) 
 \left(1-\frac{q_1} {k_1}\right) J_{n-m} ( V_1 / \omega )  \,, \nn
V_{33} & = \frac{1} {4} e^{-\frac{i\, L} {2} ( k_1-q_2)} 
\left(-\frac{ g_1 } {f_1} +\frac{g_2} {f_2}\right) 
\left(\frac{q_2} {k_1}+1\right) J_{n-m} ( V_1 / \omega )  \,, \quad
V_{34} =\frac{1} {4} e^{-\frac{i\, L} {2} ( k_1+q_2)}  
\left(-\frac{ g_1 } {f_1} +\frac{g_2} {f_2}\right)
\left(1-\frac{q_2} {k_1}\right) J_{n-m} ( V_1 / \omega )  \,,  \nn
V_{41}  & = \frac{1} {4} e^{\frac{i\, L} {2} ( k_2+q_1)} 
\left(-\frac{ g_1 } {f_1} +\frac{g_2} {f_2}\right) 
\left(1-\frac{q_1} {k_2}\right) J_{n-m} ( V_1 / \omega )  \,,  \quad
V_{42}  =\frac{1} {4} e^{\frac{i\, L} {2} ( k_2-q_1)} 
\left(-\frac{ g_1 } {f_1} +\frac{g_2} {f_2}\right) 
\left(\frac{q_1} {k_2}+1\right) J_{n-m} ( V_1 / \omega )  \,, \nn
V_{43}  & = \frac{1} {4} e^{\frac{i\, L} {2} ( k_2+q_2)}  
\left(\frac{ g_1 } {f_1} +\frac{g_2} {f_2}\right) 
\left(1-\frac{q_2} {k_2}\right) J_{n-m} ( V_1 / \omega )  \,,  \quad
V_{44}  =\frac{1} {4} e^{\frac{i\, L} {2} ( k_2-q_2)} 
 \left(\frac{ g_1 } {f_1} +\frac{g_2} {f_2}\right) 
 \left(\frac{q_2} {k_2}+1\right) J_{n-m} ( V_1 / \omega ) .
\end{align}
Here, $J_{n-m}(V_1/\omega)$ is the Bessel function that couples different Floquet sidebands. $k_1$, $k_2$, $q_1$, and $ q_2$ denote wavevectors, and $f_1$, $ f_2$, $ g_1$, and $ g_2$ stand for spinor-normalisation factors, as defined in the main text. Lastly, $L$ denotes the width of the well.
The relation above can be written for each Floquet sideband as
\begin{align}
\begin{pmatrix}
A_{1,n}^{o}\\
A_{2,n}^{i}\\
B_{1,n}^{o}\\
B_{2,n}^{i}
\end{pmatrix}
=
\sum_{m=-\infty}^{\infty}
\mathcal{S}_{nm}
\begin{pmatrix}
A_{1,m}^{i}\\
A_{2,m}^{o}\\
B_{1,m}^{i}\\
B_{2,m}^{o}
\end{pmatrix} , 
\end{align}
with $\mathcal{S}_{nm}=V U^{-1}$. This global Floquet matrix contains both propagating and evanescent sectors. The observable transport coefficients are obtained from its propagating sub-block, which is recovered by selecting only the modes with real longitudinal wavevectors.

\section{Detailed calculation of the scattering matrix for a double-Weyl mWSM ($J=2$)}\label{appendiX_Weyl_j2}

For a double-Weyl semimetal the complete wavefunction in each region is again a linear combination of four independent solutions: two propagating modes with real longitudinal wavevectors and two evanescent modes with purely imaginary longitudinal wavevectors. The continuity of the two-component spinor wavefunction and its first derivative at the two interfaces provides the boundary conditions.

At the left- and right-interfaces, we have
\begin{align}
P
\begin{pmatrix}
A_1^i\\
A_1^o\\
A_2^i\\
A_2^o
\end{pmatrix}
=
\sum_m
D \begin{pmatrix}
\alpha_1\\
\beta_1\\
\alpha_2\\
\beta_2
\end{pmatrix} \text{ and }
Z
\begin{pmatrix}
B_1^o\\
B_1^i\\
B_2^o\\
B_2^i\end{pmatrix}=
\sum_m X \begin{pmatrix}
\alpha_1\\
\beta_1\\
\alpha_2\\
\beta_2
\end{pmatrix}.
\end{align}
The explicit forms of the $4\times4$ matrices $P,\, D,\, Z,\, X$ are given below. Their entries are obtained directly from the matching of the spinor components and their derivatives:
\begin{align}
& P_1   = e^{- i \,  \frac{k_1 \, L} {2}} \, f_{11}  \, ,\quad P_2  = e^{ i \, \frac{ k_1 \, L} {2}} \, f_{21}  \, ,\quad P_3  = e^{  \frac{\zeta_1 \,L} {2}} \, f_{31}  \, ,\quad 
P_4 = e^{-\,  \frac{\zeta_1 \,L} {2}} \, f_{41} \,, \quad
P_5  = e^{-\, i \, \frac{ k_1 \, L} {2}} \, f_{12}  \, ,\quad 
P_6  = e^{ i \, \frac{ k_1 \, L} {2}} \, f_{22} \,, 
\nn & P_7 = e^{  \frac{\zeta_1 \,L} {2}} \, f_{32}  \, ,\quad 
P_8  = e^{- \,  \frac{ \zeta_1 \,L} {2}} \, f_{42} \,, \quad
 P_9 =  i \, k_1 e^{- \, i \, \frac{ k_1 \, L} {2}} \, f_{11}  \, ,
\quad P_{10}  = - i \, k_1 \, e^{ i \, \frac{ k_1 \, L} {2}} \, f_{21}  \, ,
\quad P_{11} = - \,\zeta_1 \, e^{  \frac{\zeta_1 \,L} {2}} \, f_{31}  \, ,
\nn & P_{12} = \zeta_1 \, e^{- \,   \frac{\zeta_1 \,L} {2}} \, f_{41} \,, \quad 
P_{13} =  i \, k_1 \, e^{- i \, \frac{ k_1 \, L} {2}} \, f_{12}  \, ,
\quad P_{14}  = - i \, k_1 \, e^{ i \, \frac{ k_1 \, L} {2}} \, f_{22}  \, ,\quad 
P_{15}  = -\, \zeta_1 \, e^{ \frac{\zeta_1 \, L} {2}} \, f_{32} \,,  \quad 
P_{16}  = \zeta_1 \, e^{- \,  \frac{\zeta_1 \, L} {2}} \, f_{42}\,,
\end{align}
\begin{align}
& D_1  = e^{- i \, \frac{ q_1 \, L} {2}} \, g_{11}  \, , \quad 
D_2  = e^{ i \, \frac{ q_1 \, L} {2}} \, g_{21}  \, , \quad 
D_3  = e^{  \frac{\zeta_2  \,L} {2}} \, g_{31}  \, , 
\quad D_4  = e^{-  \,  \frac{\zeta_2  \,L} {2}} \, g_{41} \,, \quad
D_5  = e^{- \,i \, \frac{ q_1 \, L} {2}} \, g_{12}  \, , \quad 
D_6  = e^{ i \, \frac{ q_1 \, L} {2}} \, g_{22}  \, , \nn &
 D_7  = e^{  \frac{\zeta_2  \,L} {2}} \, g_{32}  \, , 
 \quad D_8  = e^{- \,  \frac{\zeta_2  \,L} {2}} \, g_{42} \,, \quad
D_9  =  i \, q_1 \,e^{- i \, \frac{ q_1 \, L} {2}} \, g_{11}  \, , \quad 
D_{10}  = - i \, q_1 \, e^{ i \, \frac{ q_1 \, L} {2}} \, g_{21}  \, , \quad 
D_{11}  = - \,\zeta_2 \, e^{  \frac{\zeta_2  \,L} {2}} \, g_{31} \,,  \nn & 
D_{12}  = \zeta_2 \, e^{-  \,  \frac{\zeta_2 \,L} {2}} \, g_{41} \,, \quad
 D_{13}  =  i \, q_1\, e^{- i \, \frac{ q_1 \, L} {2}} \, g_{12}  \, , \quad 
D_{14}  = - i \, q_1 \, e^{ i \, \frac{ q_1 \, L} {2}} \, g_{22}  \, , 
\quad D_{15}  = -\,\zeta_2 \,e^{  \frac{\zeta_2  \,L} {2}} \, g_{32} \,,  
\quad D_{16}  = \zeta_2 \,e^{-\,  \frac{\zeta_2  \,L} {2}} \, g_{42}\,,
\end{align}
\begin{align}
& Z_1  = e^{ i \, \frac{ k_1 \, L} {2}} \, f_{11}  \, , \quad 
Z_2  = e^{- \, i \, \frac{ k_1 \, L} {2}} \, f_{21} \,,  
\quad Z_3  = e^{- \,  \frac{ \zeta_1 \, L} {2}} \, f_{31}  \, , 
\quad Z_4  = e^{  \frac{\zeta_1\, L} {2}} \, f_{41} \,, \quad
Z_5  = e^{ i \, \frac{ k_1 \, L} {2}} \, f_{12}  \, , 
\quad  Z_6  = e^{- i \, \frac{ k_1 \, L} {2}} \, f_{22}  \, , \nn &
Z_7  = e^{- \,  \frac{ \zeta_1 \,L} {2}} \, f_{32}  \, , 
\quad Z_8  = e^{  \frac{ \zeta_1 \, L} {2}} \, f_{42} \,, \quad
Z_9  =  i \, k_1\, e^{ i \, \frac{ k_1 \, L} {2}} \, f_{11}  \, , 
\quad Z_{10}  = - \,i \, k_1 \, e^{-\, i \, \frac{ k_1 \, L} {2}} \, f_{21}  \, , 
\quad Z_{11}  = -\,\zeta_1 \, e^{- \, \frac{\zeta_1 \,L} {2}} \, f_{31}  \, , 
\nn & Z_{12} = \zeta_1 \,e^{\frac{\zeta_1 \,L} {2}} \, f_{41} \,, \quad
Z_{13}  =  i \, k_1 \,e^{ i \, \frac{ k_1 \, L} {2}} \, f_{12} \,,  
\quad Z_{14}  = - i \, k_1 \, e^{- \, i \, \frac{ k_1 \, L} {2}} \, f_{22} \,,  
 \quad Z_{15}  = -\zeta_1 \, e^{- \,  \frac{\zeta_1 \,L} {2}} \, f_{32}  \, , \quad 
 Z_{16}  = \zeta_1 \, e^{\frac{\zeta_1 \,L} {2}} \, f_{42}\,,
\end{align}
\begin{align}
& X_1  = e^{ i \, \frac{ q_1 \, L} {2}} \, g_{11}   \,, \quad
 X_2  = e^{- i \, \frac{ q_1 \, L} {2}} \, g_{21} \,,  \quad 
 X_3  = e^{- \,  \frac{\zeta_2  \, L} {2}} \, g_{31}   \,,  \quad
 X_4  = e^{\zeta_2  \,  \frac{L} {2}} \, g_{41} \,, \quad
X_5  = e^{ i \, \frac{ q_1 \, L} {2}} \, g_{12}   \,,  \quad
X_6  = e^{- \, i \, \frac{ q_1 \, L} {2}} \, g_{22}   \,,  
\nn & X_7  = e^{-\, \frac{\zeta_2  \,L} {2}} \, g_{32}   \,,  \quad
X_8  = e^{ \frac{\zeta_2  \, L} {2}} \, g_{42} \,, 
\quad X_9  =  i \, q_1 \, e^{ i \, \frac{ q_1 \, L} {2}} \, g_{11}   \,,  
\quad X_{10}  = - i \, q_1 \, e^{- i \, \frac{ q_1 \, L} {2}} \, g_{21}   \,,  
\quad X_{11}  = - \, \zeta_2 \, e^{- \,  \frac{\zeta_2  \, L} {2}} \, g_{31}   \,, 
\nn & X_{12}  = \zeta_2 \, e^{\frac{\zeta_2  \, L} {2}} \, g_{41} \,, 
\quad X_{13}  =  i \, q_1 \, e^{ i \, \frac{ q_1 \, L} {2}} \, g_{12}   \,,  
\quad X_{14}  = - i \, q_1 \, e^{- i \, \frac{ q_1 \, L} {2}} \, g_{22}   \,,  
\quad X_{15}  = -\zeta_2 \, e^{-  \, \frac{\zeta_2 \,L} {2}} \, g_{32}   \,, 
\quad X_{16}  = \zeta_2 \, e^{- \,  \frac{\zeta_2 \,L} {2}} \, g_{42}\,.
\end{align}

From the left- and right-interfaces we obtain
\begin{align}
\begin{pmatrix}
A_1^i\\
A_1^o\\
A_2^i\\
A_2^o
\end{pmatrix}
=P^{-1}\,D
\begin{pmatrix}
\alpha_1\\
\beta_1\\
\alpha_2\\
\beta_2
\end{pmatrix} \text{ and }
\begin{pmatrix}
B_1^o\\
B_1^i\\
B_2^o\\
B_2^i
\end{pmatrix}=Z^{-1}\,X
\begin{pmatrix}
\alpha_1\\
\beta_1\\
\alpha_2\\
\beta_2
\end{pmatrix}.
\end{align}
Combining these two expressions gives a linear relation between the amplitudes on the left and right. To separate incoming from outgoing waves, we define the matrices $M=P^{-1}D$ and $N=Z^{-1}X$ (both mapping the internal amplitudes $(\alpha_1,\beta_1,\alpha_2,\beta_2)$ to the boundary amplitudes), and then rearrange rows to form the matrices $U$ and $V$ as follows:
\begin{align}
U=
\begin{pmatrix}
M_{11}&M_{12}&M_{13}&M_{14}\\
M_{41}&M_{42}&M_{43}&M_{44}\\
N_{21}&N_{22}&N_{23}&N_{24}\\
N_{31}&N_{32}&N_{33}&N_{34}
\end{pmatrix} , 
\qquad
V=
\begin{pmatrix}
M_{21}&M_{22}&M_{23}&M_{24}\\
M_{31}&M_{32}&M_{33}&M_{34}\\
N_{11}&N_{12}&N_{13}&N_{14}\\
N_{41}&N_{42}&N_{43}&N_{44}
\end{pmatrix}.
\end{align}
The boundary conditions then take the form
\begin{align}
U \begin{pmatrix}
\alpha_{1,n}\\
\beta_{1,n}\\
\alpha_{2,n}\\
\beta_{2,n}
\end{pmatrix}
= \begin{pmatrix}
A_{1,n}^{i}\\
A_{2,n}^{o}\\
B_{1,n}^{i}\\
B_{2,n}^{o}
\end{pmatrix}
, \qquad
V \begin{pmatrix}
\alpha_{1,n}\\
\beta_{1,n}\\
\alpha_{2,n}\\
\beta_{2,n}
\end{pmatrix}
= \begin{pmatrix}
A_{1,n}^{o}\\
A_{2,n}^{i}\\
B_{1,n}^{o}\\
B_{2,n}^{i}
\end{pmatrix}.
\end{align}
Eliminating the internal amplitudes $(\alpha_{1,n},\beta_{1,n},\alpha_{2,n},\beta_{2,n})$ finally yields the Floquet-scattering relation,
\begin{align}
\begin{pmatrix}
A_{1,n}^{o}\\
A_{2,n}^{i}\\
B_{1,n}^{o}\\
B_{2,n}^{i} \end{pmatrix}
= \sum_m
\mathcal{S}_{nm}
\begin{pmatrix}
A_{1,m}^{i}\\
A_{2,m}^{o}\\
B_{1,m}^{i}\\
B_{2,m}^{o}
\end{pmatrix} , 
\qquad
\mathcal{S}_{nm}=V \,U^{-1}\,.
\end{align}
As before, the propagating block of $\mathcal{S}_{nm}$ gives the physical scattering amplitudes.

\section{Detailed calculation of the scattering matrix for a triple-Weyl mWSM ($J=3$)}\label{appendiX_Weyl_j3}

For a triple-Weyl semimetal the dispersion yields six independent longitudinal solutions: two propagating modes (real wavevectors) and four evanescent modes (two complex-conjugate pairs). The Hamiltonian contains third-order derivatives along the transport direction, so the matching conditions require continuity of the wavefunction and its first and second spatial derivatives at each interface.

At the left- and right-interfaces, we have
\begin{align}
P\begin{pmatrix} A_1^i\\
A_1^o\\
A_2^i\\
A_2^o\\
A_3^i\\
A_3^o
\end{pmatrix}= \sum_m D\begin{pmatrix}
\alpha_1\\
\beta_1\\
\alpha_2\\
\beta_2\\
\alpha_3\\
\beta_3 \end{pmatrix} \text{ and }
Z \begin{pmatrix}
B_1^o\\
B_1^i\\
B_2^o\\
B_2^i\\
B_3^o\\ B_3^i \end{pmatrix}
=\sum_m X \begin{pmatrix}
\alpha_1\\
\beta_1\\
\alpha_2\\
\beta_2\\
\alpha_3\\
\beta_3 \end{pmatrix}.
\end{align}
$P,\, D,\, Z,\, X$ are $6\times6$ matrices constructed from the spinor components and their first and second derivatives. Their explicit entries are listed below:
\begin{align}
& P_1   = e^{- i \, \frac{ k_1 \, L} {2}} \, f_{11} \,,  \quad 
P_2 = e^{ i \, \frac{ k_1 \, L} {2}} \, f_{21} \,,  
\quad P_3 = e^{- i \, \frac{ k_2 \, L} {2}} \, f_{31} \,, \quad  
P_4 = e^{ i \, \frac{ k_2 \, L} {2}} \, f_{41} \,, 
\quad P_5  = e^{- i \, \frac{ k_3 \, L} {2}} \, f_{51} \,,  \quad 
P_6  = e^{ i \, \frac{ k_3 \, L} {2}} \, f_{61} \,, \nn &
P_7  = e^{- i \, \frac{ k_1 \, L} {2}} \, f_{12} \,,  \quad 
P_8  = e^{ i \, \frac{ k_1 \, L} {2}} \, f_{22} \,,  
\quad  P_9  = e^{- i \, \frac{ k_2 \, L} {2}} \, f_{32} \,,  
\quad P_{10}  = e^{ i \, \frac{ k_2 \, L} {2}} \, f_{42} \,,  \quad
 P_{11}  = e^{- i \, \frac{ k_3 \, L} {2}} \, f_{52} \,,  \quad
  P_{12} = e^{ i \, \frac{ k_3 \, L} {2}} \, f_{62} \,, \nn
& P_{13}  =  i \, k_1 \, e^{- i \, \frac{ k_1 \, L} {2}} \, f_{11} \,,  \quad
 P_{14}  = - i \, k_1 \, e^{ i \, \frac{ k_1 \, L} {2}} \, f_{21} \,,   \quad
 P_{15}  =  i \, k_2e^{- i \, \frac{ k_2 \, L} {2}} \, f_{31} \,,  \quad 
 P_{16}  = - i \, k_2e^{ i \, \frac{ k_2 \, L} {2}} \, f_{41} \,,  \quad  P_{17} 
 =  i \, k_3 \, e^{- i \, \frac{ k_3 \, L} {2}} \, f_{51} \,,  \nn &
 P_{18}  = - i \, k_3 \, e^{ i \, \frac{ k_3 \, L} {2}} \, f_{61} \,, \quad
P_{19}  =  i \, k_1 \, e^{- i \, \frac{ k_1 \, L} {2}} \, f_{12} \,,  \quad
 P_{20}  = - i \, k_1 \, e^{ i \, \frac{ k_1 \, L} {2}} \, f_{22} \,,  \quad
  P_{21}  =  i \, k_2e^{- i \, \frac{ k_2 \, L} {2}} \, f_{32} \,,  \quad 
  P_{22} = - i \, k_2e^{ i \, \frac{ k_2 \, L} {2}} \, f_{42} \,,  \nn &
  P_{23}  =  i \, k_3 \, e^{- i \, \frac{ k_3 \, L} {2}} \, f_{52} \,,\quad
  P_{24}  = - i \, k_3 \, e^{ i \, \frac{ k_3 \, L} {2}} \, f_{62} \,, \quad
P_{25}  = -k_1^2\, e^{- i \, \frac{ k_1 \, L} {2}} \, f_{11} \,,  \quad
 P_{26}  = -k_1^2\, e^{ i \, \frac{ k_1 \, L} {2}} \, f_{21} \,,  
\quad P_{27}  = -k_2^2\, e^{- i \, \frac{ k_2 \, L} {2}} \, f_{31} \,,  \nn &
 P_{28}  = -k_2^2\, e^{ i \, \frac{ k_2 \, L} {2}} \, f_{41} \,,  \quad
  P_{29}  = -k_3^2\, e^{- i \, \frac{ k_3 \, L} {2}} \, f_{51} \,,  
  \quad P_{30}  = -k_3^2\, e^{ i \, \frac{ k_3 \, L} {2}} \, f_{61} \,, \quad
P_{31} = -k_1^2\, e^{- i \, \frac{ k_1 \, L} {2}} \, f_{12} \,,  \quad
P_{32} = -k_1^2\, e^{ i \, \frac{ k_1 \, L} {2}} \, f_{22} \,, \nn & 
P_{33}  = -k_2^2\, e^{- i \, \frac{ k_2 \, L} {2}} \, f_{32} \,,  \quad 
P_{34}  = -k_2^2\, e^{ i \, \frac{ k_2 \, L} {2}} \, f_{42} \,,  \quad 
P_{35}  = -k_3^2\, e^{- i \, \frac{ k_3 \, L} {2}} \, f_{52} \,,  \quad
 P_{36}  = -k_3^2\, e^{ i \, \frac{ k_3 \, L} {2}} \, f_{62}\,,
\end{align}
\begin{align}
& D_1  = e^{- i \, \frac{ q_1 \, L} {2}} \, g_{11} \,, \quad
  D_2=e^{ i \, \frac{ q_1 \, L} {2}} \, g_{21} \,,  \quad 
 D_3  = e^{- i \, \frac{ q_2 \, L} {2}} \, g_{31} \,, \quad  D_4 
  = e^{ i \, \frac{ q_2 \, L} {2}} \, g_{41} \,, 
\quad D_5  = e^{- i \, \frac{ q_3 \, L} {2}} \, g_{51} \,,  \quad
   D_6  = e^{ i \, \frac{ q_3 \, L} {2}} \, g_{61} \,, \nn &
D_7  = e^{- i \, \frac{ q_1 \, L} {2}} \, g_{12} \,,  
\quad D_8  = e^{ i \, \frac{ q_1 \, L} {2}} \, g_{22} \,,  \quad 
D_9  = e^{- i \, \frac{ q_2 \, L} {2}} \, g_{32} \,,  
\quad D_{10}  = e^{ i \, \frac{ q_2 \, L} {2}} \, g_{42} \,,  \quad 
D_{11}  = e^{- i \, \frac{ q_3 \, L} {2}} \, g_{52} \,,  \quad
 D_{12}  = e^{ i \, \frac{ q_3 \, L} {2}} \, g_{62} \,, \nn &
D_{13}  =  i \, q_1 \, e^{- i \, \frac{ q_1 \, L} {2}} \, g_{11} \,,  \quad
 D_{14}  = - i \, q_1 \, e^{ i \, \frac{ q_1 \, L} {2}} \, g_{21} \,,  
 \quad D_{15}  =  i \, q_2 \, e^{- i \, \frac{ q_2 \, L} {2}} \, g_{31} \,,  \quad
  D_{16}  = - i \, q_2 \, e^{ i \, \frac{ q_2 \, L} {2}} \, g_{41} \,,  \quad 
  D_{17}  =  i \, q_3 \, e^{- i \, \frac{ q_3 \, L} {2}} \, g_{51} \,,  \nn & 
  D_{18}  = - i \, q_3 \, e^{ i \, \frac{ q_3 \, L} {2}} \, g_{61} \,, \quad
D_{19}  =  i \, q_1 \, e^{- i \, \frac{ q_1 \, L} {2}} \, g_{12} \,,  \quad 
D_{20}  = - i \, q_1 \, e^{ i \, \frac{ q_1 \, L} {2}} \, g_{22} \,,  \quad
 D_{21} =  i \, q_2 \, e^{- i \, \frac{ q_2 \, L} {2}} \, g_{32} \,,  
 \quad D_{22}  = - i \, q_2 \, e^{ i \, \frac{ q_2 \, L} {2}} \, g_{42} \,,  
\nn & D_{23}  =  i \, q_3 \, e^{- i \, \frac{ q_3 \, L} {2}} \, g_{52} \,,  
\quad D_{24}  = - i \, q_3 \, e^{ i \, \frac{ q_3 \, L} {2}} \, g_{62} \,, \quad
D_{25}  = -q_1^2\, e^{- i \, \frac{ q_1 \, L} {2}} \, g_{11} \,,  \quad
 D_{26}  = -q_1^2\, e^{ i \, \frac{ q_1 \, L} {2}} \, g_{21} \,,  \quad
  D_{27}  = -q_2^2\, e^{- i \, \frac{ q_2 \, L} {2}} \, g_{31} \,, \nn &
 D_{28}  = -q_2^2\, e^{ i \, \frac{ q_2 \, L} {2}} \, g_{41} \,,   
 \quad D_{29}  = -q_3^2\, e^{- i \, \frac{ q_3 \, L} {2}} \, g_{51} \,,  \quad D_{30}  = -q_3^2\, e^{ i \, \frac{ q_3 \, L} {2}} \, g_{61} \,, \quad
D_{31}  = -q_1^2\, e^{- i \, \frac{ q_1 \, L} {2}} \, g_{12} \,,  
\quad D_{32}  = -q_1^2\, e^{ i \, \frac{ q_1 \, L} {2}} \, g_{22} \,,  \nn &
 D_{33}  = -q_2^2\, e^{- i \, \frac{ q_2 \, L} {2}} \, g_{32} \,, \quad
  D_{34}  = -q_2^2\, e^{ i \, \frac{ q_2 \, L} {2}} \, g_{42} \,,  
  \quad  D_{35}  = -q_3^2\, e^{- i \, \frac{ q_3 \, L} {2}} \, g_{52} \,,  
  \quad D_{36}  = -q_3^2\, e^{ i \, \frac{ q_3 \, L} {2}} \, g_{62}\,,
\end{align}
\begin{align}
& Z_1  = e^{ i \, \frac{ k_1 \, L} {2}} \, f_{11} \,,  \quad 
Z_2 = e^{- i \, \frac{ k_1 \, L} {2}} \, f_{21} \,,  
\quad Z_3  = e^{ i \, \frac{ k_2 \, L} {2}} \, f_{31} \,, \quad 
Z_4  = e^{- i \, \frac{ k_2 \, L} {2}} \, f_{41} \,,  
\quad Z_5  = e^{ i \, \frac{ k_3 \, L} {2}} \, f_{51} \,,  \quad
 Z_6 = e^{- i \, \frac{ k_3 \, L} {2}} \, f_{61} \,, \nn &
Z_7  = e^{ i \, \frac{ k_1 \, L} {2}} \, f_{12} \,,  \quad 
Z_8  = e^{- i \, \frac{ k_1 \, L} {2}} \, f_{22} \,,  
 Z_9  = e^{ i \, \frac{ k_2 \, L} {2}} \, f_{32} \,,  
\quad Z_{10}  = e^{- i \, \frac{ k_2 \, L} {2}} \, f_{42} \,, \quad
 Z_{11}  = e^{ i \, \frac{ k_3 \, L} {2}} \, f_{52} \,,  
 \quad Z_{12}  = e^{- i \, \frac{ k_3 \, L} {2}} \, f_{62} \,, \nn &
Z_{13}  =  i \, k_1 \, e^{ i \, \frac{ k_1 \, L} {2}} \, f_{11} \,,  \quad
Z_{14}  = - i \, k_1 \, e^{- i \, \frac{ k_1 \, L} {2}} \, f_{21} \,,  \quad
 Z_{15}  =  i \, k_2e^{ i \, \frac{ k_2 \, L} {2}} \, f_{31} \,,   \quad
 Z_{16}  = - i \, k_2e^{- i \, \frac{ k_2 \, L} {2}} \, f_{41} \,,  \quad
  Z_{17}  =  i \, k_3 \, e^{ i \, \frac{ k_3 \, L} {2}} \, f_{51} \,,  \nn  &
  Z_{18}  = - i \, k_3 \, e^{- i \, \frac{ k_3 \, L} {2}} \, f_{61} \,, \quad
Z_{19}  =  i \, k_1 \, e^{ i \, \frac{ k_1 \, L} {2}} \, f_{12} \,,  
\quad Z_{20}  = - i \, k_1 \, e^{- i \, \frac{ k_1 \, L} {2}} \, f_{22} \,,  \quad
 Z_{21} =  i \, k_2 \,e^{ i \, \frac{ k_2 \, L} {2}} \, f_{32} \,,  
 \quad Z_{22} = - i \, k_2e^{- i \, \frac{ k_2 \, L} {2}} \, f_{42} \,, \nn &  
 Z_{23}  =  i \, k_3 \, e^{ i \, \frac{ k_3 \, L} {2}} \, f_{52} \,, \quad
Z_{24} = - i \, k_3 \, e^{- i \, \frac{ k_3 \, L} {2}} \, f_{62} \,, \quad
Z_{25} = -k_1^2\, e^{ i \, \frac{ k_1 \, L} {2}} \, f_{11} \,,  \quad
Z_{26}  = -k_1^2\, e^{- i \, \frac{ k_1 \, L} {2}} \, f_{21} \,,  
\quad Z_{27}  = -k_2^2\, e^{ i \, \frac{ k_2 \, L} {2}} \, f_{31} \,,  
\nn &  Z_{28} = -k_2^2\, e^{- i \, \frac{ k_2 \, L} {2}} \, f_{41} \,,  
\quad Z_{29}  = -k_3^2\, e^{ i \, \frac{ k_3 \, L} {2}} \, f_{51} \,,  \quad
 Z_{30}  = -k_3^2\, e^{- i \, \frac{ k_3 \, L} {2}} \, f_{61} \,, \quad
Z_{31}  = -k_1^2\, e^{ i \, \frac{ k_1 \, L} {2}} \, f_{12} \,,  \quad
 Z_{32}  = -k_1^2\, e^{- i \, \frac{ k_1 \, L} {2}} \, f_{22} \,,  
\nn &  Z_{33} = -k_2^2\, e^{ i \, \frac{ k_2 \, L} {2}} \, f_{32} \,, \quad
  Z_{34}  = -k_2^2\, e^{- i \, \frac{ k_2 \, L} {2}} \, f_{42} \,, 
  \quad Z_{35}  = -k_3^2\, e^{ i \, \frac{ k_3 \, L} {2}} \, f_{52} \,,  \quad
 Z_{36} = -k_3^2\, e^{- i \, \frac{ k_3 \, L} {2}} \, f_{62}\,,
\end{align}
\begin{align}
& X_1  = e^{ i \, \frac{ q_1 \, L} {2}} \, g_{11}   \,,  \quad 
X_2  = e^{- i \, \frac{ q_1 \, L} {2}} \, g_{21}   \,,  \quad X_3  = e^{ i \, \frac{ q_2 \, L} {2}} \, g_{31}   \,,  \quad X_4  = e^{- i \, \frac{ q_2 \, L} {2}} \, g_{41}   \,,  \quad X_5  = e^{ i \, \frac{ q_3 \, L} {2}} \, g_{51}   \,,  \quad X_6  = e^{- i \, \frac{ q_3 \, L} {2}} \, g_{61} \,, \nn
& X_7  = e^{ i \, \frac{ q_1 \, L} {2}} \, g_{12}   \,,  \quad 
X_8  = e^{- i \, \frac{ q_1 \, L} {2}} \, g_{22}   \,,  \quad X_9  = e^{ i \, \frac{ q_2 \, L} {2}} \, g_{32}   \,,  \quad 
X_{10}  = e^{- i \, \frac{ q_2 \, L} {2}} \, g_{42}   \,,  
\quad X_{11}  = e^{ i \, \frac{ q_3 \, L} {2}} \, g_{52}   \,,  \quad
 X_{12}  = e^{- i \, \frac{ q_3 \, L} {2}} \, g_{62} \,, \nn &
X_{13}  =  i \, q_1 \, e^{ i \, \frac{ q_1 \, L} {2}} \, g_{11}   \,,  
\quad X_{14}  = - i \, q_1 \, e^{- i \, \frac{ q_1 \, L} {2}} \, g_{21}   \,,  
\quad X_{15}  =  i \, q_2 \, e^{ i \, \frac{ q_2 \, L} {2}} \, g_{31}   \,,  
\quad X_{16}  = - i \, q_2 \, e^{- i \, \frac{ q_2 \, L} {2}} \, g_{41}   \,,  \quad X_{17} 
=  i \, q_3 \, e^{ i \, \frac{ q_3 \, L} {2}} \, g_{51}   \,,  \nn &
 X_{18}  = - i \, q_3 \, e^{- i \, \frac{ q_3 \, L} {2}} \, g_{61} \,,  \quad
X_{19} =  i \, q_1 \, e^{ i \, \frac{ q_1 \, L} {2}} \, g_{12}   \,,  
\quad X_{20}  = - \,i \, q_1 \, e^{- i \, \frac{ q_1 \, L} {2}} \, g_{22}   \,,  
\quad X_{21}  =  i \, q_2 \, e^{ i \, \frac{ q_2 \, L} {2}} \, g_{32}   \,,  
\quad X_{22}  = - \,i \, q_2 \, e^{- i \, \frac{ q_2 \, L} {2}} \, g_{42}   \,,  \nn & X_{23} =  i \, q_3 \, e^{ i \, \frac{ q_3 \, L} {2}} \, g_{52}   \,,  
\quad X_{24}  = - i \, q_3 \, e^{- i \, \frac{ q_3 \, L} {2}} \, g_{62} \,, \quad
X_{25}  = - \,q_1^2\, e^{ i \, \frac{ q_1 \, L} {2}} \, g_{11}   \,,  
\quad X_{26}  = - \,q_1^2\, e^{- \, i \, \frac{ q_1 \, L} {2}} \, g_{21}   \,,  \quad 
X_{27}  = -\, q_2^2\, e^{ i \, \frac{ q_2 \, L} {2}} \, g_{31}   \,, \nn &
 X_{28}  = -\, q_2^2\, e^{- i \, \frac{ q_2 \, L} {2}} \, g_{41}   \,,  
 \quad X_{29}  = - \,q_3^2\, e^{ i \, \frac{ q_3 \, L} {2}} \, g_{51}   \,,  
 \quad X_{30}  = - \,q_3^2\, e^{- i \, \frac{ q_3 \, L} {2}} \, g_{61} \,, \quad 
 X_{31}  = -\, q_1^2\, e^{ i \, \frac{ q_1 \, L} {2}} \, g_{12}   \,,  
\quad X_{32}  = - \,q_1^2\, e^{- i \, \frac{ q_1 \, L} {2}} \, g_{22}   \,,  
\nn & X_{33}  = - \,q_2^2\, e^{ i \, \frac{ q_2 \, L} {2}} \, g_{32}   \,,  
\quad X_{34}  = -\, q_2^2\, e^{- i \, \frac{ q_2 \, L} {2}} \, g_{42}   \,,  
\quad X_{35}  = - \,q_3^2\, e^{ i \, \frac{ q_3 \, L} {2}} \, g_{52}   \,,  \quad 
X_{36}  = -\, q_3^2\, e^{- i \, \frac{ q_3 \, L} {2}} \, g_{62} \,.
\end{align}

From the left- and right-interfaces, we obtain
\begin{align}
\begin{pmatrix}
\alpha_1\\ \beta_1\\ \alpha_2\\ \beta_2 \\ \alpha_3\\ \beta_3
\end{pmatrix}
= D^{-1} \, P
\begin{pmatrix}
A_1^i\\ A_1^o\\ A_2^i\\ A_2^o\\ A_3^i\\ A_3^o
\end{pmatrix} \text{ and }
\begin{pmatrix}
B_1^o\\B_1^i\\ B_2^o\\ B_2^i\\ B_3^o\\ B_3^i
\end{pmatrix}
= Z^{-1} \, X
\begin{pmatrix}
\alpha_1\\ \beta_1\\ \alpha_2\\ \beta_2\\ \alpha_3\\ \beta_3
\end{pmatrix}.
\end{align}
Defining $M=D^{-1} \, P$ and $N=Z^{-1} \, X$, and composing them as $K \equiv NM$, the two interface conditions combine into a single $6\times6$ relation connecting the full amplitude vectors of regions I and III directly:
\begin{align}
\begin{pmatrix}
B_1^o\\B_1^i\\B_2^o\\B_2^i\\B_3^o\\B_3^i
\end{pmatrix}
= K
\begin{pmatrix}
A_1^i\\A_1^o\\A_2^i\\A_2^o\\A_3^i\\A_3^o
\end{pmatrix}.
\end{align}
Unlike the nodal-ring semimetal and the double-Weyl mWSM, here $M$ and $N$ act on different vector spaces ($M$ maps region-I amplitudes to internal amplitudes, while $N$ maps internal amplitudes to region-III amplitudes), so the Floquet-scattering matrix cannot be obtained by simply rearranging rows of $M$ and $N$; instead, the physical amplitudes must be separated directly from $K$. Partitioning the twelve amplitudes into the six that are physically prescribed (`input') and the six that are fixed by scattering (`output'),
\begin{align}
A_{\rm in} \equiv \begin{pmatrix} A_1^i\\A_2^o\\A_3^o\end{pmatrix}, \quad
A_{\rm out} \equiv \begin{pmatrix} A_1^o\\A_2^i\\A_3^i\end{pmatrix}, \quad
B_{\rm in} \equiv \begin{pmatrix} B_1^i\\B_2^o\\B_3^o\end{pmatrix}, \quad
B_{\rm out} \equiv \begin{pmatrix} B_1^o\\B_2^i\\B_3^i\end{pmatrix},
\end{align}
we write the relation $B\text{-vec}=K\,A\text{-vec}$ in block form as
\begin{align}
B_{\rm in} = K_{II}\,A_{\rm in} + K_{IO}\,A_{\rm out}\,, \qquad
B_{\rm out} = K_{OI}\,A_{\rm in} + K_{OO}\,A_{\rm out}\,,
\end{align}
where $K_{II},K_{IO},K_{OI},K_{OO}$ are the $3\times3$ blocks of $K$ obtained by selecting the rows and columns corresponding to $A_{\rm in}, A_{\rm out}, B_{\rm in}, B_{\rm out}$. Since $B_{\rm in}$ is itself an independently prescribed input, the first relation can be solved for the undetermined amplitude,
\begin{align}
A_{\rm out} = K_{IO}^{-1}\left(B_{\rm in} - K_{II}\, A_{\rm in}\right)\,,
\end{align}
and substituting into the second relation expresses $B_{\rm out}$ in terms of the physical inputs alone. Collecting both results gives the Floquet-scattering relation
\begin{align}
\begin{pmatrix}
A_{1,n}^{o}\\
A_{2,n}^{i}\\
A_{3,n}^{i}\\
B_{1,n}^{o}\\
B_{2,n}^{i}\\
B_{3,n}^{i} \end{pmatrix}
= \sum_m
\mathcal{S}_{nm}
\begin{pmatrix}
A_{1,m}^{i}\\
A_{2,m}^{o}\\
A_{3,m}^{o}\\
B_{1,m}^{i}\\
B_{2,m}^{o}\\
B_{3,m}^{o}
\end{pmatrix} , 
\qquad
\mathcal{S}_{nm}=
\begin{pmatrix}
-K_{IO}^{-1}K_{II} & K_{IO}^{-1}\\[2pt]
K_{OI}-K_{OO}K_{IO}^{-1}K_{II} & K_{OO}K_{IO}^{-1}
\end{pmatrix}.
\end{align}
The propagating sub-block of $\mathcal{S}_{nm}$ provides the physical transmission and reflection amplitudes, while the evanescent sectors remain confined near the interfaces and do not contribute to the asymptotic transport observables.

\bibliography{ref-sb}
\end{document}